\documentclass[graybox]{svmult}

\usepackage{mathptmx}       
\usepackage{helvet}         
\usepackage{courier}        
\usepackage{type1cm}        
\usepackage{makeidx}         
\usepackage{graphicx}        
\usepackage{multicol}        
\usepackage[bottom]{footmisc}

\makeindex

\newcommand\apj{ApJ}
\newcommand\apjl{ApJL}
 
\newcommand\mnras{MNRAS} 
\newcommand\aap{A\&A} 
 
\newcommand\nat{Nature}

\begin{document}

\title*{Recent developments in planet migration theory}
\author{Cl{\'e}ment Baruteau and Fr{\'e}d{\'e}ric Masset}
\institute{Cl{\'e}ment Baruteau \at DAMTP, University of Cambridge, Wilberforce Road, Cambridge CB30WA, United Kingdom, \email{C.Baruteau@damtp.cam.ac.uk}
\and Fr{\'e}d{\'e}ric Masset \at Instituto de Ciencias F{\'i}sicas, Universidad Nacional
  Aut{\'o}noma de M{\'e}xico (UNAM), Apdo. Postal 48-3, 62251-Cuernavaca, Morelos, M{\'e}xico \email{masset@fis.unam.mx}}

\maketitle

\abstract{Planetary migration is the process by which a forming planet
  undergoes a drift of its semi-major axis caused by the tidal
  interaction with its parent protoplanetary disc. One of the key
  quantities to assess the migration of embedded planets is the tidal
  torque between the disc and planet, which has two components: the
  Lindblad torque and the corotation torque. We review the latest
  results on both torque components for planets on circular orbits,
  with a special emphasis on the various processes that give rise to
  additional, large components of the corotation torque, and those
  contributing to the saturation of this torque. These additional
  components of the corotation torque could help address the
  shortcomings that have recently been exposed by models of planet
  population syntheses. We also review recent results concerning the
  migration of giant planets that carve gaps in the disc (type~II
  migration) and the migration of sub-giant planets that open partial
  gaps in massive discs (type~III migration).}

\section{Introduction}
\label{sec:introduction}
The extraordinary diversity of extrasolar planetary systems has
challenged our understanding of how planets form and how their orbits
evolve as they form. Among the many processes contemplated thus far to
account for the observed properties of extrasolar planets, the
gravitational interaction between planets and their parent
protoplanetary disc plays a prominent role. Considered for a long time
as the key ingredient in shaping planetary systems, planet--disc
interactions, which drive the well-known planetary migration (a drift
of a planet's semi-major axis during the lifetime of the gaseous
disc), have recently been considered by many as being
over-emphasized. On the one hand, observational data show evidence for
vigorous migration in many planetary systems, as stressed by the
existence of hot Jupiters, Neptunes and Super-Earths (the recently
discovered Kepler-20 planetary system, with coplanar rocky and icy
planets alternating at periods less than 80 days \cite{Kepler20},
provides a good example), or by the existence of many mean-motion
resonances. Yet, there is also compelling evidence that other
processes are capable of altering the orbits as dramatically as
planet--disc interactions (the existence of highly-eccentric or
retrograde planets is an example). Also, although many systems seem to
have undergone orbital migration, many others display planets at
distances from their star that are of same order of magnitude as the
distances of the planets in our Solar System to the Sun. One may be
tempted to conclude from this that theories of planet--disc
interactions are overrated, to the point that they could be merely
ignored in scenarios of formation of planetary systems. Yet, as will
be detailed in the following sections, the following facts are
difficult to circumvent:
\begin{itemize}
\item Each component of the torque exerted by the disc on a planet is
  so large that it can half or double the planet's semi-major axis in
  a time that is usually two or more orders of magnitude shorter than
  the lifetime of protoplanetary discs.
\item These torque components do not cancel out.  The residual torque
  amounts to a fair fraction of each torque component, so that one
  should in general expect that planet--disc interactions have a
  strong effect on planets orbits over the disc lifetime.
\end{itemize}

The central difficulty in planetary migration theories lies precisely
in predicting the residual torque value. In addition to being a fair
difference between several large amplitude torques, it is very
sensitive to the disc properties near the planet's orbit (e.g.,
density, temperature profiles). This by no means implies that the
total torque is negligible, but it helps understand why migration
theories are slowly maturing.

One of the main purposes of this review is to provide the reader with
an up-to-date presentation of the state of planet--disc interactions,
with emphasis on the torque formulae that govern the migration of
low-mass planets. The reasons for this special emphasis are
three-fold:
\begin{itemize}
\item Low-mass planets are the most critical in planetary population
  synthesis, as they potentially undergo the fastest migration.
\item The sensitivity of detection methods has increased to a point
  where we can find a plethora of Neptune-sized planets or below,
  which exclusively underwent the migration processes typical of
  low-mass planets during their formation.
\item The subject has been the focus of significant efforts in the
  recent past.
\end{itemize}

The torque acting on a low-mass planet in circular orbit can be
decomposed into two components: (i) the differential Lindblad torque,
arising from material passing by the planet at supersonic velocities,
which is deflected by the latter and therefore exchanges angular
momentum and energy with the planet, and (ii) the corotation torque,
arising from material slowly drifting with respect to the planet, in
the vicinity of its orbit. The differential Lindblad torque has been
extensively studied from the early times of planetary migration
theories, and is known in much greater detail than the corotation
torque.  In fact the corotation region, which has been under intense
scrutiny over the last five years, has proved to have a much more
complex dynamics than previously thought.  In particular, the value of
the corotation torque depends sensitively on the radiative properties
of the gas disc, and may exhibit large values when the gas is
radiatively inefficient (as is generally expected in regions of planet
formation). In addition to this complexity, new problems emerge as the
computational resources render tractable the task of simulating a
planet embedded in realistic discs, namely three-dimensional discs
invaded by turbulence.

This review is organised as follows. After a brief description of the
physical model and notations in \S~\ref{sec:governing-equations}, we
present in \S~\ref{sec:type1} the migration of low-mass planets
(type~I migration). We detail some recent results on the differential
Lindblad torque in \S~\ref{sec:diff-lindbl-torq}, and we put special
emphasis on the recent developments on the corotation torque in
\S~\ref{sec:corotation-torque}. The migration of gap-opening planets
is then examined in \S~\ref{sec:gapopening}, with type III migration
in \S~\ref{sec:type3}, followed by type II migration in
\S~\ref{sec:type2}. Finally, in section \S~\ref{sec:applications}, we
discuss some recent themes related to planet--disc interactions, such
as the discovery of massive planets at large orbital separations, and
recent models of planetary population syntheses. Most sections end
with a brief summary of their content.

\section{Physical model and notations}
\label{sec:governing-equations}
In most of the following we shall consider two-dimensional discs,
considering vertically averaged or vertically integrated quantities
where appropriate. At the present time, most of the recent
investigation on the migration of low-mass planets has been undertaken
in two dimensions (with a list of exceptions that includes, but is not
restricted to
\cite{tanaka2002,gda2003b,gda2003,bate03,zl06,pm06,pm08,Kley09,AyliffeBate11}),
and much insight can be gained into the mechanisms of the different
components of the torque exerted by the disc on the planet through a
two-dimensional analysis. It should be remembered, however, that
two-dimensional results are plagued by the unavoidable use of a
softening length for the planet's gravitational potential, which fits
a two-fold purpose:
\begin{itemize}
\item It mimics the effects of the finite thickness of a true
  disc, by lowering the magnitude of the planet's potential well.
\item In numerical simulations, it avoids the potential divergence at
  the scale of the mesh zone.
\end{itemize}
For this reason, the reader should bear in mind that the ultimate
torque expressions should be sought by means of three-dimensional
calculations, and that two-dimensional calculations are only used in a
first step to elucidate the mechanisms that contribute to the torque.
At the time of writing this manuscript, most of the physics of the
torque in two dimensions is fairly well understood, which is why we
put special emphasis on the two-dimensional analysis.

We consider a planet of mass $M_p$ orbiting a star of mass $M_{\star}$
with orbital frequency $\Omega_p$. We denote by $q$ the
planet-to-primary mass ratio. The planet is assumed to be on a
prograde circular orbit of semi-major axis $a$, coplanar with the
disc, so that we do not consider in this work eccentric, inclined, or
retrograde planets.  The protoplanetary disc in which the planet is
embedded is modelled as a two-dimensional viscous disc in radial
equilibrium, with the centrifugal acceleration and the radial
acceleration related to the pressure gradient balancing the
gravitational acceleration due to the central star. We use $P$ to
denote the vertically integrated pressure, and $s$ to denote (a
measure of) the gas entropy, which we express as
\begin{equation}
 \label{eq:1}
 s = \frac{P}{\Sigma^\gamma},
\end{equation}
where $\Sigma$ is the surface density of the gas and $\gamma$ the
ratio of specific heats $C_p/C_v$. We denote with $T$ the vertically
averaged temperature. We assume in most of what follows that the
surface density and temperature profiles are power laws of radius,
with indices $\alpha$ and $\beta$, respectively:
\begin{equation}
  \label{eq:2}
  \Sigma\propto r^{-\alpha}
\end{equation}
and
\begin{equation}
  \label{eq:3}
  T\propto r^{-\beta}.
\end{equation}
The disc pressure scale length is $H=c_s/\Omega$, with $\Omega$ the
gas orbital frequency and $c_s$ the sound speed. We define the disc
aspect ratio by $h=H/r$. When $T$ is a power law of radius, so is $h$,
with an index $f$ dubbed the flaring index:
\begin{equation}
  \label{eq:4}
  h\propto r^f,
\end{equation}
which satisfies $\beta = 1-2f$. In almost all studies of planet--disc
interactions, the disc is modelled with a stationary kinematic
viscosity $\nu$, aimed at modelling the disc's turbulent
properties. We will consider that $\nu$ can be written as $\nu =
\alpha_{\rm v} H^2 \Omega$ \cite{ss73}, with $\alpha_{\rm v}$ denoting
the alpha viscous parameter associated with the turbulent stresses in
the disc.

We will also make use of Oort's constants. The first Oort's constant
scales with the shear of the flow,
\begin{equation}
  \label{eq:6}
  A=\frac12 r\frac{d\Omega}{dr},
\end{equation}
while the second Oort's constant scales with the vertical component of
the vorticity of the flow $\omega_z$:
\begin{equation}
  \label{eq:7}
  B = \frac{1}{2r}\frac{d(r^2\Omega)}{d r}=\frac{\omega_z}{2}.
\end{equation}
Whenever used, $A$ and $B$ are implicitly meant to be evaluated at the
planet's orbital radius.

The governing equations of the flow are the equation of continuity,
the Navier-Stokes equations and the energy equation (except when
dealing with isothermal discs, as specified below), together with the
closure relationship provided by the equation of state, which is that
of an ideal gas. We do not reproduce these governing equations here,
but refer the interested reader to, for example,
\cite{mc09,cm09,pbck10,crida08,gda2003,gda2002}.

\section{Migration of low-mass embedded planets: Type I migration}
\label{sec:type1}
Up until recently, type~I migration referred to the regime of
migration of low-mass planets that could be tackled through a linear
analysis \cite{KP93, tanaka2002}. Recently, however, \cite{pp09a} have
shown that one of the torque components, namely the corotation torque,
can become non-linear at all planetary masses, provided the disc
viscosity is sufficiently small. We shall nevertheless still qualify
type~I as the migration of low-mass planets, up to a threshold mass
that we shall specify in \S~\ref{sec:gap}. We entertain below the two
components of the tidal torque: the differential Lindblad torque and
the corotation torque. In the following, we address the properties of
type I migration by a direct inspection of the tidal torque $\Gamma$,
where the planet migration rate $\dot{a}$ is given by
\begin{equation}
2Ba\dot{a}M_p = \Gamma.
\label{eq:adot}
\end{equation}

\subsection{Differential Lindblad torque}
\label{sec:diff-lindbl-torq}

\begin{figure}
\sidecaption
 	\includegraphics[width=0.5\hsize]{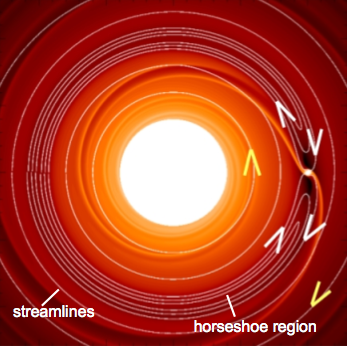}
        \caption{Disc's surface density perturbed by a low-mass
          planet.  Streamlines in the frame corotating with the planet
          are over plotted by solid curves.  The set of streamlines
          that librate with respect to the planet delimits the
          planet's horseshoe region.}
  \label{fig:hs}
\end{figure}
The differential Lindblad torque accounts for the exchange of angular
momentum between the planet and the trailing density waves (spiral
wakes) that it generates in the disc (see illustration in
Fig.~\ref{fig:hs}). The density waves propagating inside the planet's
orbit carry away negative angular momentum, and thus exert a positive
torque on the planet, named the inner Lindblad torque.  Similarly, the
spiral density waves propagating outside the planet's orbit carry away
positive angular momentum, which corresponds to a negative torque on
the planet (the outer Lindblad torque).  The angular momentum of a
planet on a circular orbit scales with the square root of its
semi-major axis. The inner Lindblad torque thus tends to make the
planet move outwards, while the outer Lindblad torque tends to make it
move inwards. The residual torque, called differential Lindblad
torque, results from a balance between the inner and outer torques. In
the absence of viscosity, the angular momentum taken away by the wakes
is conserved until wave breaking occurs, resulting in the formation of
a shock and the deposition of the wave's energy and angular momentum
to the disc \cite{gr2001}. The impact of non-linearities induced by
the planet's wakes, which in particular lead to the formation of a gap
about the planet's orbit, will be described in Section~\ref{sec:gap}.

The one-sided and differential Lindblad torques can be evaluated 
in different manners:
\begin{enumerate}
\item In a fully analytic manner upon linearization of the flow
  equations \cite{gt79}. In this framework, waves propagate away from
  Lindblad resonances with the planet \cite{ww86}, and they
  constructively interfere into a one-armed spiral pattern
  \cite{og2002}, which begins where the Keplerian flow is supersonic
  with respect to the planet \cite{gr2001}.  One-sided Lindblad
  torques are then evaluated as the sum of the torques arising at each
  Lindblad resonance. The locations of Lindblad resonances are shifted
  with respect to their nominal location (given by the condition of
  mean-motion resonance with a test particle) by pressure effects. In
  particular, resonances with high azimuthal wavenumber have
  accumulation points at $\pm 2H/3$ from the planetary orbit, instead
  of accumulation at the orbit. This provides a torque cutoff
  \cite{gt80, arty93a}, which can only be evaluated approximately
  \cite{arty93b}. This renders fully analytic methods of Lindblad
  torque calculations only approximate.
\item The torque can be evaluated by solving numerically the
  linearised equations of the flow. This approach was initially
  undertaken by \cite{KP93}, and recently revisited by \cite{pp08} and
  \cite{pbck10}.
\item An intermediate approach may be used, in which one solves
  numerically linear equations obtained by an expansion of the flow
  equations in $H/r$, where $H$ is the disc thickness
  \cite{tanaka2002}.
\end{enumerate}

\begin{figure}
\centering
 	\includegraphics[width=0.8\hsize]{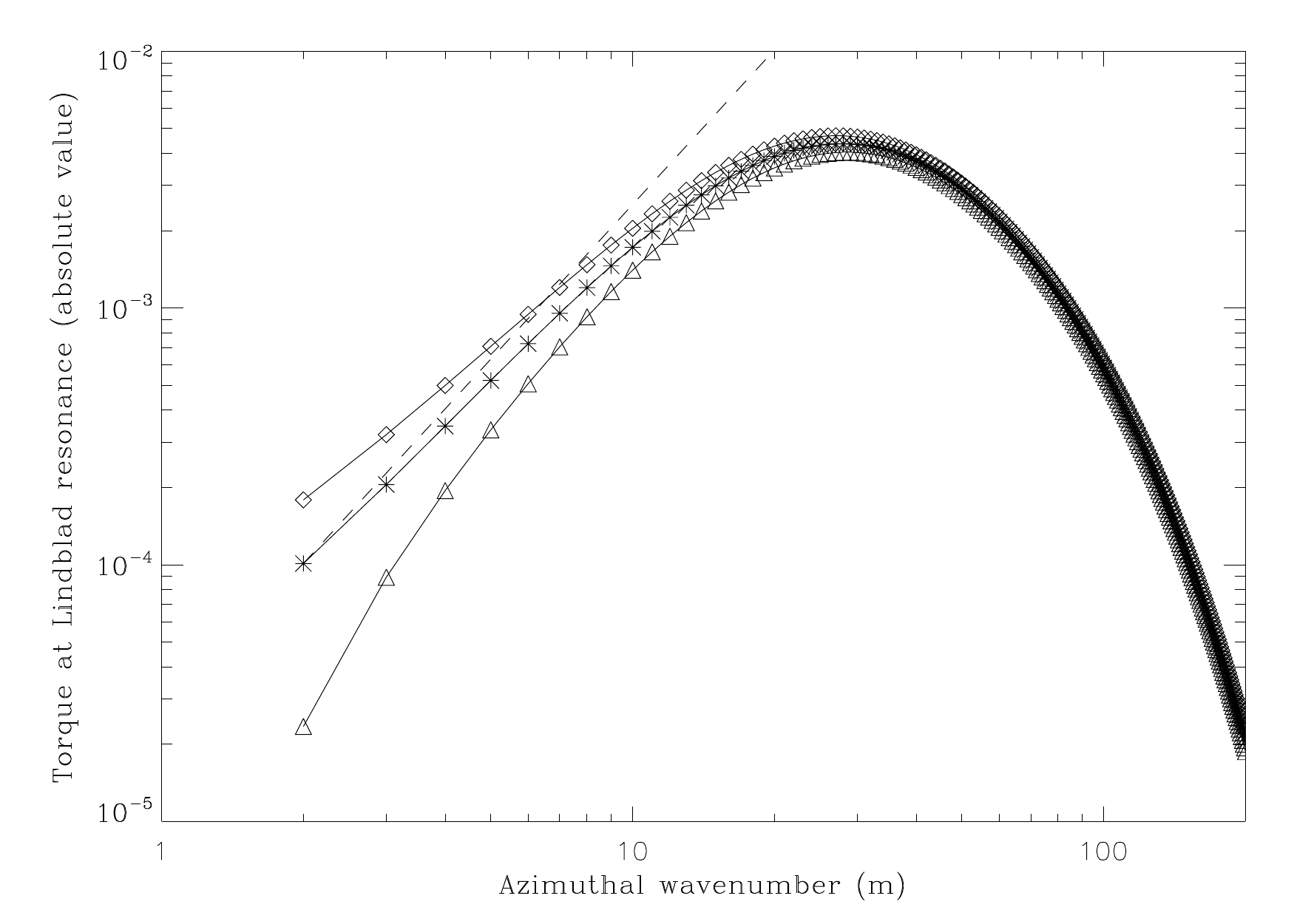}
        \caption{Torques at individual outer Lindblad resonances
          (diamonds) and inner Lindblad resonances (triangles), in
          absolute value. Results are obtained for a disc with aspect
          ratio $H/r=0.02$, and with a softening length
          $\epsilon=0.5H$ for the planet potential.  Torques are
          normalised to the torque value given by
          Eq.~(\ref{eq:8}). Stars show the average value of the inner
          and outer Lindblad torques at a given azimuthal wavenumber
          $m$, and the dashed line illustrates its $m^2$ dependence at
          small $m$.}
\label{fig:onesidetqs}
\end{figure}
Fig.~\ref{fig:onesidetqs} illustrates a number of properties of the
Lindblad torque that provide some insight into its scaling with the
disc and planet parameters. In this figure, the torque value has been
obtained through the use of analytical formulae similar to that of
\cite{w97} (Eqs.~3 to~7 therein), except for the introduction of a
softening length for the planet's potential, and for the minor
correction consisting of the introduction of a factor $\Omega/\kappa$
in the forcing terms (see \cite{mg2004}, after their Eq.~13; $\kappa$
denotes the horizontal epicyclic frequency). This figure shows that
the torque undergoes a sharp cutoff past a peak value, which is found
to be of order $m_{\rm max}\sim (2/3) (r/H)$. Also, the dashed line
shows that, up to the cutoff, the one-sided Lindblad torque
approximately scales with $m^2$ -- as expected from the WKB analysis
of \cite{gt80} -- from which we infer the one-sided Lindblad torque,
summed over $m$, to scale approximately as $m_{\rm max}^3$, {\em i.\
  e.} as $(r/H)^3$. Besides, the torque naturally scales with the
disc's surface density and with the square of the planet mass, and
dimensional arguments further imply that it ought to scale as:
\begin{equation}
  \label{eq:8}
  \Gamma_0 = \Sigma q^2\Omega^2 a^4/h^3,
\end{equation}
which is indeed the scaling of the one-sided Lindblad torque
\cite{w97}.

Fig.~\ref{fig:onesidetqs} also shows that there exists an asymmetry
between the outer and inner torques, the former being systematically
larger than the latter. The reasons for this asymmetry are examined in
depth in \cite{w97}. The relative asymmetry is found to scale with the
disc thickness: in thinner discs, the relative asymmetry of the inner
and outer torque is smaller (which can be understood as due to the
accumulation points of the inner and outer resonances lying closer to
the orbit). As a consequence, the differential Lindblad torque scales
with:
\begin{equation}
  \label{eq:9}
  \Gamma_{\rm ref} = \Sigma q^2\Omega^2 a^4/h^2.
\end{equation}
The asymmetry between the inner and outer torques also depends 
upon:
\begin{itemize}
\item The temperature gradient, since it affects the location of
  Lindblad resonances, and therefore the magnitude of the forcing
  potential at a given resonance \cite{ww86}. For instance, a steeper
  (decreasing) temperature profile decreases the disc's angular
  frequency (by increasing the magnitude of the radial pressure
  gradient), which shifts all Lindblad resonances inwards. Outer
  resonances get closer to the planet orbit, which strengthens the
  outer torque, whereas inner Lindblad resonances are shifted away
  from the planet orbit, which decreases the inner torque. The net
  effect of a steeper temperature gradient is therefore to make the
  differential Lindblad torque a more negative quantity. In the same
  vein, a shallower temperature gradient would shift all Lindblad
  resonances outwards, making the differential Lindblad torque a more
  positive quantity.  The differential Lindblad torque may become
  positive for positive temperature gradients \cite{w97}.  \smallskip
\item The surface density gradient, which affects the location of
  Lindblad resonances in the exact same way as the temperature
  gradient, but now also because the torque at a given Lindblad resonance
  directly scales with the underlying surface density \cite{ww86}. A
  steeper (decreasing) density profile naturally increases the
  magnitude of the inner torque compared to that of the outer torque,
  but this effect is mostly compensated for by an inward shift of all
  Lindblad resonances (just like when steepening the temperature
  profile, as described above).  It implies that the differential
  Lindblad torque is quite insensitive to the density gradient 
  near the planet location.
  \smallskip
\item The disc's self-gravity, which also impacts the location of
  Lindblad resonances, essentially by changing the wave's dispersion
  relation \cite{ph05, bm08b}.
\end{itemize}

An accurate determination of the asymmetry between the inner and outer
torques yields a dimensionless factor to be put in front of
$\Gamma_{\rm ref}$ to give the expression for the differential
Lindblad torque. This issue has triggered a lot of theoretical efforts
in the last three decades, and it is not completely solved yet. As of
the writing of this review, the two main results are:
\begin{itemize}
\item An expression obtained by solving numerically the linearised
  equations of the flow in two dimensions with a softened planet
  potential, in discs with arbitrary gradients of surface density and
  temperature \cite{pp08,pbck10}. It takes the form:
  \begin{equation}
    \label{eq:10}
    \frac{\Gamma_L}{\Gamma_{\rm ref}}=-(2.5+1.7\beta-0.1\alpha)\left(\frac{0.4}{\epsilon/H}\right)^{0.71},
  \end{equation}
  with $\alpha$ and $\beta$ defined in Eqs.~(\ref{eq:2})
  and~(\ref{eq:3}).  The expression in Eq.~(\ref{eq:10}) is most
  accurate for softening lengths $\epsilon \sim 0.4H$.
\item An expression obtained by a semi-analytic method for globally
  isothermal, three-dimensional discs with arbitrary gradients of
  surface density \cite{tanaka2002}:
  \begin{equation}
    \label{eq:11}
    \frac{\Gamma_L}{\Gamma_{\rm ref}}=-(2.34-0.10\alpha).
  \end{equation}
\end{itemize}
We note that Eqs.~(\ref{eq:10}) and~(\ref{eq:11}) exhibit a similar
behaviour (for the case $\beta=0$, exclusively contemplated by
\cite{tanaka2002}), that is to say a constant term of similar
magnitude, and a weak dependence on the surface density gradient. Yet,
the latter depends strongly on the softening length, as can be noticed
by comparing to the two-dimensional, unsmoothed, globally isothermal
expression also provided by \cite{tanaka2002}:
\begin{equation}
  \label{eq:12}
  \frac{\Gamma_L}{\Gamma_{\rm ref}}=-(3.20+1.47\alpha).
\end{equation}
This raises the question of whether the dependence on the temperature
gradient, in a three-dimensional disc, would be as steep as that of
Eq.~(\ref{eq:10}). Thus far this is an unanswered question, even if
recent numerical simulations seem to indicate that the dependence of
the differential Lindblad torque on the temperature gradient in a
three-dimensional disc is comparable to that of a two-dimensional disc
with a smoothing length $\epsilon \simeq0.4H$ (Casoli \& Masset, in
prep).  We also comment that self-gravity slightly enhances the
amplitude of the differential Lindblad torque by a factor
approximately equal to $(1+Q_p^{-1})$, with $Q_p$ the Toomre
Q-parameter at the planet's orbital radius \cite{bm08b}.

Unless the disc has a temperature profile that strongly increases
outward, the differential Lindblad torque is a negative quantity
which, by itself, would drive type I migration on timescales shorter
than a few $\times 10^5$ yrs for an Earth-mass object in a disc with a
mass comparable to that of the Minimum Mass Solar Nebula \cite{w97}.
Note however that Eqs.~(\ref{eq:10}) and~(\ref{eq:11}) have been
derived under the assumption that the profiles of surface density and
temperature are power laws of the radius.  Local variations in the
disc's temperature and/or density profiles, due for example to opacity
transitions \cite{mg2004} or to dust heating \cite{hp10} may change
the sign and magnitude of the differential Lindblad torque.  An
approximate generalisation of the Lindblad torque expression, valid in
non-power law discs, has been derived by \cite{2011CeMDA.111..131M}
for two-dimensional discs with a softening length $\epsilon=0.6H$.
This expression reads:
\begin{equation}
  \label{eq:13}
  \frac{\Gamma_L}{\Gamma_{\rm ref}}=
  -(2.00-0.16\alpha+1.11\beta-0.80[\beta_2^+-\beta_2^-]),
\end{equation}
where 
\begin{equation}
  \label{eq:14}
  \beta_2 = h\frac{d^2\log T }{d(\log r)^2},
\end{equation}
and where a quantity with a $\pm$ subscript is to be evaluated in
$r=a\pm H/5$. Obviously, if the temperature profile is a power-law of
radius, one has $\beta_2^+=\beta_2^-=0$, and Eq.~(\ref{eq:13}) reduces
to a standard linear combination of $\alpha$ and $\beta$.

The differential Lindblad torque has also been investigated in
strongly magnetised, non-turbulent discs. The case of a
two-dimensional disc with a toroidal magnetic field has been studied
by \cite{Terquem03} through a linear analysis. She found that the
differential Lindblad torque is reduced with respect to non-magnetised
discs (as waves propagate outside the Lindblad resonances at the
magneto-sonic speed $(c_s^2 + v_A^2)^{1/2}$, with $v_A$ the Alfv{\'e}n
speed, rather than the sound speed $c_s$). Additional angular momentum
is taken away from the planet by the propagation of slow MHD waves in
a narrow annulus near magnetic resonances. These results were
essentially confirmed by \cite{Fromangetal05} with non-linear 2D MHD
simulations in the regime of strong toroidal field (the plasma
$\beta$-parameter, $\beta=c_s^2 / v_A^2$, was taken equal to $2$ in
their study).  More recently, 2D and 3D disc models with a poloidal
magnetic field were investigated by \cite{Muto08} with a linear
analysis in the shearing sheet approximation. While the differential
Lindblad torque is reduced similarly as with a toroidal magnetic
field, extraction of angular momentum by slow MHD and Alfv{\'e}n waves
is found to occur in three dimensions only.

We sum up the results presented in this section. 
\begin{itemize}
\item The differential Lindblad torque corresponds to the net rate of
  angular momentum carried away by density waves (wakes) the planet generates
  in the disc at Lindblad resonances.
\item The sign and magnitude of the differential Lindblad torque arise
  from a slight asymmetry in the perturbed density distribution
  associated to each wake.
\item The differential Lindblad torque is a stationary quantity,
  largely independent of the disc's turbulent viscosity. Alone, it
  would drive the migration of Earth mass embedded planets in as short 
  a time as a few $\times 10^5$ yrs in typical protoplanetary discs.
\end{itemize}

\subsection{Corotation torque}
\label{sec:corotation-torque}
The other component of the tidal torque, the corotation torque, has
long been neglected in studies of planetary migration. Firstly, it was
shown to have a lower absolute value than the differential Lindblad
torque for typical (decreasing) radial profiles of the disc's surface
density \cite{tanaka2002}. Secondly, this torque component, for
reasons that will be presented at length in
section~\ref{sec:satur-prop}, should tend to zero after a finite time
in an inviscid disc (the corotation torque is said to
"saturate"). However, as indicated by \cite{wlpi91}, some amount of
turbulence should prevent the corotation torque from saturating, and
in the last decade the asymptotic value of the torque at large time in
the presence of dissipative processes has been tackled either
analytically \cite{masset01,bk2001}, by means of numerical simulations
\cite{masset02}, or both \cite{mc10,pbk11}. Besides, it was discovered
by \cite{pm06} that in radiative discs, the corotation torque could,
under certain circumstances, be so large and positive that it could
largely counteract the differential Lindblad torque, thereby leading
to outward planetary migration.  This was subsequently interpreted as
a new component of the corotation torque arising from the disc's
entropy gradient \cite{bm08a,pp08,mc09,pbck10}.

The corotation torque on low-mass planets is usually linked to the
so-called horseshoe drag, which corresponds to the exchange of angular
momentum between the planet and its horseshoe region. The planet's
horseshoe region encompasses the disc region where fluid elements are
on horseshoe streamlines with respect to the planet orbit (see
Fig.~\ref{fig:hs}). We therefore start by explaining the concept of
horseshoe dynamics and horseshoe drag.

\subsubsection{Horseshoe dynamics}
\label{sec:horseshoe-dynamics}
\begin{figure}
   \includegraphics[width=0.51\hsize]{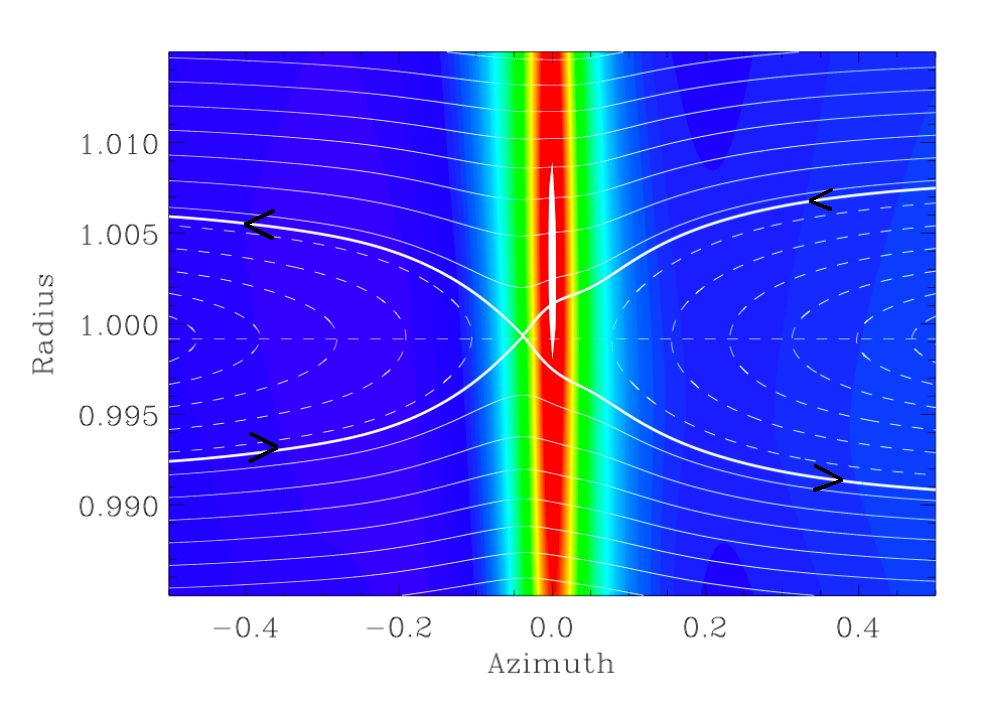}
   \includegraphics[width=0.49\hsize]{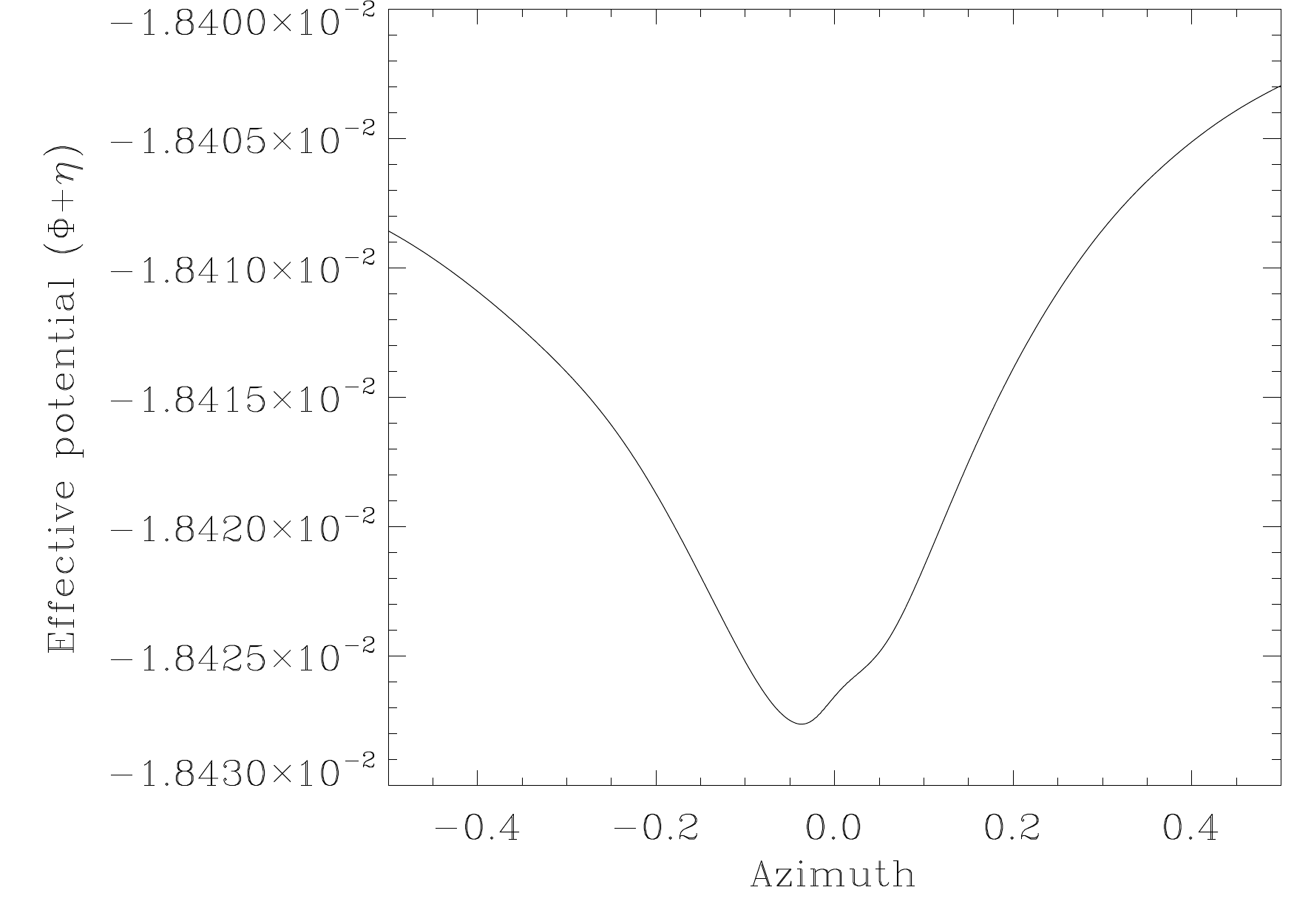}
   \caption{Left: streamlines in the vicinity of an Earth-mass planet
     embedded in a disc with $H/r=0.05$. The softening length of the
     potential is $\epsilon=0.6H$. The thick lines show the
     separatrices of the horseshoe region (the frontiers between the
     dashed streamlines that exhibit horseshoe-like motion and the
     solid ones corresponding to circulating material with respect to
     the planet orbit). A few arrows show the flow's direction with
     respect to the planet. The horizontal dashed line shows the
     corotation radius, where the angular velocities of the disc and
     the planet are equal. Right: effective potential at the
     corotation radius, as a function of azimuth.}
\label{fig:sllm}
\end{figure}

In the restricted three-body problem (RTBP), it is useful to write the
Hamiltonian of the test particle in the frame that corotates with the
secondary. In this frame, the potential does not depend explicitly on
time and the Hamiltonian $H$ is therefore conserved. It reads:
\begin{equation}
  \label{eq:15}
  H = E-\Omega_p L,
\end{equation}
where $E$ is the total energy of the test particle as seen in an
inertial frame centred on the primary, and $L$ its angular
momentum. The Hamiltonian of Eq.~(\ref{eq:15}) is usually called the
Jacobi constant. A sketch of the lines of constant Jacobi value when
the kinetic energy is zero ($E$ therefore exclusively amounts to the
potential energy), named zero velocity curves (ZVC), reveals the
existence of a horseshoe-like region encompassing the secondary's
orbit \cite{mude00}. Furthermore, in the limit of a small secondary's
mass, the trajectory of the guiding centre of the test particle in
this area is shown to have a radial displacement from the orbit that
is twice that of the associated ZVC \cite{mude00}, so that the guiding
centre of a test particle can also exhibit a horseshoe-like motion in
the vicinity of the orbit. In a similar fashion, a gas parcel in a
disc with pressure can exhibit a horseshoe-like motion, very similar
to that of the RTBP. There are, however, important differences between
these two cases. Firstly, in the RTBP, the test particle can exhibit
epicyclic motion on top of its global horseshoe-like trajectory. In
the horseshoe region of a low mass planet, where no shocks are
present, the gas parcels cannot cross each other's orbits, and
therefore essentially follow nearly circular streamlines far from the
planet. Secondly, the width of the horseshoe region is quite different
in the two cases. It is much more narrow, for the same planetary mass,
in the gaseous case than in the RTBP.  Also, it has a different
scaling with the planetary mass in the two cases: in the RTBP, the
width of the horseshoe region scales with the cube root of the
planetary mass, whereas in the gaseous case, provided the planet mass
is not too large (we will specify how large below), it scales with the
square root of the planet mass. To understand the reasons for this
difference, we depict in the left-hand panel of Fig.~\ref{fig:sllm}
the streamlines in the vicinity of the coorbital region of a low-mass
planet. 
The planet is located at radius $r=1$, and azimuth
$\varphi=0$. The separatrices of the planet's horseshoe region are depicted by
thick curves. We see that, quite to the contrary of the RTBP, there is
no equivalent to the Roche lobe region around the planet (no
circulating fluid material bound to it). Another difference is that
the fixed point (or stagnation point) at which the separatrices
intersect lies on the orbit (whereas in the RTBP, they intersect at
the Lagrange points, away from the orbit, on a line joining the
central star and the planet \cite{mude00}). The azimuth of the
stagnation point corresponds to the azimuth where, at corotation, the
effective potential (the sum of the gravitational potential and fluid
enthalpy) is minimum. This is illustrated in the right panel of
Fig.~\ref{fig:sllm}. The sign and value of the stagnation point's
azimuth is closely related to the asymmetry of the inner and outer
wakes generated by the planet \cite{pp09b}. Note that there may be
several stagnation points near the planet's corotation radius,
depending on the softening length of the planet's potential
\cite{cm09}.

Assuming that the fluid motion is in a steady state in the vicinity of
the planet, we may use a Bernoulli invariant in the corotating frame
\cite{mp03,mak2006,cm09}. This invariant can be cast as:
\begin{equation}
  \label{eq:16}
  B_J=\frac{v^2}{2} + \eta + \Phi - \frac{1}{2}r^2 \Omega_p^2,
\end{equation}
where $v$ is the fluid velocity in the corotating frame, $\eta$ the
fluid's specific enthalpy, $\Phi$ the sum of the star's and the
planet's gravitational potentials, and $\Omega_p$ the planet's angular
frequency. Equating the value of the Bernoulli invariant at the
stagnation point (where by definition $v=0$), and far from the planet
on a separatrix, one finds the following expression for the half-width
$x_s$ of the planet's horseshoe region, away from the planet:
\begin{equation}
  \label{eq:17}
  x_s \propto |\Phi_p +\eta^{'}|_s^{1/2},
\end{equation}
where $\Phi_p$ is the planetary potential and $\eta^{'}$ the
perturbation of the gas specific enthalpy introduced by the planet.
The $s$ subscript on the right hand side of Eq.~(\ref{eq:17}) means
that these variables are to be evaluated at the stagnation point.
When the planet mass is sufficiently small, the streamlines are found
to be in good agreement with those inferred from the linear expansion
of the perturbed velocity field. Note that this does not imply that
the corotation torque is in general a linear process. As was shown
indeed by \cite{pp09a}, the torque exerted by the coorbital material
on the planet eventually becomes non-linear, no matter how small the
planet mass is, provided dissipative effects are sufficiently
small. However, the fact that important non-linear processes take
place in this region hardly affects the streamlines themselves.  In
this low-mass regime, the stagnation point has therefore a location
independent of the planet mass, necessarily at the corotation radius.
In that case, the effective perturbed potential $\Phi_p + \eta^{'}$
scales with the planet mass ($M_p$), and Eq.~(\ref{eq:17}) implies
that $x_s\propto M_p^{1/2}$.  This is no longer true when the location
of the stagnation point depends on the planet mass. In particular, in
the high-mass regime, the planet gravity dominates over the perturbed
enthalpy, and the situation resembles that of the RTBP. The Bernoulli
invariant at the inner and outer stagnation points (which are
L$_1$-like and L$_2$-like, respectively) is dominated by the planetary
potential term and so scales as $M_p/R_H$, where $R_H = a (M_p /
M_{\star})^{1/3}$ is the planet's Hill radius. In the high-mass
region, the width of the horseshoe region therefore scales with the
cubic root of the planet mass, as in the RTBP \cite{mak2006}.

Lastly, another important difference between the RTBP and the case of
a low-mass embedded planet is that of the U-turn timescale.  Firstly,
it should be noted that the expression {\em U-turn timescale} is
ambiguous, for it depends on the horseshoe streamline under
consideration. The closer to corotation, the longer it takes to
perform a horseshoe U-turn, and the U-turn time reaches its minimum
value close to the separatrix. It is usually this minimum value that
is meant by the ambiguous expression {\em U-turn timescale}. Note that
the corotation torque nearly reaches a constant value (notwithstanding
saturation considerations, that we shall contemplate later) after this
timescale, as it corresponds to the fluid elements that most
contribute to the torque, because their angular momentum jump is the
largest, and because their mass flow-rate is large.

In the RTBP, the U-turn timescale is of the order of the dynamical
timescale $\tau_{\rm dyn}$, {\em i.e. } a planet orbital period. If
one regards, in a crude approximation, the case of a low-mass embedded
planet as an expurgated version of the RTBP, where most of the initial
horseshoe streamlines are made circulating, and where only those lying
close to corotation keep their horseshoe character, one expects the
U-turn timescale in this case to be significantly longer than the
dynamical timescale. This is indeed the case: the U-turn timescale is
approximately $h_p \tau_{\rm lib}$ \cite{bm08a}, with $h_p$ the disc's
aspect ratio at the planet's orbital radius, and where $\tau_{\rm
  lib}$ is the libration timescale, {\em i.e.} the time it takes to
complete a closed horseshoe trajectory. The libration timescale reads
\begin{equation}
\tau_{\rm lib} = \frac{8\pi a}{3\Omega_p x_s}.
\label{eq:taulib}
\end{equation}
An alternate, equivalent expression for the U-turn timescale is
$\tau_{\rm U-turn}\sim \tau_{\rm dyn}H/x_s$, which is corroborated by
numerical simulations in which one monitors the advection of a passive
scalar. This expression shows that the U-turn timescale can indeed be
longer than the dynamical timescale by a significant factor, as $x_s$
can be much smaller than $H$ for deeply embedded, low-mass objects.

We sum up the results presented in this section: 
\begin{itemize}
\item The coorbital region of a low-mass embedded planet 
in a gaseous disc
  exhibits a horseshoe-like region.
\item This region is much more narrow than in the restricted 
three-body problem, and its radial 
  width scales with the square root of the planetary mass.
\item The stagnation points are located at the corotation
  radius. There is no equivalent to the Roche lobe region for low-mass
  objects.
\item The horseshoe U-turn timescale is significantly longer than the
  dynamical timescale.
\end{itemize}

\subsubsection{Horseshoe drag: an overview}
\label{sec:horseshoe-drag}
Far from the horseshoe U-turns in the vicinity of the planet, a fluid
element or test particle essentially follows a nearly circular orbit,
and therefore has a nearly constant angular momentum. When it reaches
a U-turn, a fluid element is either sent inward or outward, thereby
crossing the planet orbit. It does so by exchanging angular momentum
and energy with the planet. The torque resulting from the interaction
of the planet with all the fluid elements in the course of performing
their horseshoe U-turn is called horseshoe drag \cite{wlpi91}.

\begin{figure}
  \centering
  \includegraphics[width=0.8\hsize]{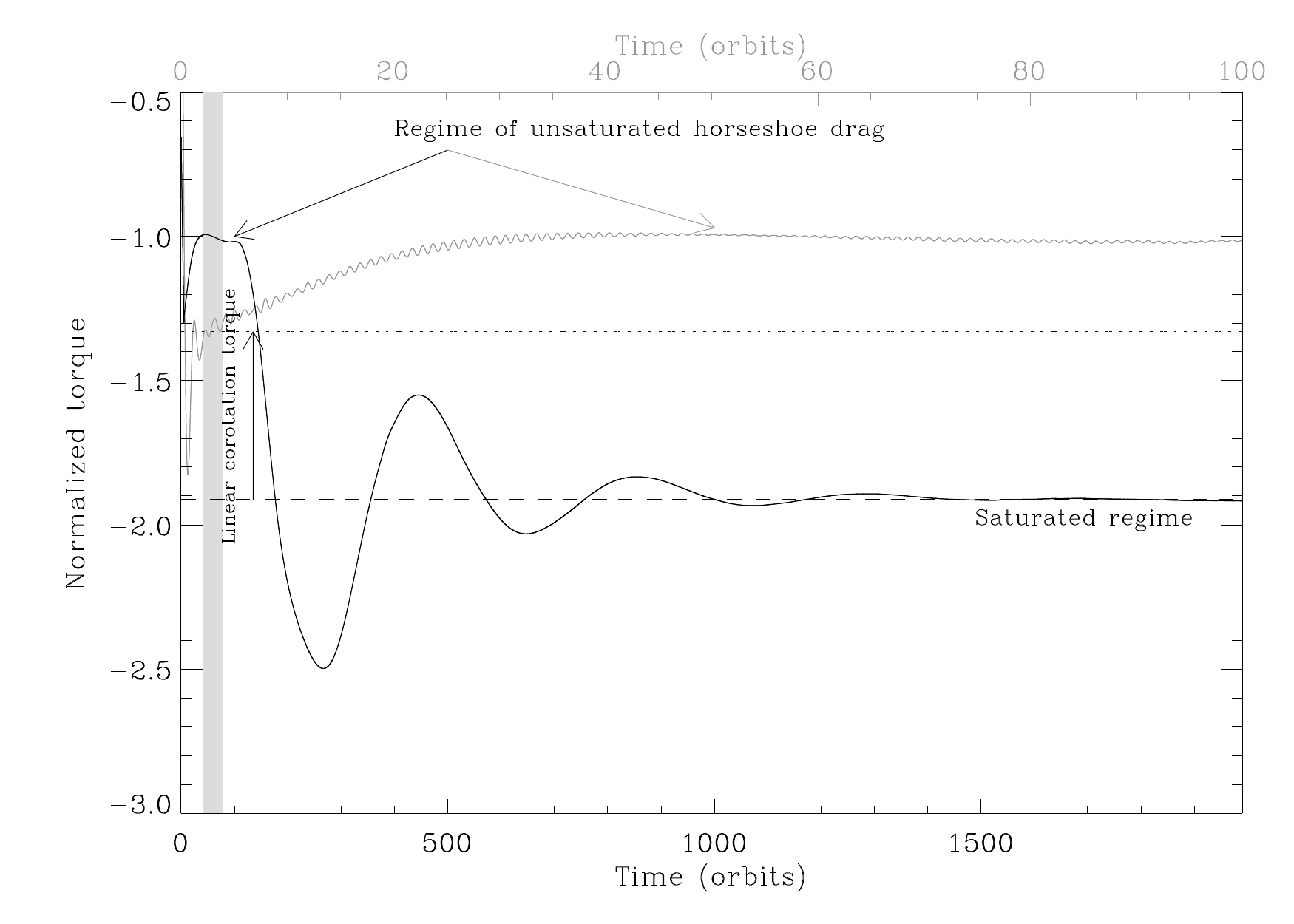}
  \caption{Time evolution of the total torque (sum of the differential
    Lindblad torque and corotation torque) on a $M_p = 10^{-6}
    M_{\star}$ planet mass embedded in an inviscid isothermal
    disc. The black curve (bottom x-axis) shows the torque evolution
    over 2000 planet orbits, while the grey curve (top x-axis) focuses
    on the evolution over the first 100 planet orbits. The dashed line
    shows the value of the differential Lindblad torque, and the
    dotted curve highlights the corotation torque predicted with a
    linear analysis.  Taken from \cite{2011CeMDA.111..131M}.}
  \label{fig:tqvstime}
\end{figure}
Upon insertion of the planet in the disc, it takes some time to
establish the horseshoe drag, namely a time of the order of the
horseshoe U-turn timescale \cite{pp09a}. This is illustrated in
Fig.~\ref{fig:tqvstime}, which displays the time evolution of the
total torque on a $M_p = 10^{-6} M_{\star}$ mass planet embedded in a
thin ($h = 0.05$) isothermal disc with uniform density profile.  The
disc is inviscid in this example. Once established (after $\sim 30$
planet orbits in our example), the horseshoe drag remains
approximately constant over a longer timescale, which corresponds to
the time it takes for a fluid element to drift from one end of the
horseshoe to the other (that is, about half a libration time, given by
Eq.~(\ref{eq:taulib})).  The value of the horseshoe drag that exists
between the horseshoe U-turn time, and half the horseshoe libration
time, is called the {\em unsaturated horseshoe drag}. Beyond this
stage, subsequent U-turns may cause further time evolution of the
horseshoe drag depending on the disc viscosity, which will be
described in \S~\ref{sec:satur-prop}. In the particular case depicted
here, where the disc is inviscid, the horseshoe drag eventually
saturates (it cancels out) after a few libration timescales. Until
\S~\ref{sec:satur-prop}, we focus on the properties of the fully
unsaturated horseshoe drag.

\subsubsection{Horseshoe drag in barotropic discs}
\label{sec:hors-drag-barotr}
\begin{figure}
  \centering
   \includegraphics[width=0.8\hsize]{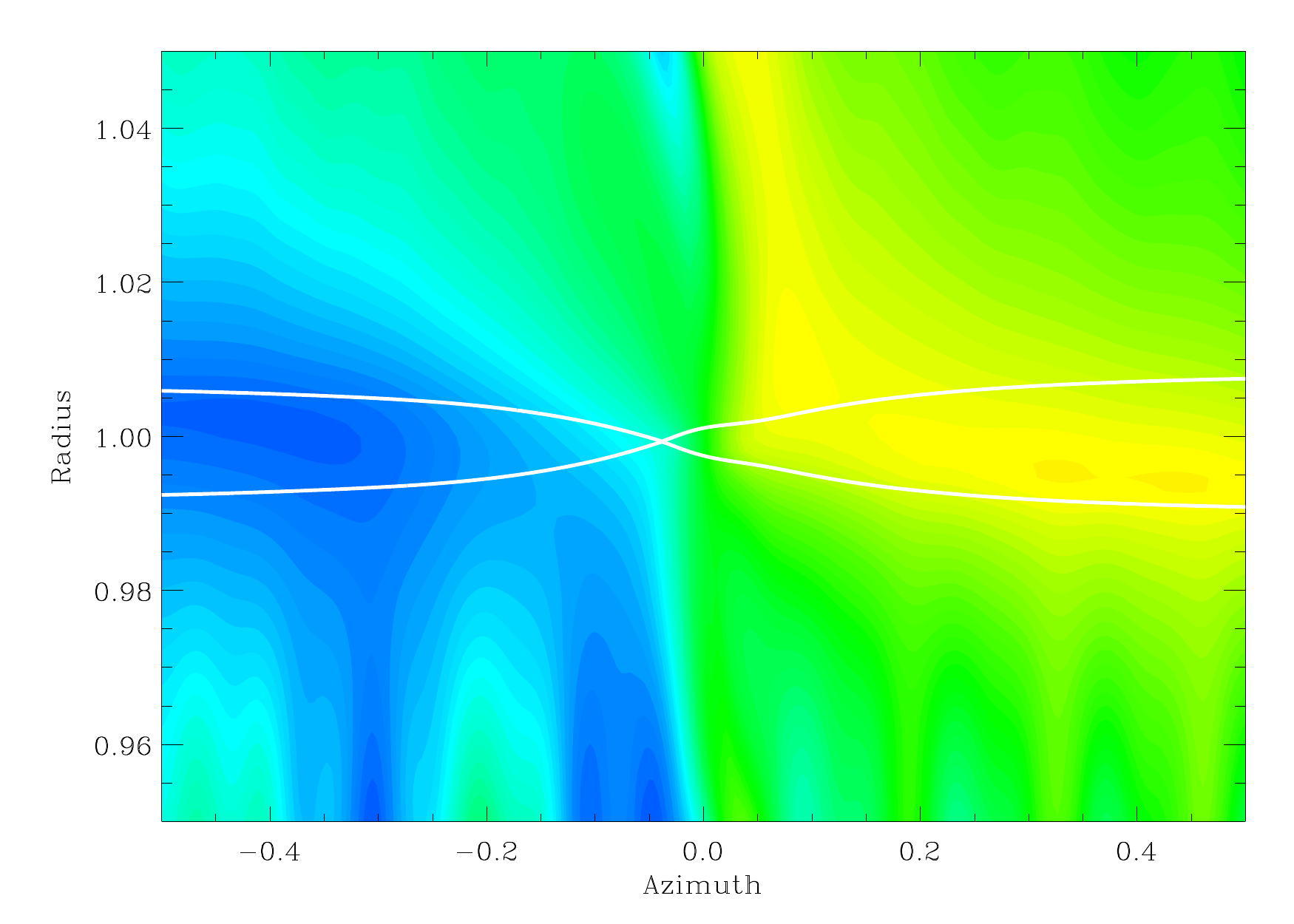}
   \caption{Perturbed surface density in the coorbital region of an
     Earth-mass planet in a disc with $h=0.05$ and $\alpha=0$ (uniform
     background surface density).  The planet is located at $r=1$,
     $\varphi=0$.  The solid curves show the separatrices of the
     planet's horseshoe region.  In order to remove the planet's wakes
     and to render this map more legible, we have subtracted the
     density perturbation obtained in a situation where no corotation
     torque is expected (namely $\alpha=3/2$, as will be shown
     below). This cancellation is imperfect, however, as the wakes of
     the two cases are not strictly identical.}
  \label{fig:pertdenshs}
\end{figure}

A hint of the torque exerted by the coorbital material on the planet
can be obtained by the examination of the perturbed surface density.
Nonetheless, this examination is rather difficult, because the density
perturbation in the planet's coorbital region is very small, typically
one or two orders of magnitude smaller than the density perturbation
associated to the wakes. Yet, as can be seen in
Fig.~\ref{fig:pertdenshs}, an approximate subtraction of the wakes
density perturbation reveals two regions of opposite signs: a region
of positive perturbed density ahead of the planet ($\phi>0$) and a
region of negative perturbed density behind the planet ($\phi<0$),
which both yield a positive torque on the planet. The sign of the
perturbed density in the coorbital region depends on the background
density profile, here it is uniform ($\alpha=0$). The largest
perturbations can be seen to originate near the downstream separatrix
in either case (the outer separatrix at negative azimuth, and the
inner separatrix at positive azimuth), but the perturbation is spread
radially and extends much beyond the horseshoe region. This is to be
expected on general grounds: in a barotropic disc, where the gas
pressure depends only on its mass density, any disturbance near
corotation excites evanescent pressure waves, which extend typically
over the disc pressure length scale (here $H=0.05a$).

\begin{figure}
  \centering
   \includegraphics[width=0.8\hsize]{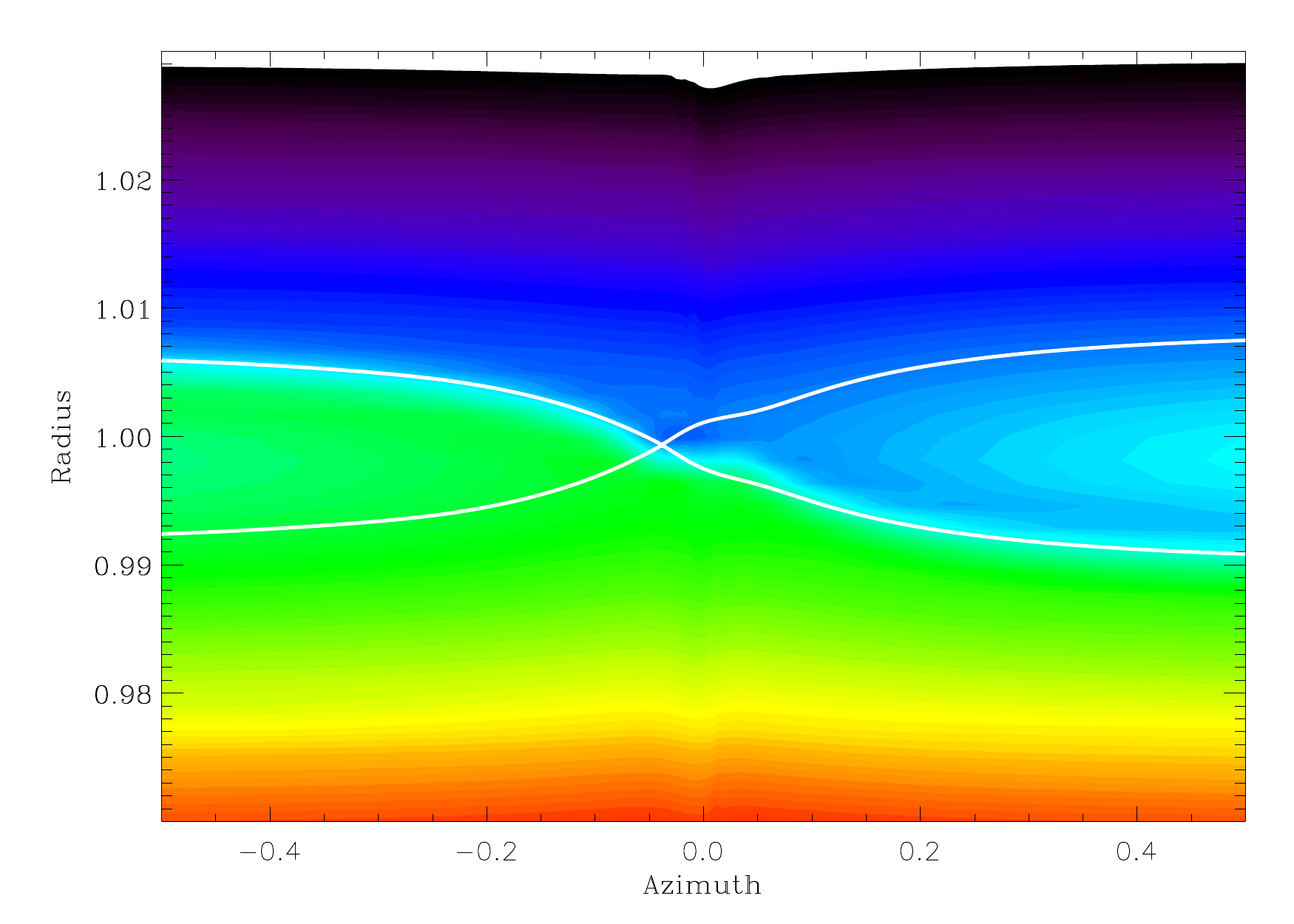}
   \caption{Advection of vortensity in the horseshoe region of an
     embedded, low-mass protoplanet. As in previous panels, the planet
     is at $r=1$, $\varphi=0$, and the separatrices of the planet's
     horseshoe region are depicted by white curves. The unperturbed
     disc's vortensity decreases with radius.  The (high) vortensity
     in the inner side of the horseshoe region (at $\varphi < 0$, $r <
     1$) is brought to the outer side (at $\varphi < 0$, $r > 1$) by
     the outward U-turns behind the planet. Similarly, the (low)
     vortensity in the outer side of the horseshoe region (at $\varphi
     > 0$, $r > 1$) is brought to the inner side (at $\varphi > 0$, $r
     < 1$) by the inward U-turns ahead of the planet.  }
  \label{fig:advvorths}
\end{figure}

Even if some insight into the corotation torque can be gained by the
examination of the perturbed density maps, a much more useful quantity
is the vortensity (the ratio of the vertical component of the
vorticity to the surface density, also known as potential vorticity),
which is materially conserved away from shocks in inviscid,
barotropic, two-dimensional discs. This is illustrated in
Fig.~\ref{fig:advvorths}.
\cite{wlpi91} has evaluated the torque exerted on the planet by test
particles embarked on horseshoe motion, by making use of the Jacobi
invariant of these particles (see
section~\ref{sec:horseshoe-dynamics}). A similar calculation can be
performed for fluid motion, provided one uses a Bernoulli invariant by
adding the enthalpy to the Jacobi constant \cite{mp03,cm09}. In both
cases one finds that the horseshoe drag has the following expression:
\begin{equation}
  \label{eq:18}
  \Gamma_{\rm HS}=8|A|B^2a\left[\int_{-x_s}^{x_s}
    \left(
      \left.\frac{\Sigma}{\omega_z}\right|_F
      -\left.\frac{\Sigma}{\omega_z}\right|_R
    \right) 
    x^2dx\right],
\end{equation}
where $x$ denotes the radial distance to the planet orbit. In
Eq.~(\ref{eq:18}), the subscript $F$ indicates that the (inverse of)
the vortensity $\Sigma/\omega_z$ has to be evaluated away from the
planet, in Front of the latter ($\phi > 0$), and the subscript $R$
indicates that it has to be evaluated at the Rear of the planet, away
from it ($\phi < 0$). The integral in Eq.~(\ref{eq:18}) is usually
called the horseshoe drag integral, and in a barotropic disc it can be
simplified to yield the following expression \cite{cm09,pbck10,mc10}:
\begin{equation}
  \label{eq:19}
  \Gamma_{\rm HS}=\frac34\Sigma {\cal V}\Omega_p^2x_s^4,
\end{equation}
where the quantity ${\cal V}$, called the (inverse) vortensity gradient 
for short, is defined by
\begin{equation}
  \label{eq:20}
  {\cal V} = \frac{d\log (\Sigma/B)}{d\log r},
\end{equation}
and can be recast as $3/2 - \alpha$ for density profiles that can be
approximated as power-law functions of radius over the radial width of
the planet's horseshoe region. In Eq.~(\ref{eq:19}), all terms are to
be evaluated at the planet's orbital radius.  This equation shows that
in two-dimensional barotropic discs, the horseshoe drag cancels out
when the surface density profile decreases locally as $r^{-3/2}$,
while it is positive for density profiles shallower than
$r^{-3/2}$. For density profiles strongly increasing outward, the
horseshoe drag can be sufficiently positive to counteract the
(negative) differential Lindblad torque, and therefore stall the
migration of low-mass planets \cite{masset06a}.  Such density jumps
may be encountered near the star's magnetospheric cavity, or near the
inner edge of a dead zone, across which the disc's effective
turbulence decreases outward (the dead zone refers to the region near
the midplane of protoplanetary discs that is sandwiched together by
partially ionized surface layers).

The horseshoe drag expression in Eq.~(\ref{eq:19}) exclusively holds
in the case of barotropic discs. Those, naturally, are an idealised
concept, and true discs have a more complex physics, which yields a
more complex expression for the corotation torque. However, in any
case, as we shall see, a common component of the corotation torque is
given by Eq.~(\ref{eq:19}), so that baroclinic effects yield
additional terms to this expression.

We sum up the results presented in this section. In two-dimensional
barotropic discs, where the gas pressure only depends on the surface
density:
\begin{itemize}
\item The horseshoe drag is powered by the advection of the fluid's
  vortensity along horseshoe streamlines inside the planet's horseshoe region.
\item It is proportional to the inverse vortensity gradient across the
  horseshoe region (that is, the quantity $3/2 + d\log \Sigma/d\log r$
  for power law discs). It can therefore be negative, zero, or
  positive depending on the surface density gradient across the
  horseshoe region. For typical discs density and temperature
  profiles, its magnitude is smaller than that of the (negative)
  differential Lindblad torque.
\end{itemize}

\subsubsection{Horseshoe drag in locally isothermal discs}
\label{sec:hors-drag-locally}
\begin{figure}
  \centering
  \includegraphics[width=0.49\hsize]{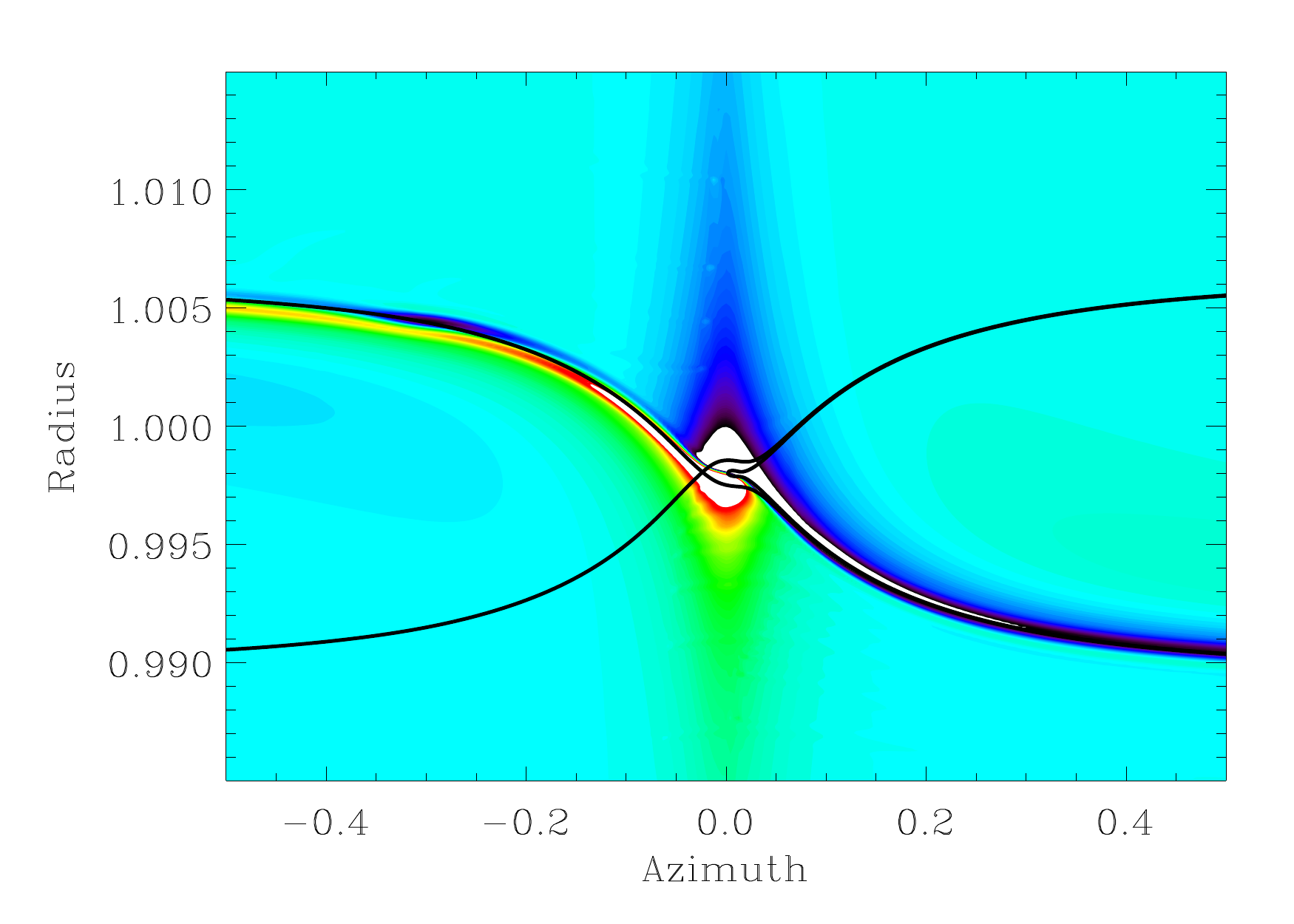}
  \includegraphics[width=0.49\hsize]{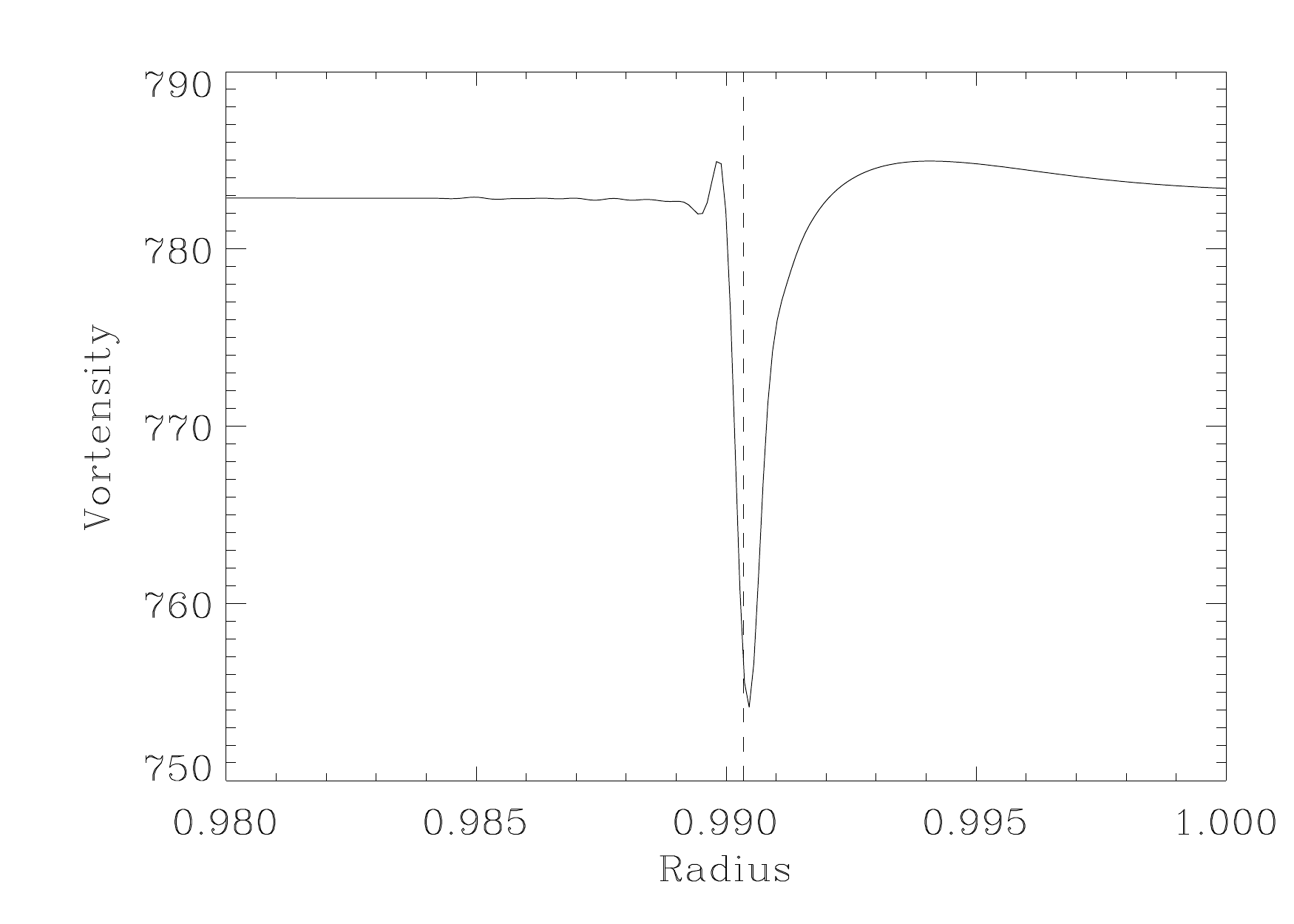}
  \caption{Vortensity field in the coorbital region of a low-mass
    planet (left) and radial profile of vortensity at $\phi=+0.5$~rad
    (right), $30$~orbital periods after the insertion of the
    planet. The disc has no background vortensity gradient, and a flat
    aspect ratio $H/r=0.05$ . Thin stripes of vortensity of opposite
    signs are clearly visible at the downstream separatrices. One can
    also see a mild production of vortensity in the wake, which fades
    away as one recedes from corotation, because of the winding of the
    wake and because fluid elements move faster away from
    corotation. Contrary to the adiabatic case that we shall present
    in \S~\ref{sec:hors-drag-adiab}, the vortensity cut is not
    singular at the separatrix, as can be seen in the right panel.
    The radial resolution in the run presented is $9.3\times
    10^{-5}a$.}
  \label{fig:vortenslociso}
\end{figure}

A long considered framework, both in analytical and numerical studies
is that of locally isothermal discs, in which the temperature is a
fixed function of radius. No energy equation is considered in this
case, but the flow is no longer barotropic: the pressure becomes a
function of the density {\em and} position (through the temperature).
The vortensity is no longer materially conserved. Its Lagrangian
derivative features a source term arising from misaligned density and
pressure gradients, or misaligned temperature and density gradients
\cite{lovelace99}:
\begin{equation}
  \label{eq:21}
  \frac{D}{Dt}\left(\frac{\vec\omega_z}{\Sigma}\right)=\frac{\nabla\Sigma\times\nabla
  P}{\Sigma^3}=\frac{\nabla\Sigma\times\nabla T}{\Sigma^2}.
\end{equation}
As the temperature gradient has a radial direction and sensibly the
same magnitude everywhere in the coorbital region, the strength of the
source term depends on the density gradient: wherever the azimuthal
density gradient is large, the source term is large. This occurs at
the tip of the horseshoe U-turns where we have a strong azimuthal
density gradient owing to the density enhancement in the planet's
immediate vicinity.  The time derivative in Eq.~(\ref{eq:21}) can be
expressed as a derivative with respect to the curvilinear abscissa $s$
along the streamline:
\begin{equation}
  \label{eq:22}
  \frac{D}{Ds}\left(\frac{\vec\omega_z}{\Sigma}\right) =\frac{\nabla\Sigma\times\nabla T}{v\Sigma^2},
\end{equation}
where $v$ is the norm of the fluid velocity in the corotating frame,
where we have used $ds=vdt$. As the fluid stagnates in the vicinity of
the stagnation point ({\em i.e.}  $v$ can be arbitrarily small,
provided one chooses a streamline sufficiently close to the stagnation
point), the source term of Eq.~(\ref{eq:22}) formally diverges in the
vicinity of the stagnation point. The total amount of vortensity
created, integrated over the horseshoe streamlines, however, remains
finite.  Fig.~\ref{fig:vortenslociso} shows a vortensity map in the
vicinity of a low-mass planet for a disc with $\alpha=3/2$ (no
background vortensity gradient) and $\beta=1$ (uniform aspect ratio).

Thus far there is no rigorous mathematical proof that the horseshoe
drag expression of Eq.~(\ref{eq:18}) still holds in locally isothermal
discs, as all demonstrations of that expression rely on the existence
of a Bernoulli invariant, which does not exist in the locally
isothermal case. Yet, data from numerical simulations suggest that
this expression is still valid in that regime. Assuming its validity
from now on, we infer that the horseshoe drag must exhibit a
dependence on the temperature gradient. The rational for this being
that the outgoing vortensity, accounted for by the horseshoe drag
integral, includes the vortensity produced in the vicinity of the
planet, which depends on the temperature gradient. Two-dimensional
numerical simulations have confirmed the existence of an additional
component of the corotation torque that depends on the temperature
gradient \cite{BaruteauPhD, cm09, pbck10}.  The sign and value of this
temperature-related corotation torque have a complex, and rather steep
dependence with the softening length of the planet's potential
\cite{cm09}. Indeed, the topology of the horseshoe region depends
heavily on the softening length: at large softening lengths, only one
X-stagnation point is observed at corotation, in the planet vicinity,
whereas at low softening lengths, two X-stagnation points are usually
observed at corotation, on either side of the planet
\cite{cm09,pp09b}. The vortensity produced along a streamline depends
on the path followed by the streamline. This situation is therefore
much more complex than in the barotropic case or the adiabatic case
that we present below, in which the existence of invariants under
certain circumstances allows to get rid of the dependence on the
actual path followed by fluid elements during their U-turns. Quite
ironically, the locally isothermal case, which has served as a
standard framework for more than two decades, is very difficult to
tackle analytically.

The steep dependence of the temperature related corotation torque on
the softening length appeals for a three-dimensional study of this
torque, which is not plagued by softening issues. Such study has been
undertaken by Casoli \& Masset (in prep.), who find a linear
dependence of the three-dimensional horseshoe drag on the temperature
gradient $\beta = -d\log T/d\log r$, as steep as the dependence on the
vortensity gradient ${\cal V}$ given by Eq.~(\ref{eq:19}).

\subsubsection{Horseshoe drag in adiabatic discs}
\label{sec:hors-drag-adiab}
In the previous sections, the set of governing equations of the fluid
did not include an energy equation, and the disc temperature, which
was set as a prescribed function of radius, did not evolve in time.
The first calculations undertaken with an energy equation were those
of \cite{mota03} and \cite{pm06}. The former were devised in the
shearing sheet framework, so that no net torque could be experienced
by the planet, owing to the symmetry properties of the shearing
sheet. Still, these authors found that radiative cooling could
significantly affect the perturbed surface density pattern associated
with the wakes, thus changing the magnitude of the one-sided Lindblad
torque. \cite{pm06} considered a planet in a three-dimensional disc,
with an energy equation and thermal diffusion, and nested grids around
the planet to achieve a very high resolution. They found that the
migration of a low-mass planet could be reversed in sufficiently
opaque discs, under the action of the corotation torque.  The same
result was in particular obtained in the adiabatic limit, which we are
now going to focus on, as thermal diffusion yields an additional
complexity, not needed at this stage. We will take thermal diffusion
and other dissipative processes into account in
\S~\ref{sec:satur-prop}.

\begin{figure}
  \centering
  \includegraphics[width=0.8\hsize]{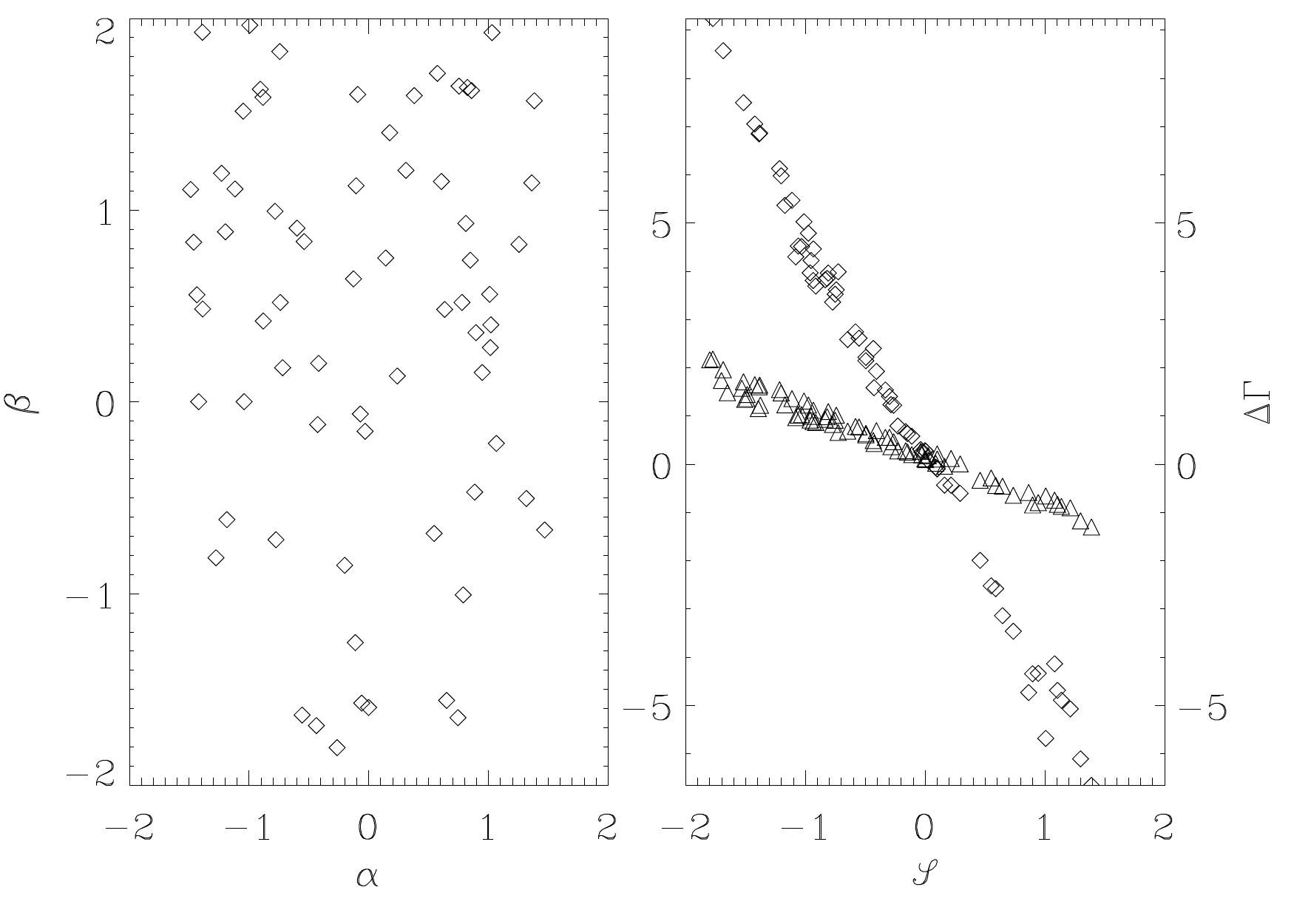}
  \caption{Torque difference between the adiabatic and locally
    isothermal cases, given by Eq.~(\ref{eq:tqexcess_def}), as a
    function of the entropy gradient (right panel), at early times 
    (linear stage, triangles) and during the horseshoe drag stage
    (diamonds). The quantity ${\cal S}$ in x-axis is defined 
    in Eq.~(\ref{eq:calS}). Results have been obtained with calculations with
    random values of the surface density slope ($-\alpha$) and
    temperature slope ($-\beta$), shown in the left plot. The total
    number of runs is $138$.}
  \label{fig:tqdiffadiso}
\end{figure}

It was soon realised that the results of \cite{pm06} were due to a new
component of the corotation torque, linked to the entropy gradient
\cite{bm08a, pp08}.  This is illustrated in
Fig.~\ref{fig:tqdiffadiso}, in which we compare the torque results for
$69$~different random disc profiles (the density slope $\alpha$ being
a random variable uniformly distributed over the interval
$[-3/2,+3/2]$, and the temperature slope $\beta$ being an independent
random variable uniformly distributed over the interval $[-2,+2]$.)
Each calculation has a smoothing length of the planet's potential
$\epsilon=0.3H$, and an aspect ratio at the planet location
$h_p=0.05$.

For each pair of $\alpha$ and $\beta$, we ran 
two calculations: a locally isothermal one,
and an adiabatic one with a ratio of specific heats $\gamma=1.4$. The
torque difference, dubbed {\em adiabatic torque excess}, is then
obtained by:
\begin{equation}
  \label{eq:23}
  \Delta\Gamma_{\rm HS}^{\rm entr}=\Gamma_{\rm ad}-\frac{\Gamma_{\rm iso}}{\gamma}.
  \label{eq:tqexcess_def}
\end{equation}
The correction of the isothermal torque $\Gamma_{\rm iso}$ by a factor
${\gamma^{-1}}$ is necessary as both the differential Lindblad torque
and the barotropic part of the horseshoe drag (the vortensity-related
corotation torque) scale with the inverse square of the sound speed,
which turns out in the adiabatic case to be $c_s^{\rm adi}=c_s^{\rm
  iso}\gamma^{1/2}$.  The right part of Fig.~\ref{fig:tqdiffadiso}
shows a clear one-to-one relationship between the adiabatic torque
excess and the entropy gradient, irrespective of the individual values
of $\alpha$ and $\beta$, which justifies the name {\em entropy-related
  corotation torque} given to the adiabatic torque excess. This torque
is shown in two different regimes: the linear regime, soon after the
insertion of the planet in the disc, and the horseshoe drag regime,
reached after a longer timescale, as discussed in previous
sections. As pointed out by \cite{pp08}, the non-linear corotation
torque can be much larger (in this example, by a factor of about $5$)
than its linear counterpart, depending on the planet's softening
length.

An interpretation of the entropy-related corotation torque was given
soon after its discovery by \cite{bm08a} and \cite{pp08}. Although
this early interpretation was shown not be quite correct when the
horseshoe drag expression in the adiabatic case was worked out
subsequently, we describe it here briefly as the mechanism on which it
is based allows for a clear understanding of the correct origin of the
entropy torque.

\begin{figure}
  \centering
  \includegraphics[width=0.49\hsize]{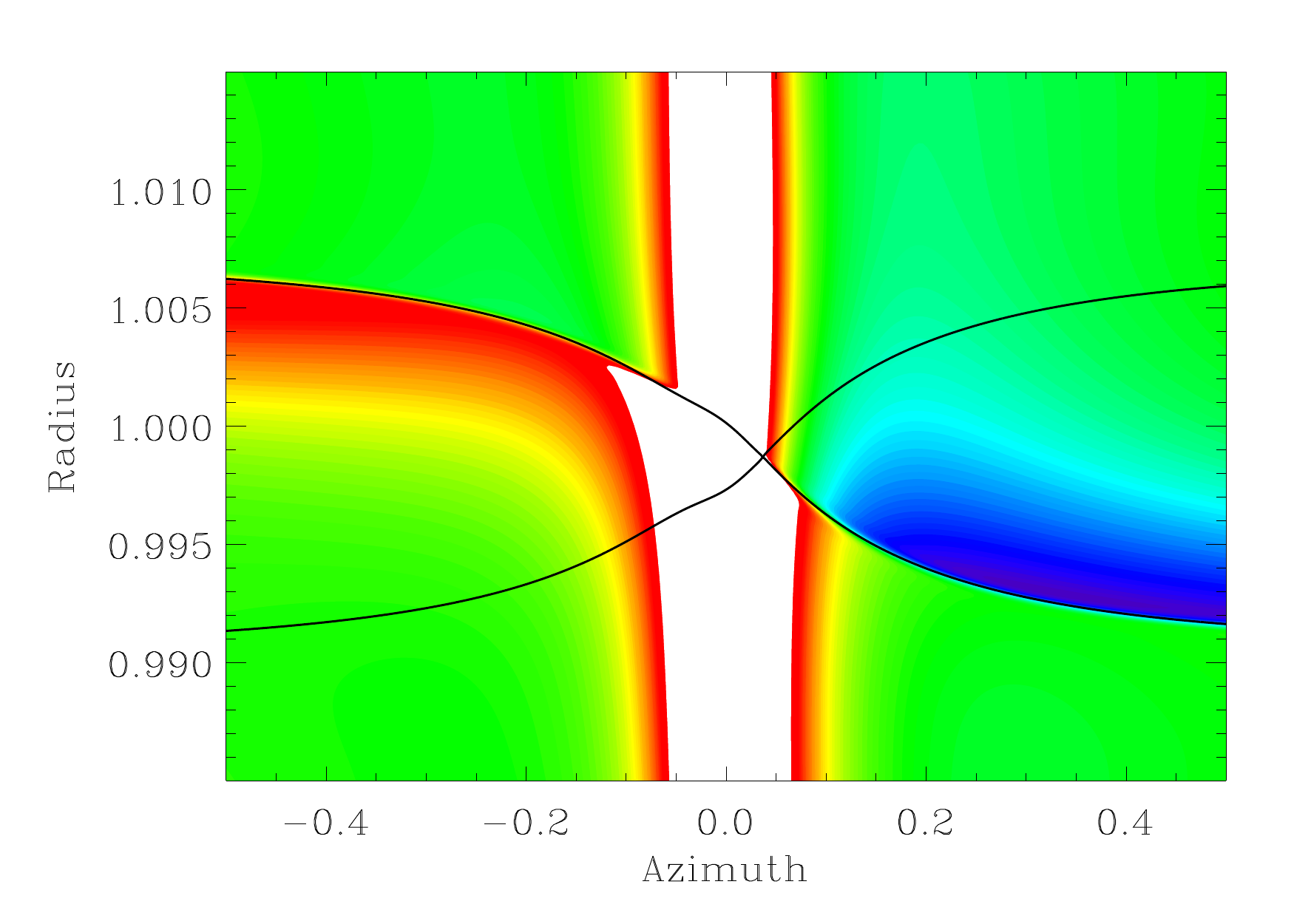}
  \includegraphics[width=0.49\hsize]{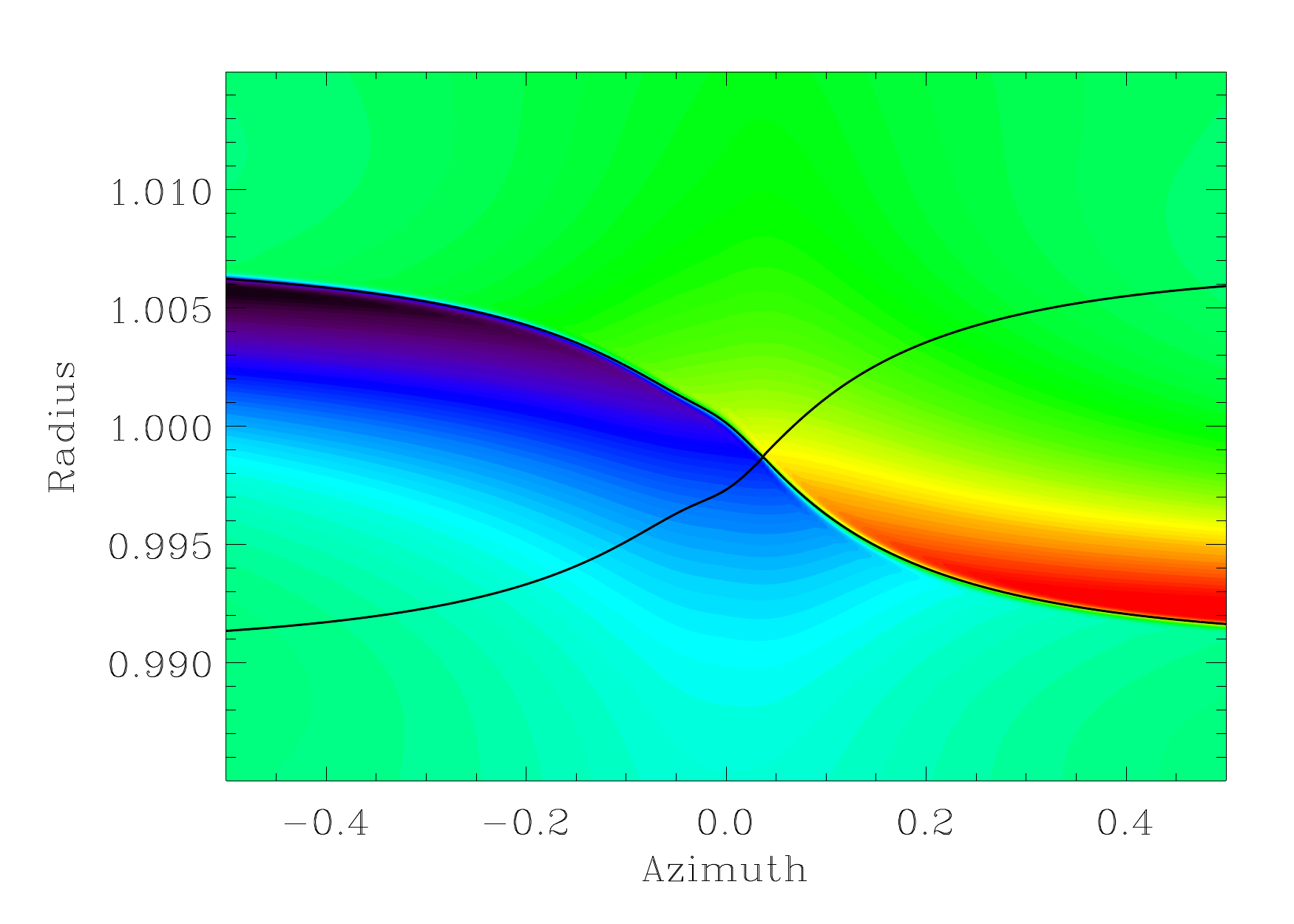}
  \caption{Surface density perturbation in the planet's coorbital
    region of an adiabatic disc (left) and perturbation of entropy
    (right). The sign of the density lobes is opposite that of the
    entropy lobes. In addition to the lobes, the left plot also shows
    the wake, saturated in this representation. The background entropy
    profile increases with radius in this example.}
  \label{fig:lobesad}
\end{figure}

In an adiabatic disc, the entropy is materially conserved along the
path of fluid elements, as long as they do not cross shocks, in a
strict analogy with vortensity for barotropic flows. If we assume, for
instance, a disc that has a positive radial entropy gradient prior to
the planet insertion, outward horseshoe U-turns (behind the planet)
bring to the outer side of the horseshoe region the low entropy of the
inside.  Similarly, inward horseshoe U-turns (ahead of the planet)
bring to the inner side of the horseshoe region the high entropy of
the outside.  As the disc maintains a pressure equilibrium, the
relative variations of the pressure across the coorbital region can be
neglected, so that the surface density features relative variations
opposite to that of the entropy, by virtue of the first order
expansion of Eq.~(\ref{eq:1}):
\begin{equation}
  \label{eq:24}
  \gamma\frac{\delta\Sigma}{\Sigma}+\frac{\delta s}{s}=\frac{\delta
    P}{P}\approx 0.
\end{equation}
As a consequence, two lobes of perturbed surface density appear in the
horseshoe region \cite{bm08a,pp08,KleyCrida08,Kley09}, that both yield
a torque of same sign. These two lobes are shown in
Fig.~\ref{fig:lobesad}.

The early interpretations identified the entropy related torque as the
torque arising from the above density lobes.  This explanation was
appealing at first, because it gives the correct sign for the entropy
related corotation torque, and the correct order of magnitude:
\cite{bm08a} performed an approximate, horseshoe-drag like integration
to evaluate the impact of these lobes on the torque, whereas
\cite{pp08} performed an approximate direct summation. Both results
were in rough agreement with the magnitude of the adiabatic torque
excess, at least for the value of the potential softening length used
in these studies. Yet, this explanation of the entropy torque quickly
turned out not to be fully satisfactory, for the following reasons:
\begin{itemize}
\item Numerical explorations performed at different smoothing values
  showed that the entropy torque was approximately scaling as
  $\epsilon^{-1}$, down to very low values of the smoothing length
  ($\epsilon \sim 0.05H$). This apparent divergence of the entropy
  torque at low smoothing was incompatible with a scaling of the
  torque in $x_s^4$, since the half-width $x_s$ of the horseshoe
  region remains finite in the limit of a vanishing softening length
  \cite{pp09b}.
  \item The saturation of the entropy torque was also problematic. We
  will examine saturation processes in detail in
  \S~\ref{sec:satur-prop}, but for our purpose it suffices to know
  that the corotation torque always saturates in inviscid discs, that
  is to say tends to zero after a few libration timescales (as we have
  seen in Fig.~\ref{fig:tqvstime}).  Therefore, one could devise a
  setup with an {\em inviscid} disc and finite thermal or entropy
  diffusion, which would forever maintain the same entropy
  perturbation within the horseshoe region (finite thermal diffusion
  is required in order to avoid phase mixing of entropy, as we shall
  see in \S~\ref{sec:satu_rad}). \label{sec:hors-drag-adiab-2} As
  expected, the entropy related corotation torque is found to saturate
  as the disc is inviscid, while the (approximate) same density lobe
  structure is maintained within the horseshoe region
  \cite{mc10}. This implies the density lobe structure would exert a
  torque at early times, but not at late times, which is
  contradictory.
\item Finally, as we have seen in \S~\ref{sec:hors-drag-barotr}, the
  density perturbation responsible for the corotation torque is not
  bound to the horseshoe region, but can extend further radially by
  the excitation of evanescent waves. In the barotropic case, the
  vortensity-related horseshoe drag is actually fully accounted for by
  an evanescent density distribution within the coorbital region.  In
  the adiabatic case, attributing the entropy torque ({\em i.e.} the
  whole difference between an adiabatic and an isothermal calculation)
  to the density lobes was therefore tantamount to assuming that the
  evanescent wave structure in the coorbital region was the same in
  the adiabatic and isothermal cases, which is not obvious.
\end{itemize}

\begin{figure}
  \centering
  \includegraphics[width=0.49\hsize]{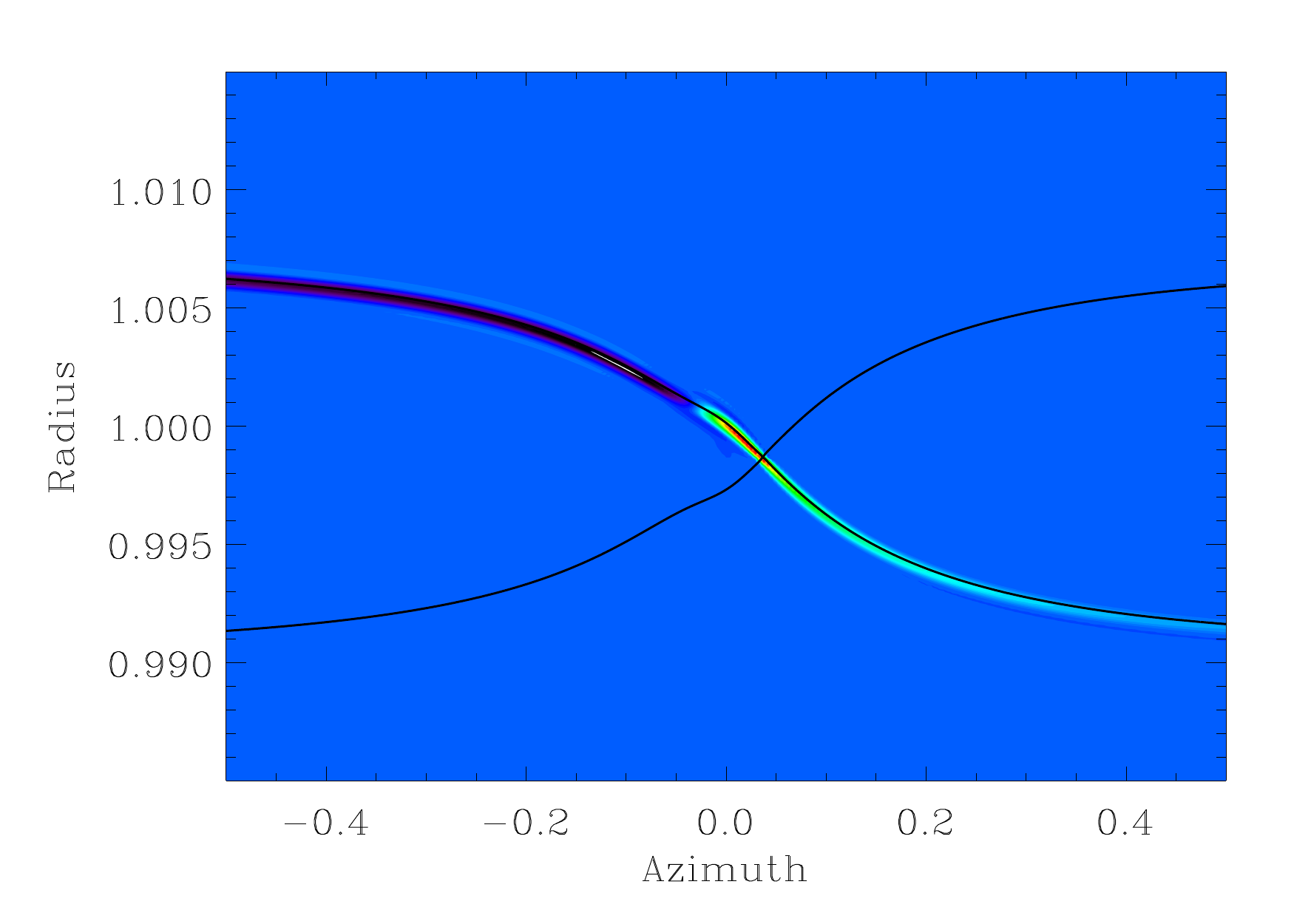}
  \includegraphics[width=0.49\hsize]{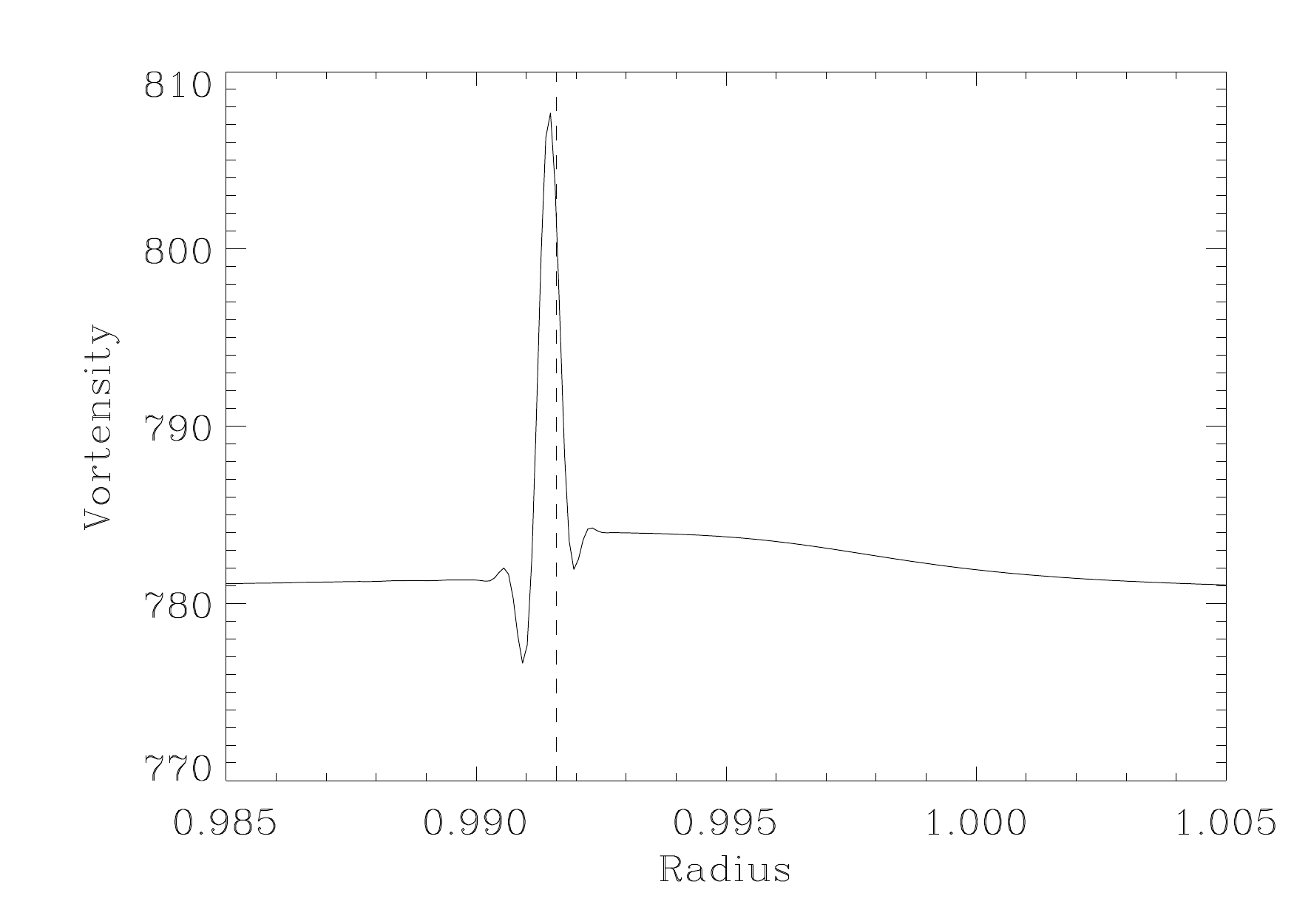}
  \caption{Vortensity map in the coorbital region of an Earth-mass
    planet embedded in an adiabatic disc with $H=0.05$ at the orbital
    radius of the planet (left), and radial profile of vortensity at
    $\phi=0.5$~rad. (right), $60$ orbital periods after the insertion
    of the planet in the disc. The grid resolution and disc gradients
    are the same as in Fig.~\ref{fig:vortenslociso} ($\alpha=3/2$,
    $\beta=0$).  The vortensity peak has a much more compact profile
    than that of Fig.~\ref{fig:vortenslociso}.  This is to be expected
    as we have a singular vortensity sheet in the adiabatic case, and
    a continuous one, peaked at the separatrix, in the locally
    isothermal case.}
  \label{fig:vortensadiab}
\end{figure}

The identification of a convenient invariant of the flow for adiabatic
discs with uniform temperature profile allowed \cite{mc09,mc10} to
demonstrate that the horseshoe drag expression was exactly the same as
that of the barotropic case, given by Eq.~(\ref{eq:18}). In this case,
the evaluation of the horseshoe drag amounts again to a budget of the
vortensity entering or leaving the vicinity of the planet on horseshoe
streamlines. An important consequence of this is that the torque due
to the density lobes must not be incorporated manually, separately,
into the corotation torque expression, and the whole problem of
determining the horseshoe drag amounts to an evaluation of the
vortensity distribution within the horseshoe region. Before we clarify
this point, we stress that the vortensity distribution within the
horseshoe region has the following features:
\begin{itemize}
\item Since the flow is baroclinic, vortensity is not materially
  conserved along streamlines. However, contrary to the locally
  isothermal case, the existence of a flow invariant in adiabatic
  discs with flat temperature profile allows an estimate of the
  vortensity acquired by a streamline during a U-turn, independently
  of its actual path \cite{mc09}.
\item The vortensity created over the interior of the horseshoe region
  is very small, and has no impact on the torque, because it has same
  sign on both sides of the planet \cite{mc09, pbck10}.
\item The main difference arises from a (formally) singular production
  of vortensity (or vorticity) on downstream separatrices, due to the
  entropy discontinuity at this location (which results from the
  entropy advection within the horseshoe region). This (formally)
  singular production of vortensity is readily apparent in the source
  term of Eq.~(\ref{eq:21}), and is illustrated in
  Fig.~\ref{fig:vortensadiab}. It can be evaluated analytically either
  using the flow invariant introduced in \cite{mc09}, or directly
  using Eq.~(\ref{eq:21}) as in \cite{pbck10}. The first approach is
  self-contained and yields the amount of singular vortensity as a
  function of the flow properties at the stagnation point. The second
  is not restricted to flat temperature profiles, but it requires
  knowledge of the fluid velocity along horseshoe streamlines, which
  depends on the exact geometry of the horseshoe region, much like in
  locally isothermal discs discussed in
  \S~\ref{sec:hors-drag-locally}.
\end{itemize}

We now clarify the contribution of the density lobes to the corotation
torque. The corotation torque is directly related to the density
perturbation within the corotation region, which can be written as
\begin{equation}
\frac{\delta \Sigma}{\Sigma} = \frac{1}{\gamma} \left( \frac{\delta P}{P} - \frac{\delta s}{s} \right),
\label{dsigma}
\end{equation}
where, counter to Eq.~(\ref{eq:24}), we shall not assume $\delta P =
0$. The pressure perturbation $\delta P$ can be shown to satisfy a
second-order partial differential equation \cite{mc09}, with solution
of the form:
\begin{equation}
\frac{\delta P}{P} = \gamma K * \frac{\delta u}{u},
\label{dp}
\end{equation}
where $u = s^{1/\gamma} \times \Sigma / \omega_z$, $K \propto
\exp(-|x|/H)$ is a Green kernel normalised to unity ($\int K(x) dx =
1$, with $x$ the radial distance to the planet orbit), and $H$ is the
local pressure scale height. Further denoting the inverse vortensity
by $l = \Sigma/\omega_z$, Eqs.~(\ref{dsigma}) and~(\ref{dp}) yield
\begin{equation}
  \frac{\delta \Sigma}{\Sigma} = K * \left( \frac{\delta l}{l}  +  \frac{1}{\gamma}\frac{\delta s}{s} \right)
  - \frac{1}{\gamma}\frac{\delta s}{s}.
\label{dsigma2}
\end{equation}
The first term on the right-hand side of Eq.~(\ref{dsigma2})
corresponds to the perturbed surface density associated to evanescent
pressure waves (like in the barotropic case, where it reduces to $K *
\delta l / l$), and the second term to the density lobes resulting
from entropy advection. Since the convolution by the unitary function
$K$ in Eq.~(\ref{dsigma2}) does not change the linear mass of the
perturbation ($\int\delta\Sigma(x)dx$), the corotation torque is the
same as if, in the expression for the density perturbation in
Eq.~(\ref{dsigma2}), the convolution product were actually discarded
\cite{mc09}. In the barotropic case for instance, this leads to the
torque expression given by Eq.~(\ref{eq:18}).  In the adiabatic case,
it shows that, counter-intuitively, the density lobes exert no
\emph{net} corotation torque. This further explains why, akin to the
barotropic case, the calculation of the corotation torque comes to
evaluating the vortensity distribution within the horse shoe
region. Since the main difference in the vortensity field between the
adiabatic and barotropic cases is the appearance of a singular sheet
of vorticity at the downstream separatrices, and given that the
magnitude of this sheet scales with the entropy gradient, this
singular vorticity sheet can be unambiguously identified as the origin
of the entropy-related torque.

Upon evaluation of the magnitude of the vorticity sheet at the
separatrices, \cite{mc09} inferred the following expression 
for the entropy-related corotation torque:
\begin{equation}
  \label{eq:25}
  \Delta\Gamma_{\rm HS}^{\rm entr}=-\frac{1.3{\cal S}}{\epsilon/H}\Sigma\Omega_p^2q^2a^4h^{-2},
\end{equation}
where the above expression has been derived in the framework of a flat
temperature profile, and assuming a ratio of specific heats
$\gamma=1.4$.  In Eq.~(\ref{eq:25}), all disc quantities are to be
evaluated at the planet's orbital radius, and the quantity ${\cal S}$
is defined by
\begin{equation}
  \label{eq:calS}
  {\cal S} = \frac{1}{\gamma}\frac{d\log s}{d\log r},
\end{equation}
and can be recast as $[\beta + (\gamma-1)\alpha)]/\gamma$ for surface
density and temperature profiles that can be approximated as power-law
functions of radius over the planet's horseshoe region.

Considering discs with arbitrary temperature profiles, \cite{pbck10}
also evaluated the production of vortensity at downstream
separatrices, which required estimating the velocity along streamlines
through a fit of numerical simulations.  Unlike \cite{mc09}, they
manually added the torque contribution from the density lobe
structure. In the end, the latter remains small compared to the torque
contribution from the singular sheet of vorticity.  This explains why,
overall, the derivations of the entropy-related corotation torque by
\cite{mc09} and \cite{pbck10} are in broad numerical agreement, within
$30$~\%.

The generalisation to an arbitrary temperature profile of
Eq.~(\ref{eq:25}) cannot be tackled fully analytically, much as in the
locally isothermal case. Yet, Eq.~(\ref{eq:23}) shows that the
adiabatic corotation torque is the sum of the entropy related term,
given by Eq.~(\ref{eq:25}), and the locally isothermal corotation
torque (corrected by a factor $\gamma$). The latter is itself made up
of two terms, as we have seen in sections~\ref{sec:hors-drag-barotr}
and~\ref{sec:hors-drag-locally}. The corotation torque is therefore,
in a general situation, the sum of three terms:

\begin{itemize}
\item The vortensity related torque, proportional to the vortensity
  gradient, and given by Eq.~(\ref{eq:19}).
\item The temperature related torque, proportional to the temperature
  gradient, discussed in section~\ref{sec:hors-drag-locally}.
\item The entropy related torque, proportional to the entropy
  gradient, given by Eq.~(\ref{eq:25}).
\end{itemize}
  
It can be observed that there are only two degrees of freedom for the
disc profiles (the density and temperature gradients $\alpha$ and
$\beta$, or the vortensity and entropy gradients ${\cal V}$ and ${\cal
  S}$, etc.), so that for a specific setup one may simplify the torque
expression as a linear combination of the two independent parameters.
This simplification is not desirable, however, because it blurs the
distinct physical origin and characteristic of each of the three
terms. Besides, one can disentangle these three terms by varying
parameters such as the smoothing length $\epsilon$ or the ratio of
specific heats $\gamma$. Any simplification of the torque expression
is thus highly setup dependent.
  
Finally, we summarise the main message to take away about the 
corotation torque.
\begin{itemize}
\item In all cases it features a term that scales with the gradient of
  vortensity across the horseshoe region, given by
  Eq.~(\ref{eq:19}). It has one or two additional terms, depending on
  whether an energy equation is taken into consideration. The first of
  those scales with the temperature gradient, and if an energy
  equation is included, there is a second one that scales with the
  entropy gradient.
\item In all cases the corotation torque comes from the {\em
    vortensity} distribution in the horseshoe region. The additional
  contributions arise from the vortensity {\em created} by the
  temperature gradient and/or the entropy gradient.
\end{itemize}

\subsection{Saturation properties of the horseshoe drag}
\label{sec:satur-prop}
We have described in Section~\ref{sec:corotation-torque} the physical
origin and properties of the corotation torque in inviscid discs, with
a special emphasis on the fully unsaturated horseshoe drag, which is
the maximum value the corotation torque may take. This value is
obtained about one horseshoe U-turn timescale after the planet
insertion in the disc, and is maintained over about half a libration
timescale.  Its sign and magnitude are determined by the gradients of
vortensity, temperature and entropy across the horseshoe region.

In the absence of diffusion processes, after about half a libration
timescale, the vortensity and entropy advected along the downstream
separatrices of the horseshoe region reach the planet again, undergo
another U-turn, and phase mixing starts to occur. Vortensity and
entropy are progressively stirred up within the horseshoe region, and
the horseshoe drag oscillates with time with a decreasing amplitude,
as shown in Fig.~\ref{fig:tqvstime}. The horseshoe drag ultimately
cancels out as both vortensity and entropy get uniformly distributed
after several libration times \cite{pbk11, mc10}. This is known as the
horseshoe drag saturation.

Diffusion processes (viscosity, thermal diffusion) may maintain
respectively the vortensity and entropy gradients across the horseshoe
region, and thus sustain the horseshoe drag to a non-vanishing
value. We review below the saturation properties of the horseshoe drag
in barotropic discs (\S~\ref{sec:satu_baro}) and in radiative discs
(\S~\ref{sec:satu_rad}).

\subsubsection{Saturation properties of the vortensity-related horseshoe drag 
in barotropic viscous discs}
\label{sec:satu_baro}

In barotropic discs, the horseshoe drag saturates as vortensity is
strictly advected along horseshoe streamlines. Viscosity acting as a
diffusion source term in the vortensity equation can sustain a
non-zero vortensity gradient across the horseshoe region. The
vortensity-related horseshoe drag then attains a steady-state value,
which arises from a net exchange of angular momentum between the
horseshoe region and the rest of the disc \cite{masset01,
  masset02}. This steady-state value depends on how the viscous
diffusion timescale across the horseshoe region ($\tau_{\rm visc}$)
compares with the horseshoe libration timescale ($\tau_{\rm lib}$) and
the horseshoe U-turn timescale ($\tau_{\rm U-turn}$). Denoting by
$\nu_p$ the kinematic viscosity at the planet location, $\tau_{\rm
  visc}\sim x_s^2 / \nu_p$.  The libration timescale is given by
Eq.~(\ref{eq:taulib}), and the U-turn timescale is typically a
fraction $H/r$ of the libration timescale \cite{bm08a}.

For the corotation torque to remain close to its maximum, fully
unsaturated value in the long term, the inequality
\begin{equation}
\tau_{\rm U-turn} \leq \tau_{\rm visc} \leq \tau_{\rm lib}/2 
\label{eq:baro_desatu}
\end{equation}
should be verified. When the second inequality is satisfied, the
vortensity at the upstream separatrices is kept stationary, which
prevents phase mixing of vortensity within the horseshoe region
\cite{bk2001,masset01}. When the first inequality is satisfied,
vortensity is approximately conserved along U-turns, which maximises
the effective vortensity gradient across the horseshoe region
\cite{masset02,mc10,pbk11}. Taking $x_s \sim 1.1 a \sqrt{q/h_p}$ (as
measured with a planet softening length $\approx 0.6H$),
inequality~(\ref{eq:baro_desatu}) may be cast as
\begin{equation}
  0.32 q^{3/2} h_p^{-7/2} \leq \alpha_{\rm v,p} \leq 0.16q^{3/2} h_p^{-9/2},
  \label{eq:baro_desatu2}
\end{equation}
where $\alpha_{\rm v,p}$ and $h_p$ denote the disc's alpha viscosity
and aspect ratio at the planet location, respectively. The alpha
viscosity for which the corotation torque takes its maximum value can
be approximated as $0.16q^{3/2} h_p^{-4}$ \cite{bl10}.

\begin{figure}
	\centering
 	\includegraphics[width=0.8\hsize]{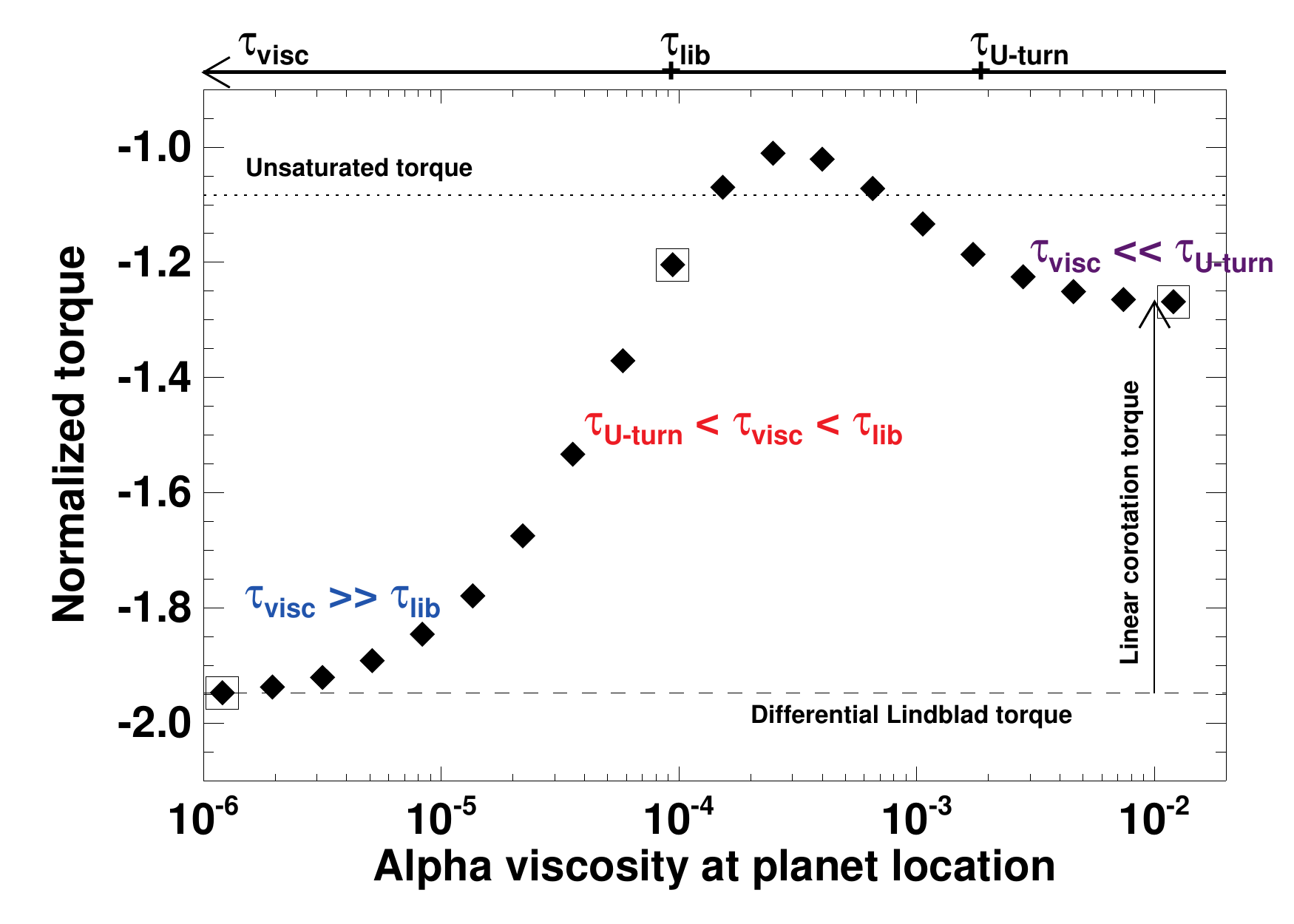}
	\centering
 	\includegraphics[width=0.32\hsize]{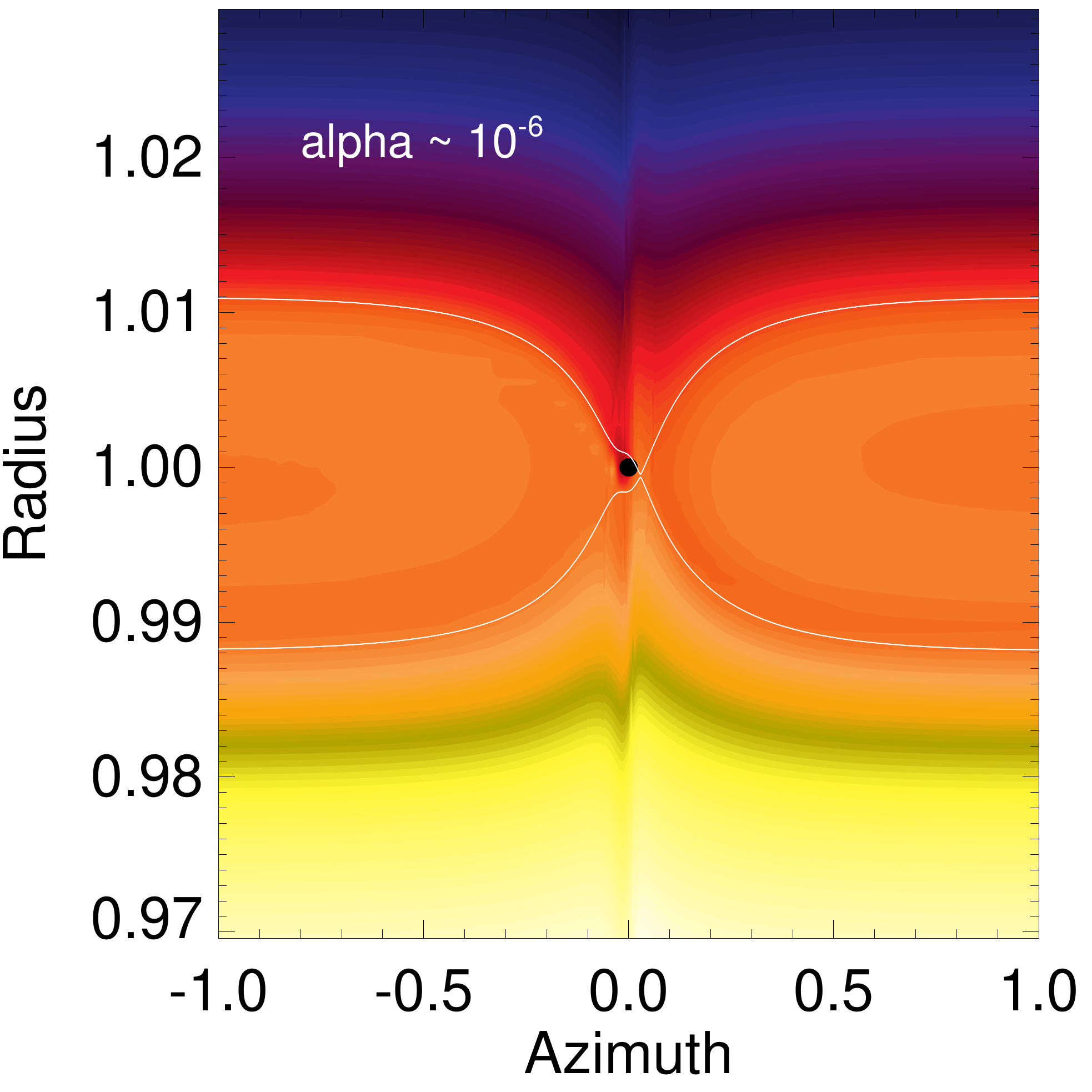}	
	\includegraphics[width=0.32\hsize]{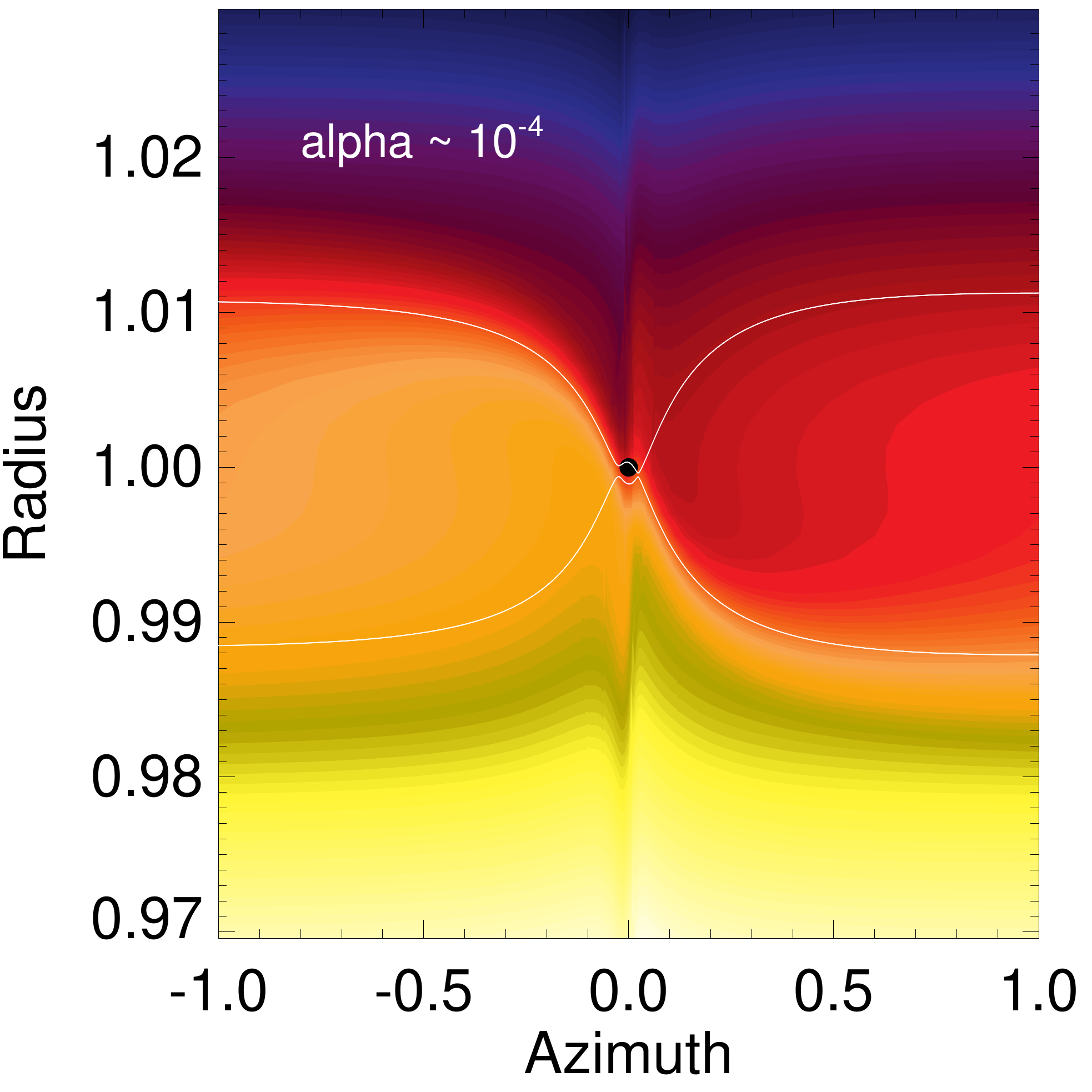}	
	\includegraphics[width=0.32\hsize]{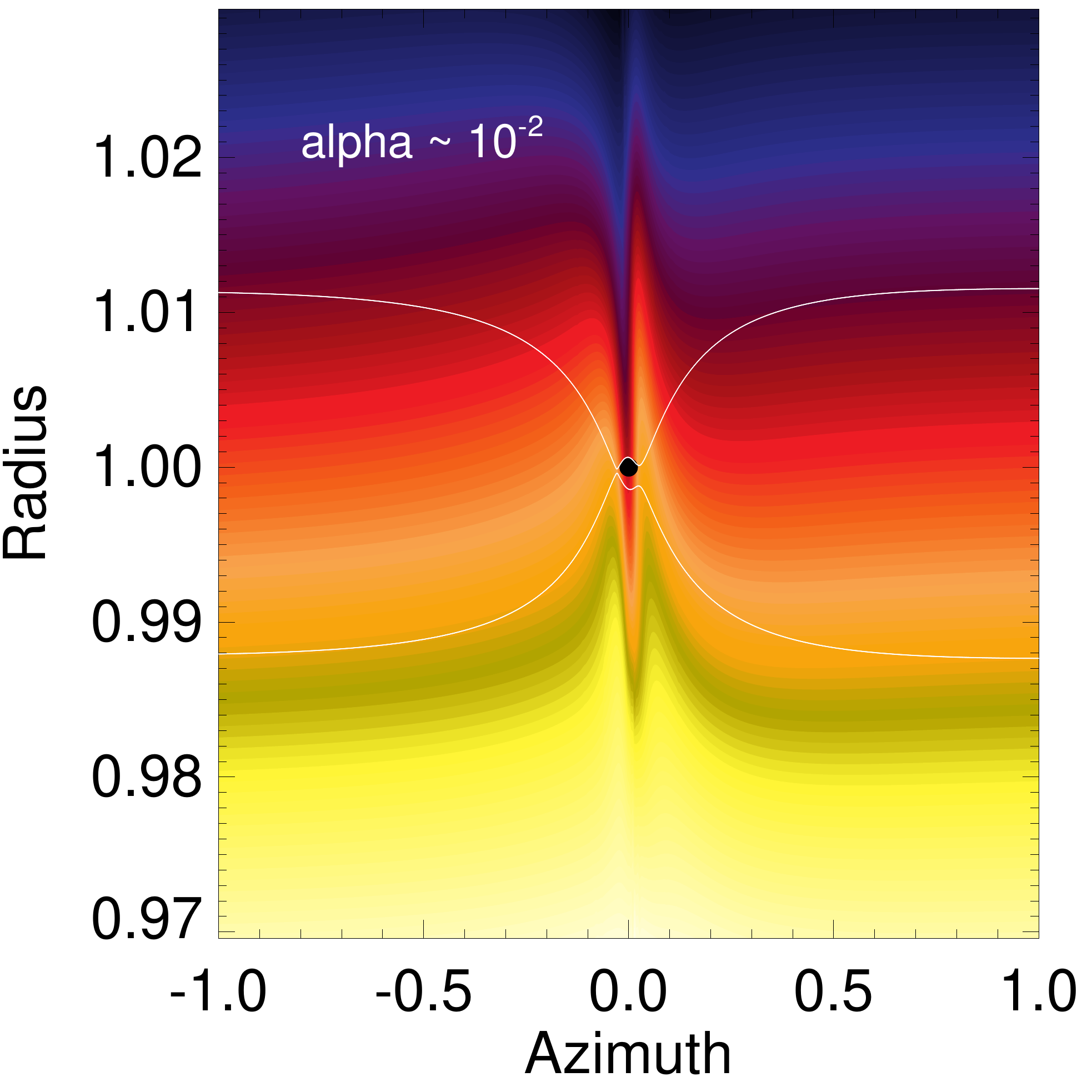}	
        \caption{Top: steady-state torque on a $M_p = 6\times 10^{-6}
          M_{\star}$ planet mass embedded in a thin ($h=0.05$) disc
          for various alpha viscous parameters at the planet
          location. In this series of runs, $\alpha=1/2$ and
          $\beta=0$.  Different saturation regimes of the corotation
          torque are illustrated, depending on how the viscous
          timescale across the planet's horseshoe region ($\tau_{\rm
            visc}$) compares with the horseshoe U-turn timescale
          ($\tau_{\rm U-turn}$) and the libration timescale
          ($\tau_{\rm lib}$). The final torque value in an inviscid
          case, which reduces to the differential Lindblad torque, is
          shown by a dashed line. The fully unsaturated total torque
          (differential Lindblad torque plus fully unsaturated
          horseshoe drag) in an inviscid run is depicted by a dotted
          line.  Bottom: vortensity distribution inside the planet's
          horseshoe region for the three alpha viscosities shown by
          squares in the top panel (viscosity increases from left to
          right). The separatrices of the horseshoe region are
          depicted by solid curves, and the planet position by a
          filled circle.}
  \label{fig:ctqvsvisc}
\end{figure}
The saturation properties of the corotation torque are illustrated in
Fig.~\ref{fig:ctqvsvisc} for a $2$~Earth-mass planet embedded in a
thin ($h_p=0.05$) viscous disc. The background temperature profile is
uniform, and the surface density decreases as $r^{-1/2}$. The top
panel displays the steady-state torque at different alpha viscosities
(a constant kinematic viscosity $\nu$ was used in the
simulations). The left-hand term in inequality~(\ref{eq:baro_desatu2})
is $\approx 1.7\times 10^{-4}$, while the right-hand term in $\approx
1.7\times 10^{-3}$, and it is clear from Fig.~\ref{fig:ctqvsvisc} that
the corotation torque is maximum between these two alpha
viscosities. When the viscosity is small enough so that $\tau_{\rm
  visc} \gg \tau_{\rm lib}$, viscosity is inefficient at restoring the
vortensity gradient across the horseshoe region, and the horseshoe
drag takes very small values (it saturates). At very large
viscosities, such that $\tau_{\rm visc} \ll \tau_{\rm U-turn}$, the
corotation torque plateaus at its value in the linear regime
\cite{pp09a}. The vortensity distribution inside the horseshoe region
for each saturation regime is shown in the bottom panels of
Fig.~\ref{fig:ctqvsvisc}. In the left panel, $\alpha_{\rm v,p} \sim
10^{-6}$, and the steady-state vortensity distribution within the
horseshoe region is uniform, resulting in a vanishing corotation
torque. In the middle panel, $\alpha_{\rm v,p} \sim 10^{-4}$ maintains
a maximum vortensity contrast between the rear and front parts of the
planet, with the consequence that the corotation torque is close to
its fully unsaturated value. In the right panel, $\alpha_{\rm v,p}
\sim 10^{-2}$ imposes the initial (unperturbed) vortensity profile
along the horseshoe U-turns, and the horseshoe drag therefore reduces
to the linear corotation torque.

All attempts to capture analytically the saturation of the corotation
torque have been carried out using a simplified streamline model that
assumes the drift of the coorbital material with the velocity of the
unperturbed disc, and which does not resolve spatially nor temporally
the U-turns. This model was proposed in \cite{masset01} and in a more
formal manner in \cite{mc10}, where a numerical implementation of it
is also described.  All analytic works on the corotation torque
saturation, whether they provide an asymptotic torque value
\cite{masset01,mc10,pbk11} or they capture the time dependence of the
torque in an inviscid disc \cite{2007LPI....38.2289W} make use of this
simplified model. Solving for the torque asymptotic value in this
simplified advection-diffusion model can be tacked in a variety of
ways.  One such way consists of neglecting the azimuthal variation of
the vortensity so as to reduce the advection-diffusion problem
essentially to a one dimensional radial problem. This is the approach
of \cite{masset01} and \cite{pbk11}. These two works are quite
different in their assumptions, and suffer from orthogonal
restrictions:
\begin{itemize}
\item \cite{masset01} exclusively contemplates the case of a disc with
  flat profiles of surface density and kinematic viscosity, so that
  his results must be rescaled by hand to apply to a general case. The
  approach used in this work considers the global angular momentum
  budget of the trapped horseshoe region, and relies upon the
  evaluation of the viscous friction of the disc on the
  separatrices. It also takes into account the viscous drift of
  material across the horseshoe region.
\item \cite{pbk11} consider a disc with an arbitrary surface density 
  gradient, and directly use the horseshoe drag integral of
  Eq.~(\ref{eq:18}). Their model assumes no radial drift of 
  disc material across the horseshoe region.
\end{itemize}
Quite remarkably, these two approaches yield the exact same result,
which can be cast either in terms of Airy functions \cite{masset01} or
in terms of Bessel functions \cite{pbk11}.

One can also solve the advection-diffusion equation satisfied by the
fluid's vortensity in two dimensions, the solution being exact in the
limit of a small viscosity. In this limit, the problem amounts to an
alternation of convolutions (viscous diffusion of vortensity between
two successive horseshoe U-turns) and reflections (mapping of
vortensity -- or, vortensity conservation -- from one tip of the
horseshoe region to the other during a U-turn).  This is the approach
of \cite{mc10}, who also discard the possible radial drift of disc
material across the horseshoe region. The dependence thus obtained
---~equation~(119) of \cite{mc10}~--- is broadly the same as that of
\cite{masset01} and \cite{pbk11}, but reproduces more closely the
results from numerical simulations. We note that the decay of the
torque value found at large viscosity (see Fig.~\ref{fig:ctqvsvisc}),
which corresponds to a decay towards the linear corotation torque
value \cite{pp09a}, has not yet been described analytically in a
self-contained manner. \cite{mc10} and \cite{pbk11} use an ad-hoc
reduction factor, either with one free parameter \cite{mc10} or two
free parameters \cite{pbk11}, the value of the free parameters being
inferred from numerical simulations.

In summary, in barotropic discs, vortensity essentially obeys an
advection-diffusion equation in the coorbital region. When the viscous
diffusion timescale across the horseshoe region is:
\begin{itemize}
\item Long compared to the libration period, vortensity is
  progressively stirred up and the corotation torque ultimately
  saturates (tends to zero).
\item Short compared to the libration period, but long compared to the
  horseshoe U-turn time, the corotation torque is close to its fully
  unsaturated, maximum value.
\item Short compared to the horseshoe U-turn time, the corotation
  torque tends to its value predicted in the linear regime. 
\end{itemize}

\subsubsection{Saturation properties of the horseshoe drag 
in radiative discs}
\label{sec:satu_rad}
Much as in barotropic discs, the estimate of the asymptotic corotation
torque value in radiative discs amounts to the determination of the
vortensity distribution within the horseshoe region at later times.
There is an additional complexity, however, due to the fact that this
is no longer an advection-diffusion problem, but an
advection-diffusion-creation problem, as vortensity is created during
the U-turns (see section~\ref{sec:hors-drag-adiab}, and in particular
Fig.~\ref{fig:vortensadiab}). Furthermore, the amount of vortensity
created depends on the entropy distribution, as was explained in
section~\ref{sec:hors-drag-adiab}. This analysis was undertaken by
\cite{pbk11} in the case of a unitary thermal Prandtl number (the
viscosity $\nu$ and thermal diffusion $\chi$ have same value). A
corotation torque expression was proposed by these authors, as a
result of a fit of numerical simulations. Under the assumption of a
unitary Prandtl number, the parameter space to be explored is
one-dimensional, and for a (common) value of $\nu$ and $\chi$, the
radiative torque is found to saturate more easily than the barotropic
torque. This is interpreted by the authors as due to the fact the
entropy-related corotation torque is essentially due to a unique
streamline, where the advection speed is maximal (that of the
separatrices).

To relax the assumption of a unitary Prandtl number, one may assume
that the torque dependence upon viscosity or thermal diffusion have
the same shape, which allows to propose a formula with two independent
parameters $\nu$ and $\chi$, which can then be validated by checking
its accuracy with numerical simulations. This is the approach adopted
by \cite{pbk11}. Another solution consists in using a streamline model
such as the one outlined in section~\ref{sec:satu_baro}. This is the
approach of \cite{mc10}.  As the vortensity is now determined by an
advection-diffusion-creation problem, one needs to amend the
barotropic model of section~\ref{sec:satu_baro} by adding the creation
of vortensity during the U-turns, which is determined by the entropy
field.  Therefore, prior to the determination of the vortensity
distribution, an analysis of the entropy distribution at later times
is required. This preliminary determination can be done easily,
because the entropy obeys an advection-diffusion problem formally
equivalent to the vortensity distribution in the barotropic case, in
which one replaces the vortensity with the entropy, and the viscosity
$\nu$ with the thermal diffusion $\chi$. Once the entropy distribution
within the horseshoe region is known, the vortensity distribution at
late times is obtained, which allows, upon the use of the horseshoe
drag expression of Eq.~(\ref{eq:18}), for an expression of the
corotation torque as a function of viscosity and thermal diffusivity,
and which can be checked {\em a posteriori} against numerical
simulations.

The corotation torque expressions, as a function of viscosity and
thermal diffusivity, are given by Eqs.~(161-164) of \cite{mc10}, or by
Eqs.~(52-53) of \cite{pbk11}.

\subsection{Type I migration in turbulent discs}
We have examined in the previous sections the properties of
planet--disc interactions assuming viscous discs, described with a
stationary kinematic viscosity aimed at modelling their turbulent
transport properties. Because the corotation torque may play a
dramatic role in the orbital evolution of low-mass planets, and its
magnitude is intimately related to diffusion processes taking place
within the planet's horseshoe region, it is relevant to determine how
turbulence may impact type I migration.

Turbulence in protoplanetary discs can have a variety of origins.
These include hydrodynamic instabilities, such as Rossby-wave
instabilities \cite{lovelace99}, the global baroclinic instability
\cite{kb2003, Lyra11}, the sub-critical baroclinic instability
\cite{LesurPap10}, planetary gap instabilities \cite{minkai11_edge,
  minkai11_vortex} (which we will discuss in \S~\ref{sec:type3}), or
the Kelvin-Helmholtz instability triggered by the vertical shear of
the gas as dust settles into the mid plane \cite{jhk06}. Convective
instability might also be relevant in the inner parts of massive
discs, and it would be interesting to examine its impact on type I
migration. Perhaps the most likely source of turbulence in
protoplanetary discs is the magnetohydrodynamic (MHD) turbulence
resulting from the magnetorotational instability (MRI) \cite{bh91}. It
relies on the coupling of the ionised gas to the weak magnetic field
in the disc. Ionisation may occur in the vicinity of the central
object due to the star's irradiation, or further out in the disc
layers, most probably through the UV background or cosmic rays. It is
currently debated which regions of planet formation near the disc mid
plane are sufficiently ionised ('active') to trigger the MRI, and
which ones remain neutral (which is usually referred to a 'dead
zone'). In the latter case, some transport of angular momentum would
still be present through the propagation of waves induced by MHD
turbulence in the disc's upper layers \cite{flemingstone03}. The alpha
viscous parameter associated to MHD turbulence is typically in the
range $[5\times 10^{-3}-5\times10^{-2}]$ in active regions, while
being about two orders of magnitude smaller in dead zones.

\begin{figure}
  \centering
  \resizebox{\hsize}{!}
  {
  \includegraphics[width=\hsize]{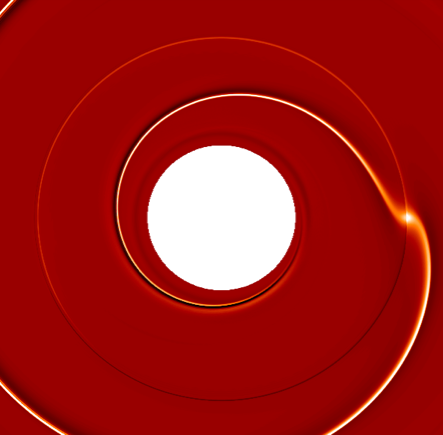}
  \hspace{1cm}
  \includegraphics[width=0.9628\hsize]{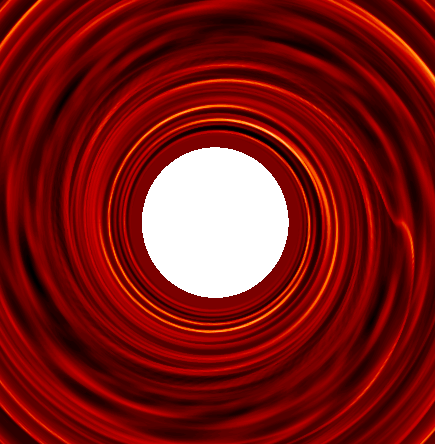}
  \hspace{1cm}
  \includegraphics[width=0.5575\hsize]{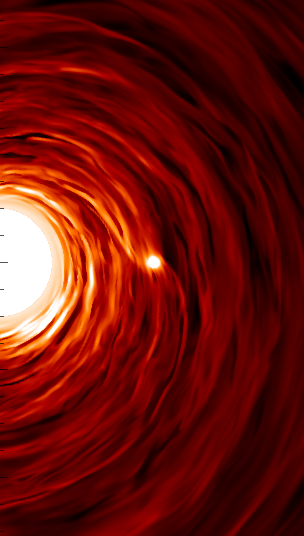}
  }
  \caption{\label{fig:turb}Perturbation of the disc's density by an
    embedded planet. Left: adiabatic two-dimensional disc. The density
    lobes within the coorbital region, which arise from the advection
    of entropy, help identify the (tiny) radial width of the planet's
    horseshoe region. Middle: case of an isothermal 2D disc with
    turbulence induced by stochastic stirring. Right: case of an
    isothermal 3D disc invaded by MHD turbulence due to the MRI (the
    density in the disc mid plane is displayed). In the middle and
    right panels, the turbulent density perturbation is comparable to
    the perturbed density associated to the planet's wakes.  Images
    taken from \cite{BaruteauPhD, bl10, bfnm11}, respectively.}
\end{figure}

The properties of type I migration in weakly magnetised turbulent
discs have been investigated in a couple of studies. \cite{qmwmhd4}
performed 3D simulations of locally isothermal discs fully invaded by
MHD turbulence. They found that the running time-averaged torque on a
fixed protoplanet experiences rather large-amplitude oscillations over
the reduced temporal range over which the simulation could be run, and
that its final value differs quite substantially from the torque value
expected in viscous disc models. Similar results were obtained by
\cite{nelson05}, who allowed the planet orbit to evolve. A primary
reason for the observed difference between the viscous torque and the
time-averaged turbulent torque is that the 3D MHD simulations were not
converged in time. This was suggested by \cite{bl10}, who considered
2D isothermal discs subject to stochastic forcing, using the
turbulence model originally developed by \cite{lsa04}.  They showed
indeed that when time-averaged over a sufficiently long time period,
which may be as long as a thousand orbits, both the differential
Lindblad torque and the corotation torque behave very similarly as in
equivalent viscous disc models. These results were essentially
confirmed by the 3D MHD simulations by \cite{bfnm11}, who adopted a
locally isothermal disc model with a mean toroidal magnetic field, in
which non-ideal MHD effects and vertical stratification were neglected
(see illustration in Fig.~\ref{fig:turb}). Similar agreement was
obtained by \cite{Uribe11} with vertical stratification. Nonetheless,
\cite{bfnm11} found an additional corotation torque with moderate
magnitude in their 3D MHD simulations, related to the presence of a
mean toroidal magnetic field. The existence and properties of this
additional corotation torque have been explored by Guilet, Baruteau \&
Papaloizou (in prep.) in 2D weakly magnetised, non-turbulent disc
models, in which the effects of turbulence are modelled by viscous and
magnetic diffusivities. They find that the additional corotation
torque can take large values, and even exceed the differential
Lindblad torque, depending on the disc's viscous and magnetic
diffusivities, and the amplitude of the background magnetic field.

The aforementioned results were for embedded planets with a horseshoe
radial width that is a moderate fraction of the disc's pressure
scaleheight, the latter being the typical size of turbulent eddies.
The existence of horseshoe dynamics and a corotation torque is unknown
for planets with a horseshoe region width that is a small fraction of
the turbulent eddy size. In this case, it is possible that turbulence
acts more as a source of random advection of vortensity through the
horseshoe region, rather than diffusion.

\section{Migration of gap-opening planets: Type II and III migration}
\label{sec:gapopening}
The disc response to a low-mass planet has been studied in details in
\S~\ref{sec:type1}, where we have focused on the two components of the
type I migration torque. The aim of this section is to examine the
range of planet masses that is relevant to type I migration in
\S~\ref{sec:gap}, and to give a concise description of planet--disc
interactions for planets that are massive enough to significantly
perturb the disc's mass distribution (\S~\ref{sec:type3}
and~\ref{sec:type2}).

\subsection{Shock formation and gap-opening criterion}
\label{sec:gap}
The wakes generated by a planet in a disc carry angular momentum as
they propagate away from the planet. This angular momentum is
eventually deposited in the disc through some wave damping processes,
which leads to redistributing the disc mass. An efficient wave damping
mechanism relies on the non-linear wave evolution of the wakes into
shocks \cite{gr2001}. The (negative) angular momentum deposited by the
inner wake decreases the semi-major axis of the fluid elements in the
disc region inside the planet's orbit (the inner disc). Similarly, the
(positive) angular momentum deposited by the outer wake increases the
semi-major axis of the fluid elements in the outer disc.

The distance $d_s$ from the planet where the planet-generated 
wakes become shocks is given by \cite{gr2001, Dong11b}:
\begin{equation}
  d_s \approx 0.8 \left( \frac{\gamma+1}{12/5} \, \frac{q}{h^3} \right)^{-2/5} H(a),
  \label{eq:ds}
\end{equation}
where $\gamma$ is the gas adiabatic index, and $a$ denotes the
planet's semi-major axis. As the magnitude of the one-sided Lindblad
torque peaks at $\sim 4H(a)/3$ from the planet's orbit, a linear
description of the differential Lindblad torque thus fails when $|d_s|
\sim 4H(a)/3$. This condition can be recast as $q \sim 0.3h^3$ for
$\gamma=5/3$. When $|d_s| \leq 2H(a)/3$, wakes turn into shocks within
their excitation region. Fluid elements just outside the planet's
horseshoe region are pushed away from the planet orbit after crossing
the wakes, which directly affects the planet's coorbital region by
inducing asymmetric U-turns \cite{Oleron}.  Horseshoe fluid elements
therefore get progressively repelled from the planet orbit after each
U-turn, and the planet slowly depletes its coorbital region. The
equilibrium structure (width, depth) of the annular gap the planet
forms around its orbit is determined by a balance between gravity,
viscous and pressure torques \cite{crida06}.

\begin{figure}
\sidecaption
  \resizebox{0.6\hsize}{!}
  {
    \includegraphics{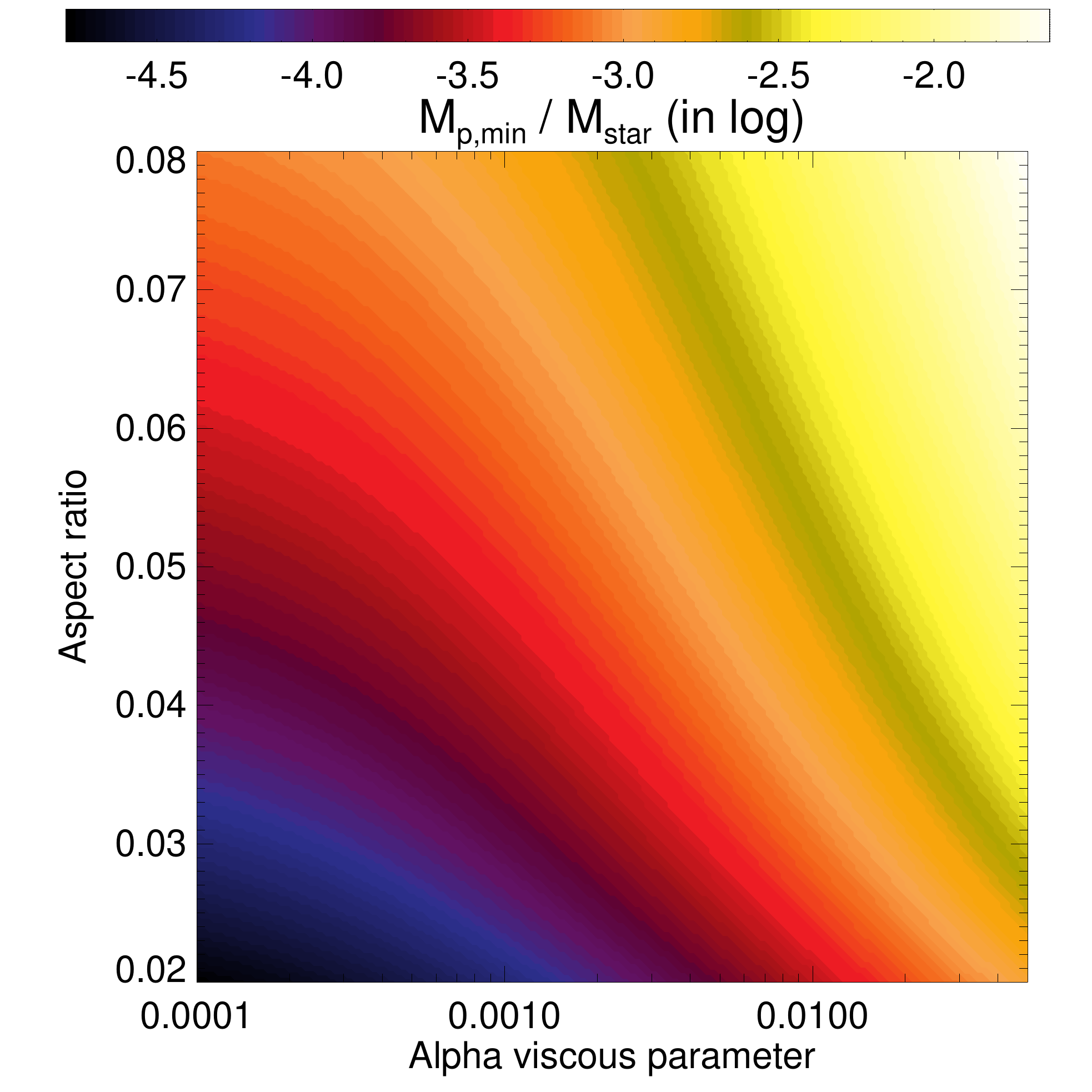} 
  }
  \caption{\label{fig:gapcrit}Minimum planet-to-primary mass ratio
    leading to gap-opening as a function of the disc's alpha viscous
    parameter (x-axis) and aspect ratio (y-axis) at the planet's
    semi-major axis. This minimum mass is calculated numerically using
    criterion~(\ref{eq:gapcriterion}).}
\end{figure}

Shock formation and its damping efficiency are very sensitive to  
the disc's viscosity, and gap-opening results from a balance
between (i) a planet mass large enough to induce shocks where the  
wake excitation takes place, and (ii) a viscosity small enough to 
maximise the amount of angular momentum deposited by the shocks 
in the planet's immediate vicinity: 
\begin{enumerate}
\item The first condition reads $|d_s| \leq 2H(a)/3$, which corresponds 
to $q \geq 1.5 h^3$ for $\gamma=5/3$. It means that the planet's Bondi radius 
$r_B = GM_p / c_s^2$, where the pressure distribution is most strongly perturbed 
by the planet, as well the planet's Hill radius become comparable to the local 
pressure scaleheight. This is known as the thermal criterion for gap opening 
\cite{PPIII}. 
\item The second condition, known as the viscous criterion, can be
  expressed as $q \geq 40 / {\cal R}$, where ${\cal R} = a_p^2
  \Omega_p / \nu$ is the Reynolds number\footnote{Although
    traditionally dubbed Reynolds number essentially for dimensional
    considerations, this ratio has little to do with the dimensionless
    ratio that must be considered to assess whether a flow is laminar
    or turbulent. If one regards the planet as an obstacle in the
    sheared Keplerian flow, it would be more appropriate to consider
    as a characteristic scale the size of its Roche lobe or $\sim
    x_s$, and as a characteristic velocity $2|A|x_s$.}
  \cite{LinPap79,bryden99}.
\end{enumerate} 
The above two conditions for gap-opening have been revisited by
\cite{crida06}, who provide a unified criterion that takes the form
\begin{equation}
  1.1\left(\frac{q}{h^3}\right)^{-1/3} + \frac{50\alpha_{\rm v} h^2}{q} \leq 1,
\label{eq:gapcriterion}
\end{equation}
where we have written the disc's kinematic viscosity $\nu =
\alpha_{\rm v} h^2 a^2 \Omega$ \cite{ss73}, and where in
Eq.~(\ref{eq:gapcriterion}) $h$ and $\alpha_{\rm v}$ are to be
evaluated at the planet's semi-major axis. An illustration of the
smallest planet mass opening a gap according to
criterion~(\ref{eq:gapcriterion}) is shown in Fig.~\ref{fig:gapcrit},
where it is clear that the gap-opening mass increases with increasing
disc viscosity and aspect ratio. Assuming $h \approx 0.05$, which may
be typical of planet forming regions, the gap-opening mass is in the
Saturn-mass range for regions with low turbulent activity (dead zones,
with typically $\alpha_{\rm v} \sim 10^{-4}$), and is in the
Jupiter-mass range in regions where $\alpha_{\rm v} \sim 10^{-2}$.
Note that when disc self-gravity is included, the gap-opening
criterion of Eq.~(\ref{eq:gapcriterion}) should involve the effective
planet mass, that is the sum of the planet and circumplanetary disc
masses, rather than the planet mass alone.

Prior to a depletion of their coorbital region due to shock formation
at the wake's excitation region, planets with increasing mass
experience a flow transition in their immediate vicinity. The flow
passes from a low-mass planet configuration described in
section~\ref{sec:horseshoe-dynamics} and Fig.~\ref{fig:sllm}, to a
high-mass configuration, where fluid elements may become trapped
inside a circumplanetary disc around the planet. This flow transition
is accompanied by a rapid increase in the half-width $x_s$ of the
horseshoe region, from a fluid-dominated regime (where $x_s \propto
(q/h)^{1/2}$) to a gravity-dominated regime (where $x_s \sim R_H
\propto q^{1/3}$) \cite{mak2006}. This rapid increase yields a
significant increase in the corotation torque, as the latter scales as
$x_s^4$. This effect is found to be most significant for
planet-to-primary mass ratios $q\sim 0.6h^3$ \cite{mak2006}, which
corresponds to 20 Earth-mass planets in $h=0.05$ discs. It may
contribute to further slowing down, or even reversing the migration of
growing planets before they carve a gap around their orbit
\cite{gda2003, mak2006}.

\subsection{Partial gap-opening: type III migration in massive discs}
\label{sec:type3}
\begin{figure}
\centering
 \resizebox{\hsize}{!}
  {
  	\includegraphics[width=0.45\hsize]{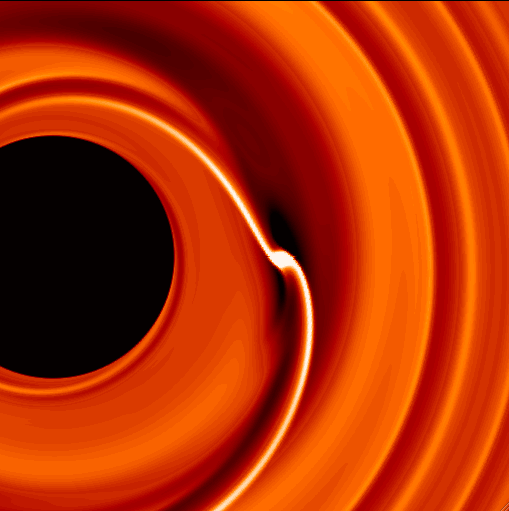}	
	 \hspace{1cm}
	 \includegraphics[width=0.55\hsize]{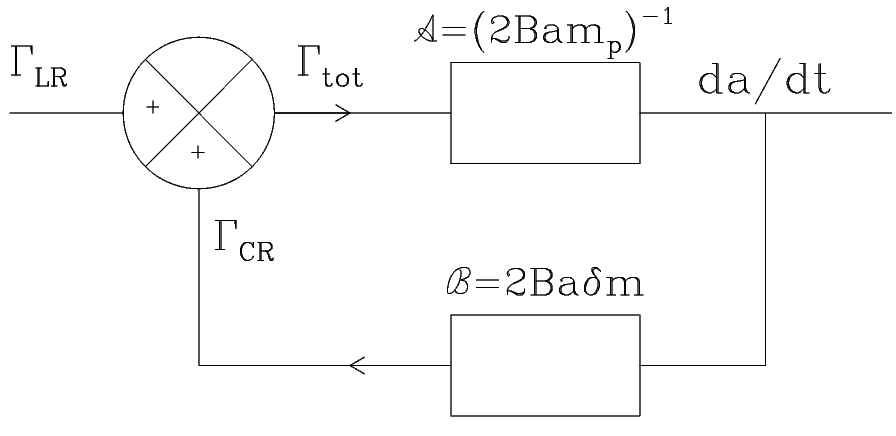}
  }
  \caption{Left: illustration of the flow asymmetry ahead of and
    behind a Saturn-mass planet undergoing rapid inward runaway
    migration. Right: type~III planetary migration seen as a feedback
    loop. The latter remains stable if the open-loop transfer function
    ${\cal A}\times {\cal B} < 1$, or $\delta m < M_p$. From
    \cite{Oleron}.  }
  \label{fig:type3}
\end{figure}

So far, we have addressed the properties of planet migration through a
direct analysis of the tidal torque, the latter being directly
proportional to the migration rate, see Eq.~(\ref{eq:adot}).  This
approach is valid for low-mass planets that do not open a gap, for
which migration has a negligible feedback on the tidal torque (note
that a weak, negative feedback slightly decreases the magnitude of the
entropy-related horseshoe drag \cite{mc09}).  Nevertheless, migrating
planets that open a partial gap around their orbit experience an
additional corotation torque due to fluid elements flowing across the
horseshoe region \cite{mp03}. If for instance the planet migrates
inwards, fluid elements circulating near the inner separatrix of the
horseshoe region enter the horseshoe region, and execute an outward
U-turn when they reach the vicinity of the planet.  Upon completion of
the U-turn, these fluid elements leave the horseshoe region as the
planet keeps migrating, and end up circulating in the outer
disc. Consequently, the mass distribution within the horseshoe region
may become asymmetric, as the horseshoe region adopts approximately a
trapezoidal shape in the azimuth-radius plane \cite{mp03}. As a
consequence, in the case of an inward migrating planet, there is more
mass behind the planet than ahead of it, owing to the partial
depletion of the asymmetric horseshoe region. This point is
illustrated in the left panel of Fig.~\ref{fig:type3}. Similarly, if
the planet migrates outwards, fluid elements circulating near the
outer separatrix may embark on single inward U-turns across the
horseshoe region.

Assuming steady migration at a moderate rate (this point will be
clarified below), the additional corotation torque experienced by the
planet due to the orbit-crossing flow is, to lowest order in $x_s/a$,
\begin{equation}
  \Gamma_{\rm cross} = 2\pi a\dot{a}\Sigma_s \times 4Ba x_s,
  \label{gammacross}
\end{equation}
where $\Sigma_s$ is the surface density at the inner (outer) horseshoe
separatrix for a planet migrating inwards (outwards). The term $2\pi
a\dot{a}\Sigma_s$ in the right-hand side of Eq.~(\ref{gammacross}) is
the mass flux across the horseshoe region. The second term ($4Ba x_s$)
is the amount of specific angular momentum that a fluid element near a
horseshoe separatrix exchanges with the planet when performing a
horseshoe U-turn. Note that the above expression for $\Gamma_{\rm
  cross}$ assumes that all circulating fluid elements entering the
coorbital region embark on horseshoe U-turns, whereas a fraction of
them may actually become trapped inside the planet's circumplanetary
disc.  Since $\Gamma_{\rm cross}$ is proportional to, and has same
sign as $\dot{a}$, migration may become a runaway process. We now
discuss under which circumstances a runaway may happen.

The planet and its coorbital material (which encompasses the horseshoe
region, with mass $M_{\rm hs}$, and the circumplanetary disc, with
mass $M_{\rm cpd}$) migrate at the same drift rate, $\dot{a}$, which
we assume to be constant. The rate of angular momentum change of the
planet and its coorbital region includes (i) the above contribution
$\Gamma_{\rm cross}$ to the corotation torque, and (ii) the tidal
torque that, for planets opening a partial gap, essentially reduces to
the differential Lindblad torque $\Gamma_{\rm LR}$:
\begin{equation}
  2Ba\dot{a}(M_p + M_{\rm cpd} + M_{\rm hs}) = 2\pi a\dot{a}\Sigma_s\times 4Bax_s + \Gamma_{\rm LR}.
  \label{eq:runaway1}
\end{equation}
Eq.~(\ref{eq:runaway1}) can be written as
\begin{equation}
  2Ba\dot{a}\tilde{M}_p = 2Ba\dot{a}\delta m + \Gamma_{\rm LR},
  \label{cmd1}
\end{equation}
where $\tilde{M}_p = M_p + M_{\rm cpd}$ corresponds to an
\emph{effective} planet mass, and where $\delta m = 4\pi a x_s
\Sigma_s - M_{\rm hs}$ is called the coorbital mass deficit
\cite{mp03}. It represents the difference between (i) the mass the
horseshoe region would have if it had a uniform surface density equal
to that of the separatrix-crossing flow, and (ii) the actual horseshoe
region mass.

The migration rate of a partial gap-opening planet, given by
Eq.~(\ref{cmd1}), can be described as a feedback loop \cite{Oleron}.
This is illustrated in the right panel of Fig.~\ref{fig:type3}, where
the loop input is the differential Lindblad torque, and its output is
the migration rate. When $\delta m < \tilde{M}_p$, the feedback loop
remains stable. The drift rate in this case is not strictly a type~I
nor a type~II migration rate. It is rather a type I rate enhanced by
coorbital effects. No special name has been assigned to this kind of
migrating regime.  When $\delta m > \tilde{M}_p$, the feedback loop
gets unstable, and migration enters a runaway regime, which can be
either inward or outward. The drift rate as a function of the disc
mass undergoes a bifurcation \cite{Oleron}, and this regime is called
runaway type~III migration \cite{mp03,Peplinski2,Peplinski3}.

\begin{figure}
  \centering
  \resizebox{0.8\hsize}{!}
  {
    \includegraphics{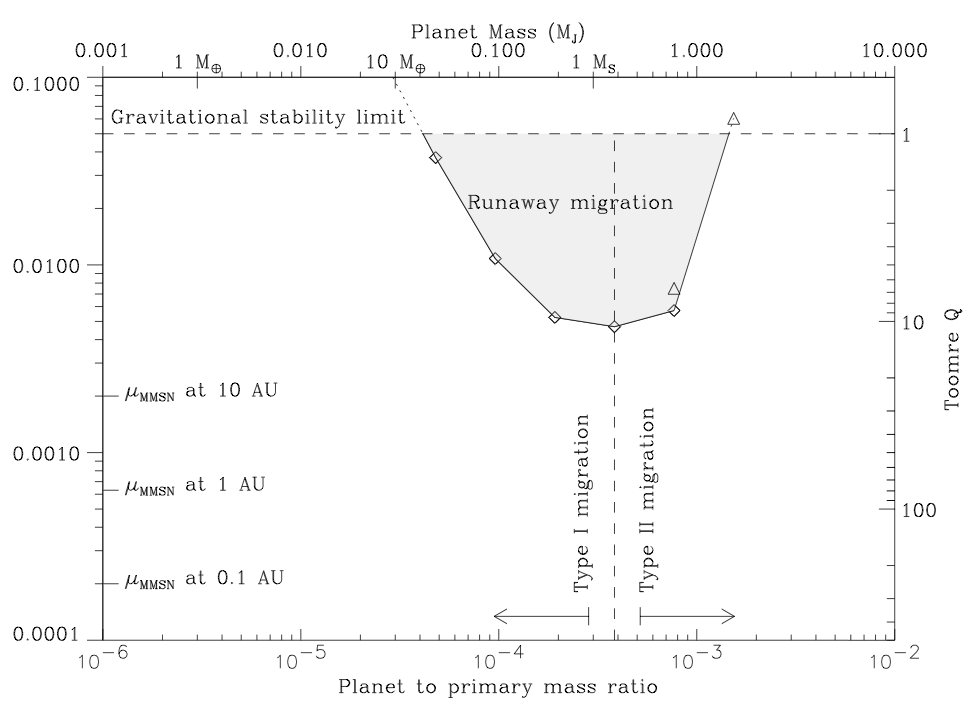} 
  }
  \caption{\label{fig:occur}Occurrence for type~I, ~II and ~III
    (runaway) migrations with varying the planet-to-primary mass ratio
    (bottom $x-$axis) and the disc-to-primary mass ratio at the planet
    location (left $y-$axis). The disc's aspect ratio is $h = 0.05$
    and its alpha viscosity is $\alpha_{\rm v} = 4\times 10^{-3}$. The
    right $y-$axis shows the Toomre Q-parameter at the planet
    location. The upper part of the plot is limited by the
    gravitational instability limit (dashed line). From \cite{mp03}.}
\end{figure}

Runaway migration is based on the planet's ability to build up a
coorbital mass deficit by opening a gap. It does not apply to low-mass
planets, for which $\delta m \ll \tilde{M}_p$.  It does not apply to
high-mass planets neither, which open a wide, deep gap, so that the
surface density of the separatrix-crossing flow is too small to
produce a significant mass deficit. It rather concerns
intermediate-mass planets, marginally satisfying the gap-opening
criterion in Eq.~(\ref{eq:gapcriterion}), in massive discs (the larger
the disc mass, the larger the density of the orbit-crossing flow). Its
occurrence is illustrated in Fig.~\ref{fig:occur} for a disc with
aspect ratio $h=5\%$ and alpha viscosity $\alpha_{\rm v} = 4\times
10^{-3}$, where we see that runaway migration may be particularly
relevant to Saturn-mass planets in massive discs (with a Toomre-Q
parameter at the planet's orbital radius typically less than about
10).  Bear in mind, however, that the occurrence for runaway migration
is sensitive to the values of $h$ and $\alpha_{\rm v}$, since they
affect the planet's ability to open a partial gap.  Also, note that
the criterion for runaway migration features the \emph{effective}
planet mass $\tilde{M}_p$, sum of the planet mass and the
circumplanetary disc mass.  The occurrence for runaway migration is
therefore sensitive to the mass distribution inside the
circumplanetary disc, which may be significantly affected by the
assumed physical modelling, e.g., whether gas accretion on the planet
is taken into account \cite{dangelo08}, the inclusion of self-gravity
\cite{ZhangLin08}, or the treatment for the gas thermodynamics
\cite{Peplinski1}. It may also be affected by grid resolution effects
in numerical simulations \cite{gda2005, Peplinski1}.

The simple model described above helps understand the condition for
migration to enter a runaway regime.  However, since it assumes steady
migration (that is, constant $\dot{a}$), this model is no longer valid
when migration actually enters the runaway regime, where the migration
rate increases exponentially over a time comparable to the horseshoe
libration period. A more general approach can be found in \cite{mp03,
  Oleron, Peplinski2}. As long as the orbital separation by which the
planet migrates over a libration period remains smaller than the
radial width of the planet's horseshoe region, $\Gamma_{\rm cross}$
remains approximately proportional to $\dot{a}$ (slow runaway
regime). At larger migration rates (fast runaway regime), $\Gamma_{\rm
  cross}$ reaches a maximum and slowly decreases with increasing
$\dot{a}$ \cite{mp03} (see also Fig.~16 in \cite{Oleron}). The precise
dependence of $\Gamma_{\rm cross}$ with $\dot{a}$ in this fast runaway
regime is intrinsically related to the evolution of the mass coorbital
deficit, and therefore to the planet's migration history.  The orbital
evolution of planets subject to runaway type III migration is
therefore difficult to predict. Numerical simulations find that,
depending on the resolution of the gas flow surrounding the planet,
the timescale for inward runaway type III migration can be as short as
a few $10^2$ orbits \cite{mp03,cbkm09}.

The sign of $\Gamma_{\rm cross}$ is dictated by the initial drift of
the planet. Runaway migration can therefore be directed inwards or
outwards, depending on the sign of $\dot{a}$ before the runaway takes
place. In particular, migration may be directed outwards if, despite
the coorbital region being partly depleted, a (positive) horseshoe
drag remains strong enough to counteract the (negative) differential
Lindblad torque. Outward runaway migration could thus be an attractive
mechanism to account for the recent discovery of massive planets at
large orbital separations (which we will further discuss in
\S~\ref{sec:far}). Simulations \cite{mp03, Oleron, Peplinski3} however
show that, despite the expected increase in the mass of the
orbit-crossing flow as the planet moves outwards (for background
surface density profiles shallower than $r^{-2}$), the mass coorbital
deficit cannot be retained indefinitely. The increase in the mass of
the circumplanetary disc, and the strong distortion of the flow within
it at large migration rates, lead the planet to eventually lose its
coorbital mass deficit, and the sense of migration is found to
reverse.

Type III migration has been recently revisited in low-viscosity discs
($\alpha_{\rm v} \leq {\rm a\;few} \times 10^{-4}$).  Depending on the
disc mass, the edges of the planet-induced gap may be subject to two
kinds of instabilities. In low-mass discs, gap edges are unstable to
vortex-forming modes \cite{lovelace99, li2000, minkai11_vortex}.  They
lead to the formation of several vortices sliding along the gap edges,
which merge and form large-scale vortices. When they pass by the
planet, these vortices may embark on horseshoe U-turns and exert a
large corotation torque on the planet, with the consequence that the
planet can be scattered inwards or outwards \cite{minkai10}. When the
fluid's self-gravity is taken into account, only a fraction of the
large-scale vortices actually embark on horseshoe U-turns, the rest of
the vortices keeps on sliding along the gap edges.  This provides a
periodic, intermittent corotation torque on the planet.  Depending on
the relative strengths of the vortices embarking on inward and outward
U-turns, this mechanism acts much like an intermittent type III
migration regime. In massive self-gravitating discs (stable against
the gravitational instability), vortex-forming modes are replaced by
global edge modes, which excite spiral density waves
\cite{minkai11_edge}.  A decreasing radial profile of the Toomre-Q
parameter favours edge modes at the gap's outer edge. The periodic
protrusion of edge mode-induced density waves near the gap's outer
edge provides a periodic source of (positive) corotation torque on the
planet, and induces an intermittent type III migration
regime. Numerical simulations by \cite{minkai12} show that edge modes
can sustain outward migration, until the planet leaves its gap.

We briefly sum up the main results of this paragraph:
\begin{itemize}
\item Migrating planets experience an additional corotation torque due
  to fluid elements flowing across the horseshoe region, and embarking
  on horseshoe U-turns. It is proportional to the planet's migration
  rate at small migration rates, which gives a positive feed back on
  migration. When the feedback loop diverges, the migration type is
  known as type III migration.
\item The planet and its circumplanetary disc feel an effective
  corotation torque that is proportional to the coorbital mass
  deficit, defined through Eq.~(\ref{cmd1}).  The occurrence of a
  runaway feedback ({\em i.e.} of type~III migration) corresponds to
  the coorbital mass deficit exceeding the mass of the planet and its
  circumplanetary disc. This applies to planets opening a partial gap
  around their orbit in massive protoplanetary discs.
\item The orbital evolution of planets undergoing type III migration
  is sensitive to the time evolution of the coorbital mass deficit,
  which makes it difficult to predict. Numerical simulations show that
  runaway migration operates on very short timescales, typically in
  100 to 1000 planet orbits.
\end{itemize} 

\subsection{Deep gap-opening: Type II migration}
\label{sec:type2}

\begin{figure}
\sidecaption
 	\includegraphics[width=0.5\hsize]{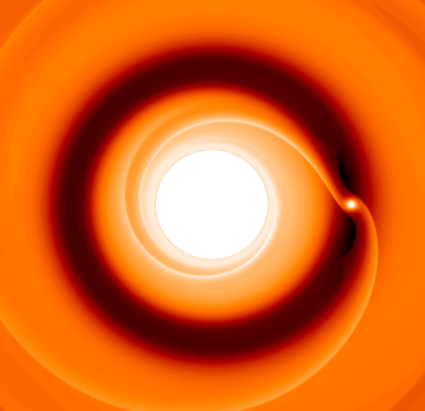}
        \caption{Gap opened by a Jupiter-mass planet orbiting a Sun-like star.}
  \label{fig:gap}
\end{figure}
Planets massive enough to clear their coorbital region and open a deep
gap around their orbit (see illustration in Fig.~\ref{fig:gap}) enter
the migration regime called Type II migration. Such planets satisfy
the gap-opening criterion given by
Eq.~(\ref{eq:gapcriterion}). Assuming, for instance, a protoplanetary
disc with aspect ratio $\sim 5\%$ and alpha viscosity $\alpha_{\rm v}
\sim 10^{-2}$, type II migration typically applies to planets more
massive than Jupiter orbiting Sun-like stars. Compared to the type I
and type III migration regimes described previously, the amplitude of
the corotation torque is much reduced due to the clearing of the
planet's coorbital region, and the differential Lindblad torque
balances the viscous torque exerted by the disc. The net torque on the
planet can be written as a fraction $C_{\rm II}$ of the viscous torque
due to the outer disc \cite{cm07}. This factor $C_{\rm II}$ features
the time-dependent fraction of gas $f_{\rm gas}$ remaining in the
planet's coorbital region.

The particular case with $f_{\rm gas}$ going to zero corresponds to
what is usually referred to as standard type II migration regime. Its
timescale can be approximated as
  \begin{equation}
    \tau_{\rm II} \approx \frac{2r_{\rm o}^2}{3\nu(r_{\rm o})} 
    \left( 1 + \frac{M_p}{4\pi \Sigma(r_{\rm o}) r_{\rm o}^2} \right),
  \label{tauII_a}	 
  \end{equation}
  where $\nu$ denotes the disc's kinematic viscosity, $\Sigma$ the
  surface density of the disc perturbed by the planet, and $r_{\rm o}$
  is the location in the outer disc where most of the planet's angular
  momentum is deposited. It can be approximated as the location of the
  outer separatrix of the planet's horseshoe region, which, for
  gap-opening planets, is $r_{\rm o} \approx a + 2.5 R_{\rm H}$
  \cite{mak2006}.  The first term on the right-hand side of
  Eq.~(\ref{tauII_a}) corresponds to the viscous drift timescale at
  radius $r_{\rm o}$, and the second term features the ratio of the
  planet mass to the local disc mass at radius $r_{\rm o}$. Two
  migration regimes can therefore be distinguished:
  \begin{enumerate}
  \item \emph{Disc-dominated type II migration}.  When the planet mass
    is much smaller than the local disc mass (by which we refer to the
    quantity $4\pi \Sigma(r_{\rm o}) r_{\rm o}^2$), the planet behaves
    much like a fluid element that the disc causes to drift
    viscously. The planet's migration timescale then matches the
    disc's viscous drift timescale $\approx 2r_{\rm o}^2 / 3\nu(r_{\rm
      o})$ \cite{lp86}. In this migration regime, called
    disc-dominated type II migration, the planet remains confined
    within its gap. Should the planet migrate slightly faster than the
    disc near its orbit, the increased inner Lindblad torque due to
    the planet getting closer to the gap's inner edge would push the
    planet outward. Conversely, should the planet migrate at a slower
    pace than the (local) disc, the increased outer Lindblad torque
    would push the planet back inward. The timescale for the
    disc-dominated type II migration regime, $\tau_{\rm II, d}$, can
    be recast as
  \begin{equation}
    \tau_{\rm II, d}
    \approx  4.7\times 10^{4} {\rm yrs}\times\left( \frac{\alpha_{\rm v}}{10^{-2}} \right)^{-1}\left( \frac{h}{0.05} \right)^{-2} \left( \frac{M_{\star}}{M_{\odot}} \right)^{-1/2}\left( \frac{r_{\rm o}}{\rm 5\;AU} \right)^{3/2},
  \label{tauIId}
\end{equation}
where $\alpha_{\rm v}$ and $h$ are to be evaluated at $r_{\rm o}$.
\smallskip
\item \emph{Planet-dominated type II migration}.  When the planet mass
  becomes comparable to, or exceeds the local disc mass, the orbital
  evolution of a gap-opening planet is no longer dictated by the disc
  alone. The inertia of the planet slows down its orbital migration
  \cite{sc95,ipp99}, and in the limit when the planet mass is large
  compared to the local disc mass, the planet enters the so-called
  planet-dominated type II migration regime, whose timescale
  $\tau_{\rm II, p}$ reads
  \begin{equation}
    \tau_{\rm II, p} \approx \tau_{\rm II, d} \times \left( \frac{M_{\rm p}}{4\pi \Sigma(r_{\rm o}) r_{\rm o}^2} \right),
\label{tauIIp}
\end{equation}
with $\tau_{\rm II, d}$ given by Eq.~(\ref{tauIId}), and 
where the planet to local disc mass ratio reads
\begin{equation}
  \frac{M_p}{4\pi \Sigma(r_{\rm o}) r_{\rm o}^2} 
  \approx  200
  \left( \frac{M_p}{M_{\odot}} \right) 
  \left( \frac{\Sigma(r_{\rm o})}{150\;{\rm g\;cm}^{-2}} \right)^{-1}
  \left( \frac{r_{\rm o}}{\rm 5\;AU} \right)^{-2}.
  \label{msat}
\end{equation}
\end{enumerate}
 
In self-gravitating discs, the distinction between the planet- and
disc-dominated type II migration regimes should involve the comparison
between the local disc mass and the effective planet mass
$\tilde{M}_p$, that is the sum of the planet and circumplanetary disc
masses.  Related to this point, we comment that the planet and its
circumplanetary disc migrate at the same drift rate. When its
self-gravity is included, the protoplanetary disc torques both the
planet and the circumplanetary disc. However, if self-gravity is
discarded, as is usually the case in numerical simulations, the
protoplanetary disc can only torque the planet, and the
circumplanetary disc remains a passive spectator of the migration. In
this case, the planet must exert an additional effort to maintain the
planet and circumplanetary disc joint migration. Put another way, the
circumplanetary disc artificially slows down migration when
self-gravity is discarded. To avoid this artificial slowdown,
\cite{cbkm09} showed that, in simulations discarding self-gravity, the
calculation of the torque on the planet must exclude the material
inside the circumplanetary disc. In addition, migration rates with and
without self-gravity can be in close agreement, provided that the mass
of the circumplanetary disc is added to that of the planet when
calculating the gravitational potential felt by the protoplanetary
disc.
 
In the early stages of their formation and orbital evolution, most
massive gap-opening planets should be subject to disc-dominated type
II migration, and migrate on a timescale comparable to the disc's
viscous timescale. Note from Eq.~(\ref{tauII_a}) that this corresponds
to the shortest migration timescale a gap-opening planet can
get. Depletion of the protoplanetary disc, or substantial migration
towards the central object, should, however, slow down migration as
the planet's inertial mass becomes comparable to the local disc
mass. It is nonetheless interesting to note from Eq.~(\ref{tauIId})
that in the early stages of the disc evolution, type II migration can
be relatively fast in the disc's turbulent parts. This may make
difficult the maintenance of massive planets at reasonably large
orbital separations from their host star. Additional mechanisms, like
the effect of stellar irradiation on the disc's density and
temperature profiles near the planet's orbit (formation of shadow
regions near the gap's inner edge, and irradiated 'puffed up' regions
near the gap's outer edge) \cite{JangCondell08} could help slow down
type II migration.

Gap formation and type II migration are intimately related to the disc
viscosity in laminar viscous disc models, and a few studies have
investigated their properties in MHD turbulent discs \cite{qmwmhd2,
  Winters03, qmwmhd3}. These studies have considered three-dimensional
magnetised disc models, where vertical stratification and non-ideal
MHD effects were discarded for simplicity. They found that the
structure of the annular gap opened by a massive planet in fully MHD
turbulent discs is essentially in line with the predictions of viscous
disc models with a similar alpha viscous parameter near the planet
location. Gaps in turbulent discs tend, however, to be wider than in
viscous discs \cite{qmwmhd2, Winters03}. Some other differences arise
between turbulent and viscous disc models, particularly in the
vicinity of the planet, where magnetic field lines are compressed and
ordered at the location of the wakes and the circumplanetary disc. The
connection between the circumplanetary and protoplanetary discs
through magnetic field lines can cause magnetic braking of the
circumplanetary material \cite{qmwmhd3}, which may help increase gas
accretion onto the planet \cite{Winters03, qmwmhd3}.

Before leaving this section, we comment that the overall properties of
planet--disc interactions with gap-opening planets remain essentially
unchanged when taking the disc's vertical stratification into account,
and therefore the two-dimensional approximation is valid. Nonetheless,
different structures in the flow circulating around the planet in two-
and three-dimensions, and the related accretion rate onto the planet,
may affect the planet's migration rate (see e.g., \cite{gda2005}).

We summarise the results described in this section:
\begin{itemize}
\item Planets massive enough to open a wide and deep annular gap
  around their orbit are subject to Type II migration.
\item When the local disc mass (roughly speaking, the mass interior to
  the planet orbit) remains large compared to the planet mass, the
  planet migration timescale corresponds to the viscous drift
  timescale (disc-dominated type II migration).
\item When the planet mass becomes comparable to, or exceeds the local 
disc mass, migration is slowed down by the planet's inertia (transition to 
planet-dominated type II migration).
\end{itemize} 

\section{Planet migration theories and exoplanets' observed diversity}
\label{sec:applications}
The properties of planet--disc interactions have been examined in
details in the previous sections, with a particular emphasis on the
expected migration rate for planets with different masses. The
migration rate is intimately related to the disc's physical properties
(e.g., mass, sound speed, cooling properties, turbulent stresses) near
the planet location, which underlines that the modelling of
protoplanetary discs plays as much of an important role as
planet--disc interaction theories in predicting the evolution of
planetary systems. We continue our exposition with a brief discussion
of several aspects of planet--disc interactions which could account
for the observed diversity of (extrasolar) planets.

\subsection{Massive planets at large orbital separations}
\label{sec:far}
Amongst the recent discoveries of exoplanets, particularly exciting is
the observation by direct imaging of about 10 massive exoplanets
located at separations ranging from 10 to 200 AU from their host star
(e.g., \cite{Marois08, Lagrange10}).  Most of these planets are so far
observed to be the only planetary companions of their host star. Yet,
it is possible that their present location results from a scattering
event with another massive companion on a shorter-period orbit. A
remarkable exception is the HR 8799 planetary system. It comprises
four planets with masses evaluated in the range $[7-10]$ Jupiter
masses, and estimated separations of 14, 24, 38 and 68 AU
\cite{Marois10}. The planets are close to being in mutual mean-motion
resonances, and it seems likely that planet--disc interactions could
have played a major role in shaping this planetary system. We discuss
below the relevance of planet--disc interactions to account for
massive planets at large orbital separations.

\subsubsection{Outward migration of a pair of massive resonant planets}
In the standard core-accretion scenario for planet formation, it is
difficult to form Jupiter-like planets in isolation further than $\sim
10$ AU from a Sun-like star \cite{Pollack96, IdaLin1}. As we have seen
in \S~\ref{sec:gapopening}, planets in the Jupiter-mass range orbiting
Solar-type stars are expected to open an annular gap around their
orbit. If a partial gap is opened, outward runaway type III migration
could occur under some circumstances, but as we have discussed in
\S~\ref{sec:type3}, numerical simulations indicate that it is
difficult to sustain this outward migration in the long term. If the
planet opens a deep gap, inward type II migration is expected. It is
therefore unlikely that a single massive planet formed through the
core-accretion scenario within $\sim 10$ AU of its host star could
migrate to several tens of AUs.

\begin{figure}
 	\includegraphics[width=0.5\hsize]{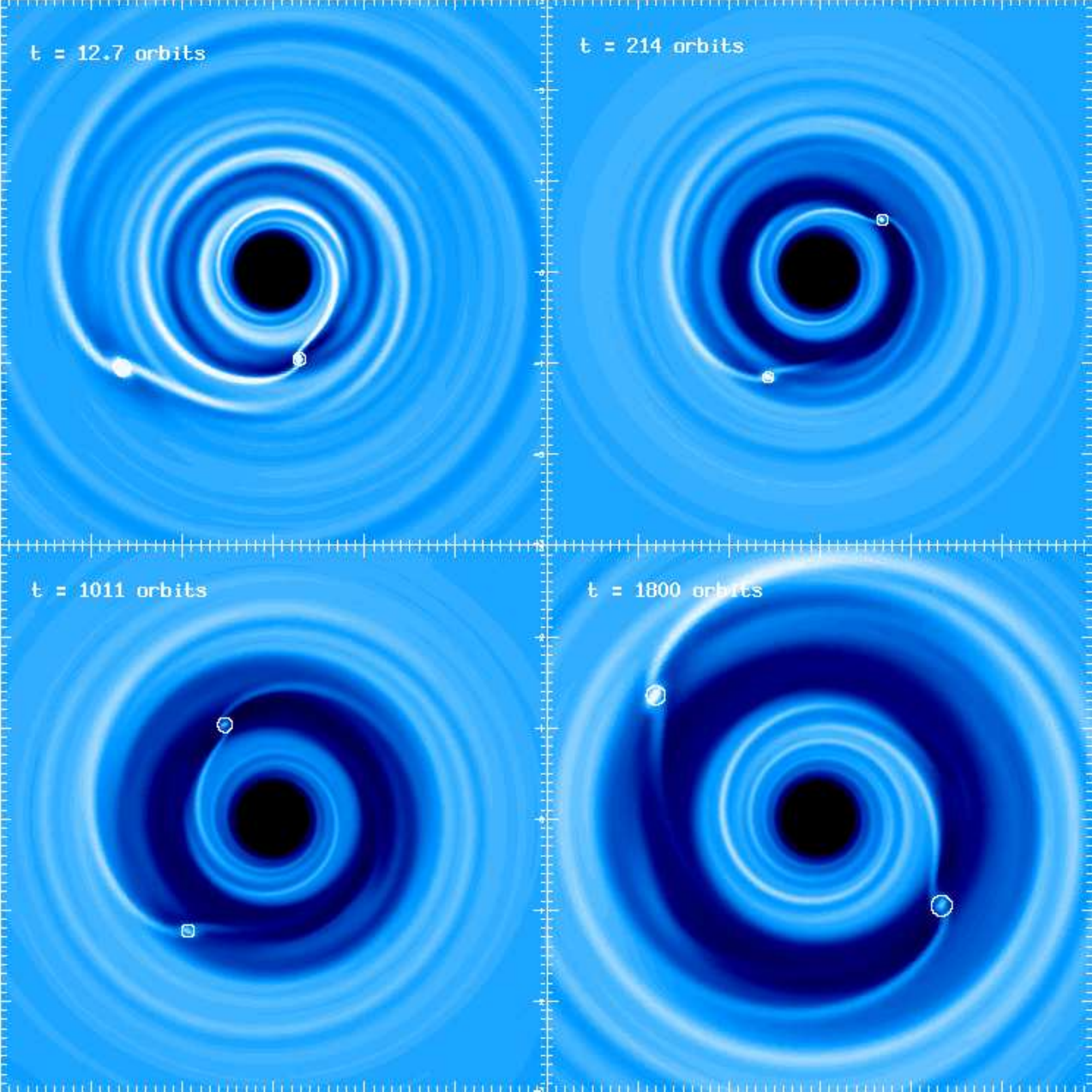}
	\includegraphics[width=0.5\hsize]{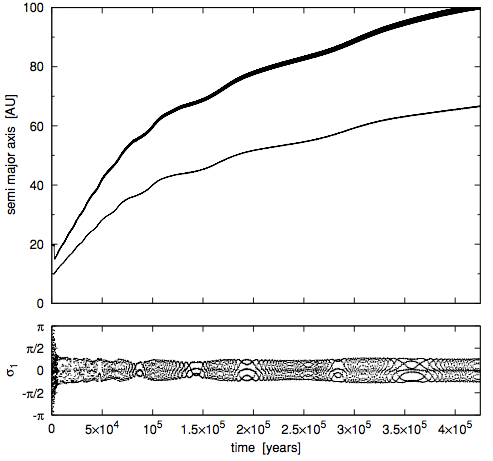}
        \caption{Illustration of the joint outward migration of a pair
          of resonant massive planets. The left panel shows the
          evolution of the disc's surface density perturbed by a
          Jupiter-mass planet (inner planet) and a Saturn-mass planet
          (outer one). After an episode of rapid convergent migration
          (top-left quadrant) resulting in their capture into
          mean-motion resonance, planets open overlapping gaps
          (top-right quadrant), which leads to their joint outward
          migration (lower quadrants). The right panels illustrate the
          outcome of the same mechanism applied to an inner 3-Jupiter
          mass planet, and an outer 2-Jupiter-mass planet orbiting a
          $2 M_{\odot}$ mass star (taken from \cite{Crida09}). The
          time evolution of the planets semi-major axis is shown in
          the top-right panel, and that of their $2:1$ critical
          resonance angle is in the bottom-right panel.  }
  \label{fig:massetsnell}
\end{figure}

A notable exception to this generally expected trend has been recently
proposed by \cite{Crida09}, based on a migration mechanism originally
studied by \cite{massnel2001}. This mechanism relies on the joint
migration of a pair of resonant massive planets embedded in a common
gap. In this mechanism, the innermost planet is massive enough to open
a deep gap and migrate inwards on a timescale comparable to that of
type II migration.  The outermost, less massive planet migrates
inwards at a larger pace while carving a partial gap around its
orbit. If both planets open overlapping gaps, and maintain a
mean-motion resonance between their orbits, their joint migration
\emph{could} proceed outwards. The global picture is the following: as
the inner planet is more massive, the torque it experiences from the
inner disc (inner Lindblad torque) is larger than the (absolute value
of the) torque the outer planet experiences from the outer disc (outer
Lindblad torque). To maintain joint outward migration in the long
term, the fluid elements outside the common gap must be funnelled to
the inner disc by embarking on horseshoe trajectories.  Otherwise,
material would pile up at the outer edge of the common gap, much like
a snow-plough, and the torque balance as well as the sense of
migration would eventually reverse. An illustration of the joint
outward migration mechanism is shown in Fig.~\ref{fig:massetsnell}.

The migration reversal described above requires an asymmetric density
profile within the common gap. It is thus sensitive to the disc's
aspect ratio and viscosity, which enter the gap-opening criterion. It
is also sensitive to the mass ratio of the two planets.  If the
outer-to-inner planets mass ratio is too small, the density contrast
within the common gap will be too large to affect the evolution of the
innermost planet (the gas density near the outer planet's orbit
remains too large to significantly decrease the outer Lindblad torque
acting on the inner planet). Conversely, if the outer-to-inner planets
mass ratio is too large, the Lindblad torques imbalance will favour
joint inward migration. By changing the planets mass ratio during
joint migration, gas accretion onto the planets could affect the
possibility of sustaining outward migration in the long term. This
issue requires further investigation, and accurate modelling of the
gas accretion processes onto Saturn sized planets.

We also mention that the joint outward migration scenario has been
recently discussed in the context of the Solar System \cite{Walsh11}.
Inward migration of Jupiter in the primordial Solar nebula down to
$\approx 1.5$ AU, followed by joint outward migration with Saturn to
the current location of both planets (the "Grand tack") would truncate
the disc of planetesimals interior to Jupiter's orbit at about 1
AU. The subsequent formation of the terrestrial planets is found to
occur with the correct mass ratio between Earth and Mars, and would
also account for the compositional structure of the asteroid belt
\cite{Walsh11}.

\subsubsection{Migration of planets formed by gravitational instability}
\begin{figure}
 	\includegraphics[width=0.5\hsize]{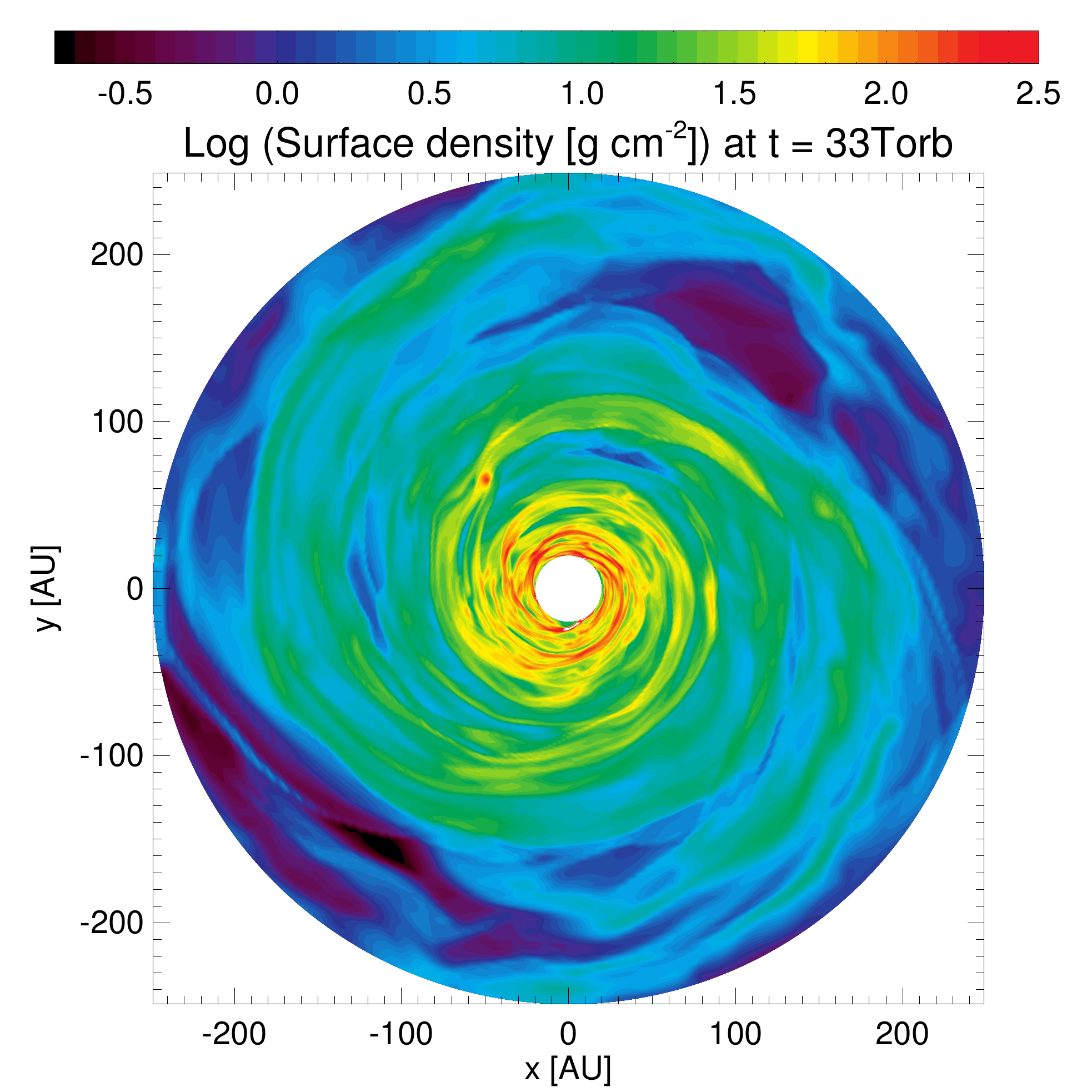}
	\includegraphics[width=0.5\hsize]{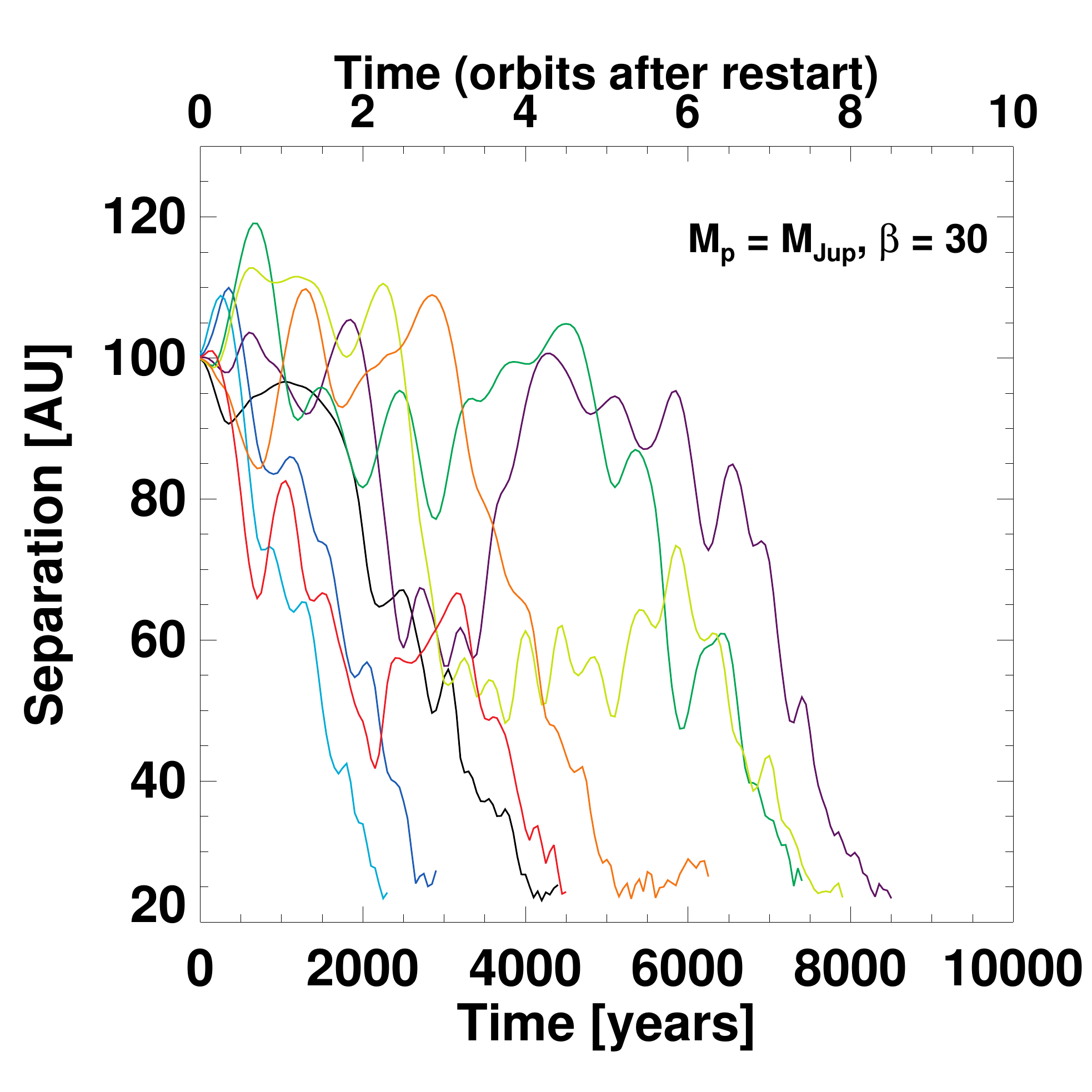}
        \caption{Jupiter-mass planet embedded in a gravito-turbulent
          disc. After setting up a quasi steady-state
          gravito-turbulent disc with (gravito-turbulent) shock
          heating balancing disc cooling (parametrised here by a
          simple $\beta$-cooling function, see \cite{bmp11}),
          simulations were restarted with inserting a Jupiter-mass
          planet at 100 AU. The left panel shows the disc's surface
          density three orbits after restart, the planet being located
          at $x \sim -50$ AU, $y \sim 60$ AU.  The right panel
          displays the time evolution of the planet's orbital
          separation in 8 different restart simulations with varying
          the azimuth of the planet prior to its insertion in the
          disc.  Taken from \cite{bmp11}.}
  \label{fig:bmp11}
\end{figure}

An alternative to the core-accretion formation scenario involves the
fragmentation of massive protoplanetary discs into clumps through the
gravitational instability. Gravitational instability (GI) may
typically occur at separations larger than 30 to 50 AU from a central
(Sun-like) star, if the Toomre-$Q$ parameter approaches unity and the
disc's cooling timescale becomes of order the dynamical timescale
(e.g., \cite{Rafikov05,2001ApJ...553..174G}). While several massive
planets could form by fragmentation of a massive disc at several tens
of AUs from their star, they are unlikely to stay in place. The tidal
interaction with the gravito-turbulent disc they are embedded in
should rapidly bring planets formed by GI to the disc's inner regions
\cite{bmp11, Michael11, zhu12}, in a timescale comparable to that of
type I migration \cite{bmp11}.  The orbital evolution of a single
Jupiter-mass planet embedded in a gravito-turbulent disc (where the
planet is supposed to have formed by GI) is illustrated in
Fig.~\ref{fig:bmp11}, where we see that the planet migrates from 100
to 20 AU in typically less than $10^4$ yrs.

Investigation is under way to determine the evolution of planets
formed by GI when they reach the inner parts of protoplanetary
discs. The latter should be too hot to be gravitationally unstable,
and other sources of turbulence, such as the magnetorotational
instability, could prevail, changing the background disc profiles as
well as the amount of turbulence. It is thus possible that the rapid
type I migration of planets formed by GI slows down in the disc inner
parts and results in the formation of a gap. Gap-opening may also
occur if significant gas accretion occurs during the planets fast
inward migration \cite{zhu12}. Planet--planet interactions, which may
result in scattering events, mergers or captures in resonance, should
also play a prominent role in shaping planetary systems formed by GI.

\subsection{Planet population syntheses}
Planetary astrophysics is undergoing an epoch of explosive growth,
driven by the observational discoveries of more than 750 exoplanets
over the past two decades. Outstanding progress in detection
techniques have uncovered planetary systems very different from
ours. Since the discovery of the first Hot Jupiter \cite{MQ95}, radial
velocity surveys have made possible the detection of Earth-like
planets, some in the habitable zone of their star
\cite{Pepe11}. Transit space missions CoRoT and KEPLER are digging out
hundreds of close-in extrasolar planets, some in exotic environments
(like Kepler-16 b, the first circumbinary exoplanet discovered
\cite{Kepler16b}). Direct imaging has revealed the existence of
massive giant planets located at several tens of AU from their star.

Such diversity provides an exciting opportunity to test our theories
for the formation and evolution of planetary systems.  By coupling
theoretical models of planet formation and migration, and of disc
evolution, planet population syntheses estimate the statistical
distribution of exoplanets according to their mass, semi-major axis,
and eccentricity, which they compare to observed distributions
\cite{IdaLin1, IdaLin4, Mordasini09b, sli09, HellaryNelson12}.  At the
moment, models of planet population syntheses are not able to
reproduce the statistical properties of extrasolar planets. For
instance, they predict a deficit of super-Earths and Neptune-like
planets with orbital periods less than 50 days, while observations
have revealed a significant number of exoplanets in this range of mass
and period \cite{Howard10}.  The origin for this discrepancy can be
found in uncertain prescriptions for the minimum core mass for the
onset of gas accretion \cite{HellaryNelson12}, as well as in the
modeling of type I migration.  The difficulty raised by the
excessively rapid inward type I migration, predicted by the long-time
reference torque formula by \cite{tanaka2002}, was circumvented by
introducing a reduction factor in front of this torque formula.
Population syntheses models tried to constrain this factor to
reproduce the statistical properties of detected exoplanets. This
reduction factor was found to range from $0.01$ to $0.1$.  We note
however that the introduction of this reduction factor to provide
planetary population synthesis in agreement with the statistics of
detected extrasolar planets is {\em ad hoc}, and that there is no
reason to expect that the type~I migration drift rates are
systematically overestimated in theoretical studies by a factor $10$
to $100$. Rather, as we emphasized throughout this manuscript, type~I
migration is very sensitive to the disc's density and temperature
profiles near the planet orbit.  Large slopes of mass density and/or
temperature, over a limited radial range, may reverse the tidal torque
exerted on the planet. This, in turn, may create ``planetary traps''
at the points where the tidal torque cancels out (much like what was
contemplated by \cite{masset06a} for the case of a positive surface
density gradient), which can stop incoming protoplanets, depending on
their mass. The number and location of these traps may vary as the
disc evolves.  This view of type~I migrating objects subject to
several traps on their way to the star \cite{2010ApJ...715L..68L,hp10}
sounds more compatible with the state of migration theories than an
{\em ad hoc} reduction factor.  This has motivated recent works to
produce accurate, yet simple formulae for type I migration \cite{mc10,
  pbk11,2011CeMDA.111..131M}. These formulae include a description of
the corotation torque in discs with arbitrary viscosity and thermal
diffusion, and corrections to the Lindblad torque for discs with non
power-law profiles. Their incorporation into models of planet
population synthesis will hopefully provide a better comprehension of
the diversity of observed exoplanets.

\section{Conclusions}
We have reviewed the recent progress made in understanding
planet--disc interactions, and the resulting planets' orbital
migration. We have particularly focused on the migration of growing
protoplanets (type I migration), which has been the subject of
intensive investigation over the past five years. Being for a while
the second-place actor of planet migration theories, the corotation
torque has been shown to play a prominent role in realistic
protoplanetary discs, where it can slow down, stall, or reverse type I
migration. This review is especially aimed at giving a comprehensive,
detailed description of the mechanisms responsible for the corotation
torque. The type II and type III migration regimes for gap-opening
planets are also reviewed and discussed in the context of observed
exoplanets. Being aimed at migration of planets on circular orbits,
this review has set aside interesting recent developments on the tidal
interactions of eccentric or inclined planets with their discs. We
have also focused essentially on the mechanisms that drive the
migration of a single planet in a disc, and we have therefore excluded
most results about the migration of several planets.  For a recent
review covering these topics, the reader is referred to
\cite{KN_review12}. This list of restrictions of the present review
stresses that the research on planet--disc interactions is a very
active branch of planet formation, with a growing body of avenues. We
finally reiterate the plea made in the introduction: planetary
migration is not overrated. The tremendous value of each of the tidal
torque components exerted on a given planet, associated to the great
sensitivity of these torques to the underlying disc structure, appeals
for a detailed knowledge of the properties of protoplanetary discs,
and significant efforts toward an accurate determination of each
torque component. This also reasserts tidal interactions as a
prominent process in shaping forming planetary systems.

\section*{Acknowledgments}
It is a pleasure to thank J{\'e}r{\^o}me Guilet, Sijme-Jan
Paardekooper, and Stephen Thomson for their detailed reading of this
manuscript, as well as John Papaloizou for stimulating discussions.

\bibliographystyle{spmpsci}

\end{document}